\documentclass[sigconf]{acmart}
\renewcommand\footnotetextcopyrightpermission[1]{} 
\setcopyright{none}

\usepackage[normalem]{ulem}

\usepackage{dblfloatfix}

\usepackage[us,12hr]{datetime}
\newcommand{\ignore}[1]{}
\usepackage[normalem]{ulem}

\usepackage{moresize}

\usepackage{soul,color}
\usepackage{colortbl} 
\soulregister\cite7
\soulregister\ref7
\soulregister\pageref7

\DeclareRobustCommand{\hlcyan}[1]{{\sethlcolor{SkyBlue}\hl{#1}}}
\DeclareRobustCommand{\hlgreen}[1]{{\sethlcolor{YellowGreen}\hl{#1}}}

\sethlcolor{Apricot}

        \DeclareRobustCommand{\hlcyan}[1]{{\sethlcolor{white}\hl{#1}}}
        \DeclareRobustCommand{\hlgreen}[1]{{\sethlcolor{white}\hl{#1}}}
        
        \sethlcolor{white}

\usepackage{amsmath}
\usepackage{amsfonts}
\usepackage[normalem]{ulem}
\usepackage{flushend}
\usepackage{pbox}
\usepackage{rotating}
\usepackage{url}
\usepackage{breakurl}
\usepackage{courier}
\usepackage{pgfplots}
\usepackage{enumitem}
\usepackage{booktabs}
\usepackage{epstopdf}
\usepackage{graphicx}
\usepackage{subfig}

\newcommand{\macb}[1]{\textbf{\textsf{#1}}.}

\usepackage{color}
\definecolor{bluekeywords}{rgb}{0.13,0.13,1}
\definecolor{greencomments}{rgb}{0,0.5,0}
\definecolor{redstrings}{rgb}{0.9,0,0}
\definecolor{grey}{rgb}{0.4,0.4,0.4}
\definecolor{lightgrey}{RGB}{230,230,230}

\usepackage{multirow}
\usepackage{listings}
\usepackage[font={small,sf,bf}]{caption}
\usepackage[utf8]{inputenc}
\usepackage{cleveref}
\crefname{section}{§}{§§}
\Crefname{section}{§}{§§}

\newcommand{\Mod}[1]{\ \text{mod}\ #1}

\usepackage[absolute,overlay]{textpos}
\newcommand{\goal}[1]{\textcolor{red}{[GOAL: #1]}}
\newcommand{\htor}[1]{\textcolor{green}{[HTOR: #1]}}

\newcounter{maccounter}
\newcommand{\maciej}[1]{
\protect\stepcounter{maccounter}
\textcolor{blue}{[MACIEJ \#\arabic{maccounter}: #1]}}

 \newcommand{\commt}[1]{\textcolor{blue}{[COMMENT: #1]}}
 \newcommand{\gcommt}[1]{\textcolor{black}{\newline[COMMENT: #1]}}
 \newcommand{\change}[1]{\textcolor{BrickRed}{\newline[CHANGE: #1]}}

\renewcommand{\goal}[1]{}
\renewcommand{\htor}[1]{}
\renewcommand{\maciej}[1]{}

      \renewcommand{\commt}[1]{}
      \renewcommand{\gcommt}[1]{}
      \renewcommand{\change}[1]{}

\definecolor{BrickRed}{RGB}{80,25,33}

\newboolean{publicversion}
\setboolean{publicversion}{true}

\ifthenelse{\boolean{publicversion}}{
  \pagenumbering{gobble}

}{
  \pagenumbering{arabic}

}

\newtheorem{thm}{Theorem}

\usepackage{cellspace}
\usepackage{makecell}

\begin{document}

\title[Slim NoC: A Low-Diameter On-Chip Network Topology]{Slim NoC: A Low-Diameter On-Chip Network Topology\\for High Energy Efficiency and Scalability}

\author{Maciej Besta$^1$, Syed Minhaj Hassan$^2$, Sudhakar Yalamanchili$^2$,\break Rachata Ausavarungnirun$^3$, Onur Mutlu$^{1,3}$, Torsten Hoefler$^1$}
       \affiliation{\vspace{0.3em}$^1$ETH Zurich, $^2$Georgia Institute of Technology, $^3$Carnegie Mellon University\\
}

\begin{abstract}
Emerging chips with hundreds and thousands of cores require networks with
unprecedented energy/area efficiency and scalability. To address this, we
propose Slim NoC (SN): a new on-chip network design that delivers significant
improvements in efficiency and scalability compared to the state-of-the-art.
The key idea is to use two concepts from graph and number theory,
degree-diameter graphs combined with non-prime finite fields, to enable the
smallest number of ports for a given core count. SN is inspired by state-of-the-art
off-chip topologies; it identifies and distills their advantages for NoC
settings while solving several key issues that lead to significant overheads
on-chip. SN provides NoC-specific layouts, which further enhance
area/energy efficiency. We show how to augment SN with state-of-the-art router
microarchitecture schemes such as Elastic Links, to make the network even more
scalable and efficient.  Our extensive experimental evaluations show that SN
outperforms both traditional low-radix topologies {(e.g., meshes and tori) and modern
high-radix networks (e.g., various Flattened Butterflies)} in
area, latency, throughput, and static/dynamic power consumption for both
synthetic and real workloads. SN provides a promising
direction in scalable and energy-efficient NoC topologies.
\end{abstract}


%
%


\maketitle
\pagestyle{plain}

{\noindent \textbf{This is a full version of a paper published at\\ ACM ASPLOS'18 under the same title}}

\section{Introduction}
\label{sec:intro}


\commt{(§1) In this section, we made the text much less dense and more tractable. We also
add more concrete examples of the use cases of Slim NoC, its key advantages, etc.}


\change{(§1) We added one more example of a massively parallel chip, namely
PEZY-SC2.}

\hl{Massively parallel manycore networks are becoming the
base of today's and future computing systems. Three examples of such systems are: (1)
SW26010, a 260-core processor used in the world's fastest ($\approx$93
petaflops in the LINPACK benchmark~\cite{dongarra1979linpack}) supercomputer Sunway TaihuLight~\cite{fu2016sunway}; (2)
PEZY-SC2~\cite{pezy}, a Japanese chip with 2048 {nodes} used
in the ZettaScaler-2.2 supercomputer; (3) Adapteva
Epiphany~\cite{olofsson2016epiphany}, a future processor with 1024 cores.}

To accommodate such high-performance systems, one needs
high-performance and energy-efficient networks on a chip (NoCs). A desirable network
is both \emph{high-radix} (i.e., its routers have many ports) and
\emph{low-diameter} (i.e., it has low maximum distance between nodes) as such networks offer
low latency and efficient on-chip wiring density~\cite{dally07}.
%
Yet, combining high radix and low diameter in a NoC leads to long
inter-router links that may span the whole die. To fully utilize such long
links, buffers must be enlarged in proportion to the link length. Large buffers are 
power-hungry~\cite{bless-rtr} and they may hinder the scalability of different
workloads. Moreover, using more ports further increases the router buffer area,
leaving less space for cores and caches for a fixed die size. We aim to solve
these issues and preserve the advantages of high radix and low diameter to
enable more energy-efficient and scalable NoCs.

\change{(§1) We provide a more clear description of the key idea of Slim Fly.}


\hl{In this work, we first observe that some 
state-of-the-art
low-diameter \emph{off-chip} networks \emph{may} be excellent NoC candidates
that can effectively address the above area and power issues. We investigate
Dragonfly (\texttt{DF}) and Slim Fly (\texttt{SF}), two modern topologies designed for datacenters and
supercomputers, that offer cost-effectiveness and performance advantages
for various classes of distributed-memory workloads~\cite{besta2015accelerating, schmid2016high, solomonik2017scaling, besta2015active, besta2014fault, gerstenberger2013enabling}. Dragonfly~\cite{dally08} is a high-radix
\mbox{diameter--3} network that has an intuitive layout and it reduces the
number of long, expensive wires. It is less costly than torus, Folded
Clos~\cite{leiserson1985fat, Scott:2006:BHC:1135775.1136488}, and Flattened
Butterfly~\cite{dally07} (\texttt{FBF}).
%
%
Slim Fly~\cite{slim-fly}\footnote{We consider the MMS variant of
Slim Fly as described by Besta and Hoefler~\cite{slim-fly}. For details, please
see the original Slim Fly publication~\cite{slim-fly}.} has lower cost and
power consumption. 
%
%
First, \texttt{SF} lowers diameter
and thus average path length so that \emph{fewer costly switching resources are
needed}. Second, to ensure high performance, \texttt{SF} is based on graphs
that approximate solutions to the degree-diameter problem, well-known in graph
theory~\cite{mckay98}.
%
%
These properties suggest that \texttt{SF} may also be a good NoC candidate.}

\change{(§1) We offer a more precise explanation of what the following means:
``inappropriate metrics for NoCs'', ``NoC-optimized cost models''.}

Unfortunately, as we show in \mbox{\cref{sec:motivation}}, naively using
\texttt{SF} or \texttt{DF} as NoCs consistently leads to significant overheads
in performance, power consumption, and area. We analyze the reasons behind
these challenges in \mbox{\cref{sec:motivation}}. To overcome them, and thus enable
\emph{high-radix} and \emph{low-diameter} NoCs, we propose Slim NoC (\texttt{SN}), a new family
of \emph{on-chip} networks inspired by \texttt{SF}.


\change{(§1) We inserted more forward-references to facilitate the navigation
of the paper. We emphasize the key ideas and the direction of our work more.}

\begin{figure*}[t]
\centering
 \subfloat[Latency (an adversarial traffic pattern)]{
  \includegraphics[width=0.225\textwidth]{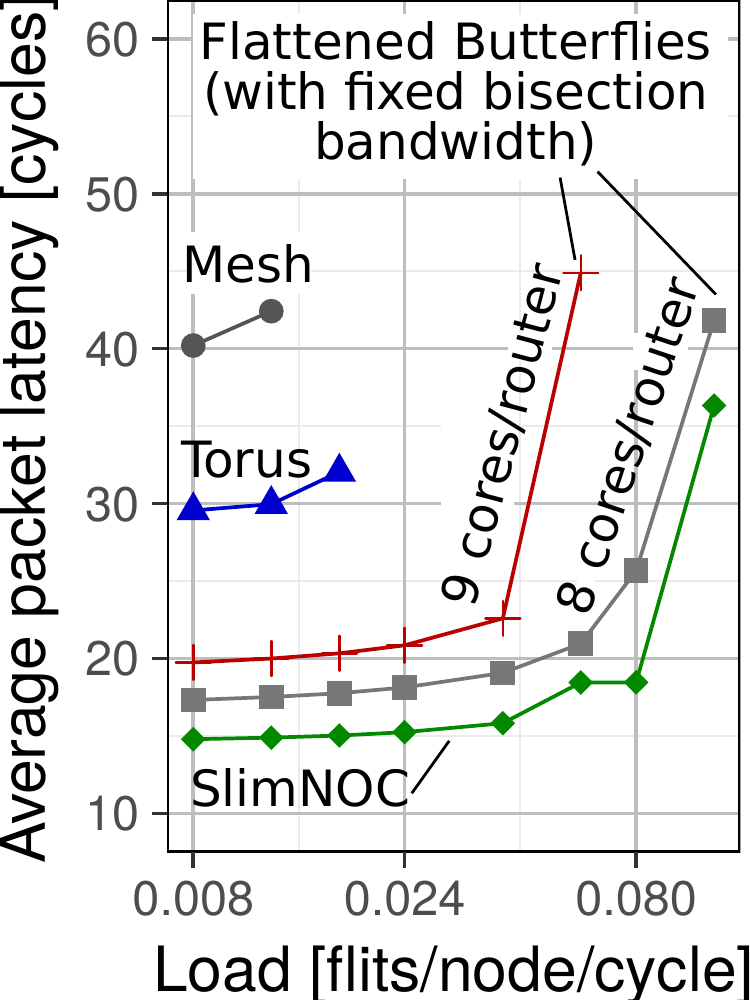}
 }\hspace{0.3em}
 \subfloat[Throughput per power (45nm)]{
  \includegraphics[width=0.1875\textwidth]{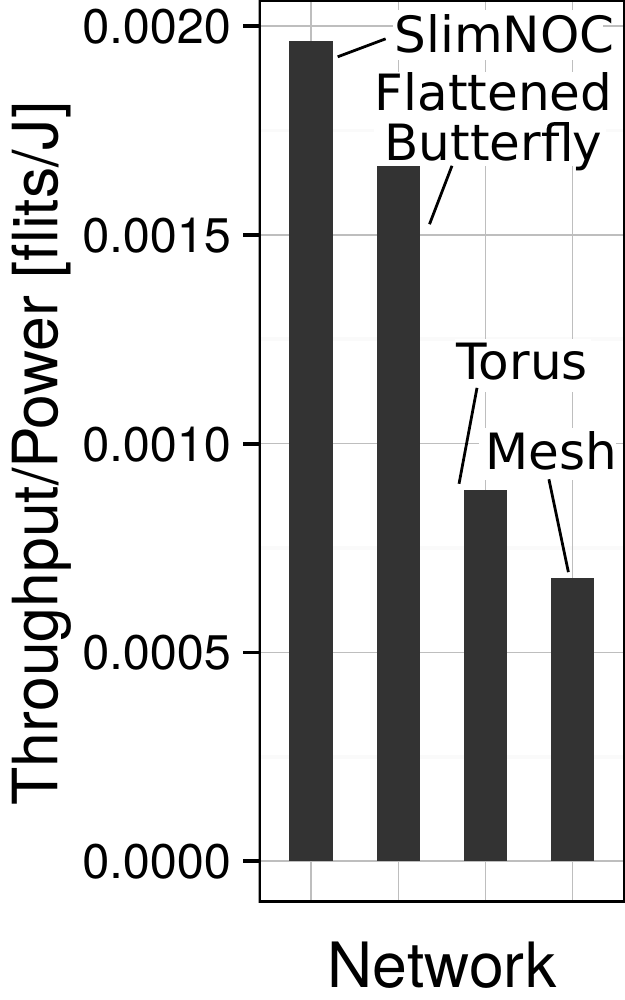}
  }\hspace{0.3em}
 \subfloat[Throughput per power (22nm)]{
  \includegraphics[width=0.1875\textwidth]{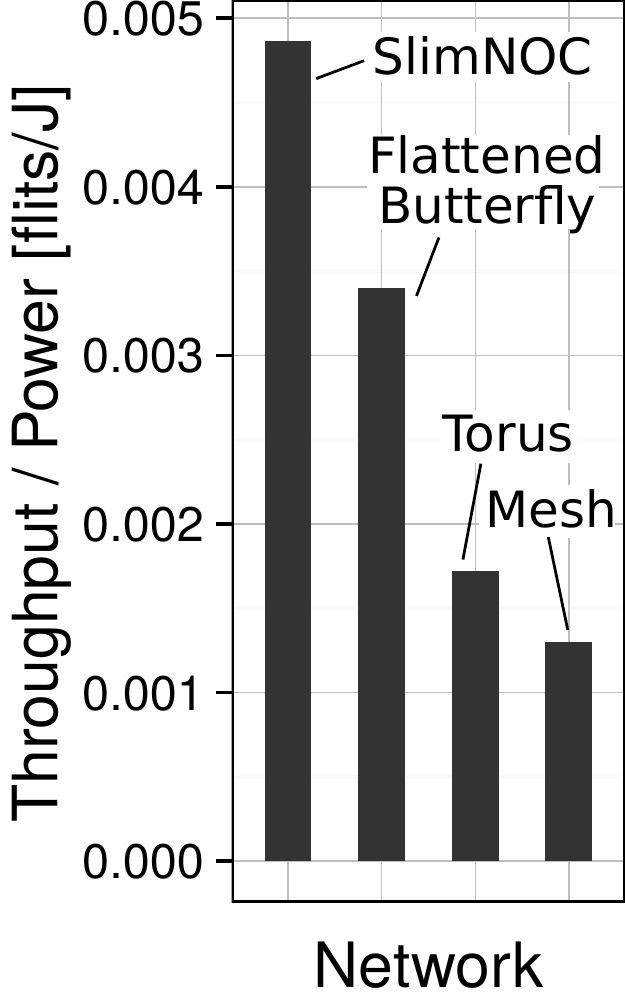}
 }
\caption{\hlgreen{Example advantages of Slim NoC over high- and
low-radix NoC topologies (1296 cores); see~\mbox{\cref{sec:targets_}} for
detailed methodology. 
%
%
We omit performance data for points after network saturation.}}
\label{fig:motivation-positive}
\end{figure*}

  The design of Slim NoC is based on two key observations. First, we
observe that \texttt{SF} uses degree-diameter graphs that are challenging to
lay out to \emph{satisfy NoC constraints} such as the same number of routers or
{nodes} on each side of a die. To solve this problem, our key idea is to use
\emph{non-prime} finite fields to generate diameter--2 graphs for \texttt{SN}
that fit within these constraints (\mbox{\cref{sec:key_ideas}}) and thus can be
used to manufacture chips with tens, hundreds, and thousands of cores. Second,
we observe that most off-chip topologies optimize cost and layouts in a way
that does \emph{not} address NoC limitations, resulting in large buffers.
%
%
We solve this problem with NoC-optimized \texttt{SN} layouts and cost models that
consider average wire lengths and buffer sizes (\mbox{\cref{sec:layoutModel}}).
The resulting \texttt{SN} design outperforms state-of-the-art NoC topologies,
as our experiments show (\mbox{\cref{sec:evaluation}}).

  To make \texttt{SN} even more scalable and efficient, we augment it with
orthogonal state-of-the-art router microarchitecture schemes: central buffer
routers~\cite{6558397} (to decrease buffer area), Elastic Links~\cite{4798250}
and ElastiStore~\cite{Seitanidis:2014:EEB:2616606.2616900} (to increase
performance and power efficiency), and SMART
links~\cite{Chen:2013:SSR:2485288.2485371} (to reduce latency).
Doing so leads to a low-diameter and energy-efficient NoC that outperforms other
designs, as shown in Figure~\ref{fig:motivation-positive}.
Example advantages of \texttt{SN} over an \texttt{FBF}, a mesh, and a torus (all using the
\emph{same microarchitectural schemes} as \texttt{SN}) are: (1) latency is
lower by $\approx$10\%, $\approx$50\%, and $\approx$64\%, respectively, (2) throughput/power is higher by
 $\approx$18\%, $>$100\%, and $>$150\% (at 45nm), and $\approx$42\%, $>$150\%, and $>$250\% (at
22nm).

\change{(§1) We added more key results at the end of the introduction to provide a
better high-level summary of Slim NoC advantages. We also improved the naming
of the used parameters to make things easier to follow; for example, we
disposed of the terms ``fixed and variable'' when referring to cycles.}

\hl{We comprehensively compare \texttt{SN} to five NoC topologies (2D mesh,
torus, two Flattened Butterfly variants, and briefly to hierarchical
NoCs~\cite{yuan2011nonblocking, leiserson1985fat}) using various comparison metrics (area, buffer
sizes, static and dynamic power consumption, energy-delay, throughput/power,
and performance for synthetic traffic and real applications). We show that
\texttt{SN} improves the energy-delay product by $\approx$55\% on average (geometric mean) over \texttt{FBF}, the
best-performing previous topology, on real applications, while consuming up to
$\approx$33\% less area.  We also analyze \texttt{SN}'s sensitivity to many parameters,
such as (1) layout, (2) concentration, (3) router cycle time, (4)
network size, (5) technology node, (6) injection rate, (7) wire type, (8)
buffer type, (9) bisection bandwidth, (10) router microarchitecture
improvement, (11) traffic pattern, and find that \texttt{SN}'s benefits are
robust.}

\hlgreen{Our comprehensive results show that \texttt{SN} outperforms state-of-the-art
topologies for both large and small NoC sizes.}

\section{Background}
\label{sec:background}

To alleviate issues of \texttt{SF} for \emph{on-chip} networks, we
first provide background on the \texttt{SF} topology
(\mbox{\cref{sec:background_slim-fly}}). We then analyze \texttt{SF}'s
performance when used as an \emph{on-chip} network
(\mbox{\cref{sec:motivation}}). 


\goal{Introduce the section}
%
%



We broadly introduce the elements and concepts we use in Slim NoC.
For the reader's convenience, we first summarize all symbols in the paper in 
Table~\ref{tab:symbols}.

\begin{table*}[b]
\centering 
\setlength{\tabcolsep}{2pt}
\small
\sf
\begin{tabular}{@{}l|ll@{}}
\toprule
\multirow{7}{*}{\begin{turn}{90}\shortstack{{Network} {structure}}\end{turn}} & $N$&The number of \hl{nodes} in the whole network\\
                   & $p$&The number of \hl{nodes} attached to a router (\emph{concentration})\\
                   & $k'$&The number of channels to other routers (\emph{network radix})\\
                   & $k$&\emph{Router radix} ($k = k' + p$)\\
                   & $N_r$&The number of routers in the network\\
                   & $D$&The diameter of a network\\ 
                   & $q$&A parameter that determines the structure of an \texttt{SN} (see \cref{sec:background_slim-fly})\\ \midrule
\multirow{8}{*}{\begin{turn}{90}\shortstack{Physical layout}\end{turn}} & $M$&The average Manhattan distance between connected routers\\
                   & $x_i$&$x$ coordinate of a router $i$ ($1 \leq i \leq N_r$) and its attached \hl{nodes}\\
                   & $y_i$&$y$ coordinate of a router $i$ ($1 \leq i \leq N_r$) and its attached \hl{nodes}\\
                   & $|$VC$|$ & The number of virtual channels per physical link\\ 
                   & $[G|a,b]$ & \makecell[l]{A router label in the subgroup view: $G$ is a subgroup type, $a$ is\\a subgroup ID, $b$ is the position in a subgroup; Figure~\ref{fig:mms_example} shows details}\\
                   & $H$ & \makecell[l]{The number of hops (between routers adjacent to\\each other on a 2D grid) traversed in one link cycle} \\
                   \midrule
\multirow{9}{*}{\begin{turn}{90}\shortstack{Buffer models}\end{turn}} & $\delta_{ij}$&The size of an edge buffer at a router $i$ connected to $j$ [flits]\\    
                   & $\delta_{cb}$&The size of a single central router buffer [flits]\\
                   & $\Delta_{eb}$& \makecell[l]{The total size of router buffers in a network with edge buffers [flits]}\\
                   & $\Delta_{cb}$& \makecell[l]{The total size of router buffers in a network with central buffers [flits]}\\
                   & $b$&The bandwidth of a link [bits/s]\\
                   & $L$&The size of a flit [bits]\\
                   & $T_{ij}$&Round trip time on the link connecting routers $i$ and $j$ [s]\\
                   & $W$&\makecell[l]{The maximal number of wires that can be placed over one router\\and its attached nodes}\\ \bottomrule
\end{tabular}
\caption{\hlgreen{Symbols used in the paper.}}
\label{tab:symbols}
\end{table*}

\begin{figure*}[t!]
\centering
\begin{tabular}{cc}
\subfloat[\vspace{-0.125em} (\cref{sec:background_slim-fly}) \hl{A comparison between Slim Fly and Dragonfly~\cite{dally08}}.]{
  \includegraphics[width=0.59\textwidth]{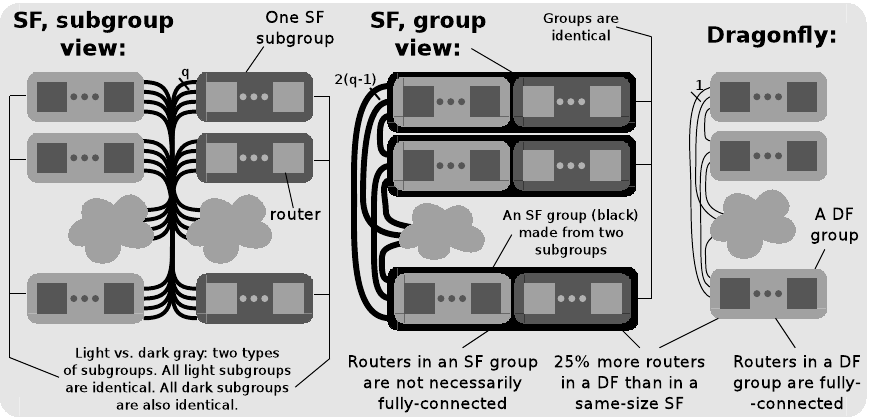}
  \label{fig:sf_vs_df_intuition}
 }
&
 \subfloat[\vspace{-0.125em}(\cref{sec:layoutModel}) \hl{Labeling and indices of Slim Fly routers}.]{
  \includegraphics[width=0.385\textwidth]{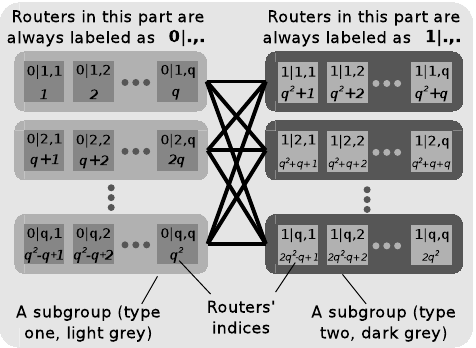}
  \label{fig:sf_numbering}
 }
\end{tabular}
\caption{\hl{An illustration of the Slim Fly structure and the labeling as well as indices used in the Slim NoC design.}}
\label{fig:mms_example}
\end{figure*}


\subsection{The Slim Fly Topology: The Main Inspiration}
\label{sec:background_slim-fly}

\change{(§3.1) We provided more intuition on and references to graph theory works that underlie
Slim Fly.}

%
\texttt{SF}~\cite{slim-fly} is a cost-effective topology for large
computing centers that uses mathematical optimization to minimize the network
diameter $D$ for a given radix $k$ while maximizing the
number of attached nodes $N$ (i.e., network scalability) and maintaining high bandwidth.
There are two key reasons for \texttt{SF}'s advantages.
First, it \emph{lowers diameter} ($D=2$): this ensures the lowest latency for
many traffic patterns, and it reduces the number of required network resources
(packets traverse fewer routers and cables), lowering cost and static/dynamic
power consumption~\cite{slim-fly}. 
Second, it uses graphs that approach the Moore Bound {(MB)}~\cite{mckay98}, a notion from
graph theory that indicates the \emph{upper bound on the number of vertices in a graph
with a given $D$ and $k$}.
%
%
This maximizes scalability
and offers high resilience to link failures because 
the considered graphs are good \emph{expanders}~\cite{Pippenger:1992:FCN:140901.141867}.

\change{(§3.1, ``SF Structure Intuition'') We provide more
intuition on the Slim Fly structure.}

\macb{SF Structure Intuition}
\hl{
\texttt{SF} has a highly symmetric structure; see
Figure~\ref{fig:sf_vs_df_intuition}. It consists of identical \emph{groups} of
routers (see the middle part of Figure~\ref{fig:sf_vs_df_intuition}). Every two
such groups are connected with the \emph{same} number of cables. Thus, the
\texttt{SF} network is \emph{isomorphic} to a fully-connected graph where each vertex is a
collapsed \texttt{SF} group. Still, routers that constitute a group are \emph{not
necessarily} fully-connected. Finally, each group consists of two
\emph{subgroups} of an identical size. These subgroups usually differ in their
cabling pattern~\cite{slim-fly}.
}

\change{(§3.1, ``SF Structure Details'') We moved the less necessary details of the ``q'' parameter of Slim Fly
into a footnote.}

\macb{SF Structure Details}
\hl{
%
%
Routers in \texttt{SF} are grouped into \emph{subgroups} with the
same number of routers (denoted as $q$). There are two types of
subgroups, each with the same pattern of intra-group links. Every two
subgroups of different types are connected with the same number of cables (also
$q$). No links exist between subgroups of the same type. Thus, subgroups
form a fully-connected \emph{bipartite} graph where an edge is formed by $q$
cables.
%
%
Subgroups of different types can be merged pairwise into identical 
\emph{groups}, each with $2q$ routers.
Groups form a fully-connected graph where an edge consists of $2(q-1)$
cables.
%
%
The value of $q$ determines other \texttt{SF} parameters}\footnote{$q$ can be
any \emph{prime power} such that $q = 4w + u$; $w \in \mathbb{N}, u \in
\{\pm 1, 0\}$.  An \texttt{SF} with a given $q$ has 
the number of routers $N_r = 2q^2$, the network radix $k' = \frac{3q-u}{2}$,
and the number of nodes $N = N_r p$. The concentration $p$ is
$\left\lfloor \frac{k'}{2} \right\rfloor + \kappa$; $\kappa$ is a user-specified parameter that determines a desired
tradeoff between higher node density (larger $\kappa$) and lower contention
(smaller $\kappa$).}, \hl{including the number of routers $N_r$, the network radix
$k'$, the number of nodes $N$, and the concentration $p$.
}


\goal{Describe the structure intuitively by comparing to Dragonfly}
\macb{SF vs.~DF}
\hlgreen{
Intuitively, \texttt{SF} is similar to the \emph{balanced}
\texttt{DF}~\cite{dally08} that also consists of groups of routers. Yet, only
one cable connects two \texttt{DF} groups (see
Figure~\ref{fig:sf_vs_df_intuition}), resulting in higher $D$ and a lower
number of inter-group links. 
Moreover, each \texttt{DF} group is a fully-connected graph, which is not
necessarily true for \texttt{SF}.
Finally, \texttt{SF} reduces the number of routers by $\approx$25\% and
increases their network radix by $\approx$40\% in comparison to a \texttt{DF}
with a comparable $N$~\cite{slim-fly}.
}

\hlgreen{
As shown in past work~\cite{slim-fly}, \texttt{SF} maximizes scalability for a fixed $k$ and $D = 2$,
while maintaining high bandwidth.
%
%
{Other} advantages of \texttt{SN} (low cost, power consumption, latency, high resilience) 
stem from the
properties of the underlying degree-diameter graphs. 
%
%
For these, we refer the reader to the original \texttt{SF} work~\cite{slim-fly}.
%
}

\change{(§3.1) We removed the last paragraph of this subsection for better readability.}

\begin{figure*}[t]
\centering
 \subfloat[Wire length]{
  \includegraphics[width=0.225\textwidth]{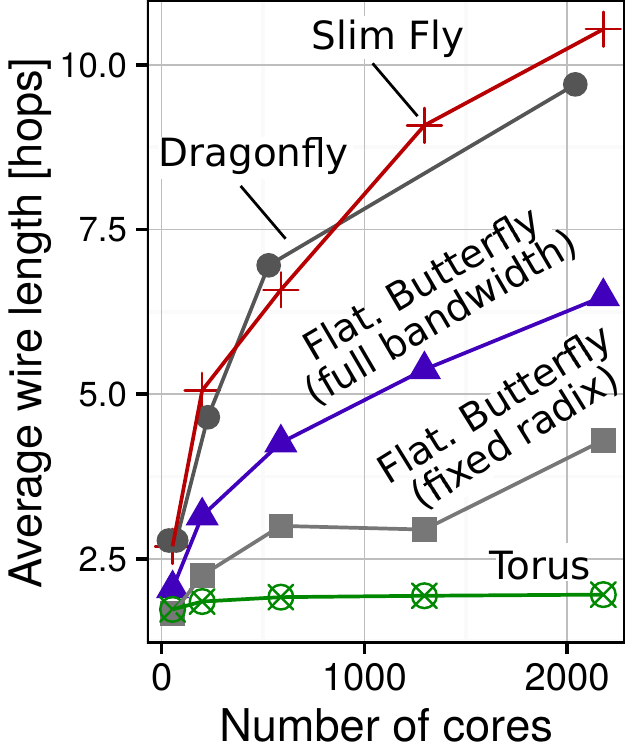}
  \label{fig:motiv_wires}
 }
 \subfloat[Normalized area]{
  \includegraphics[width=0.225\textwidth]{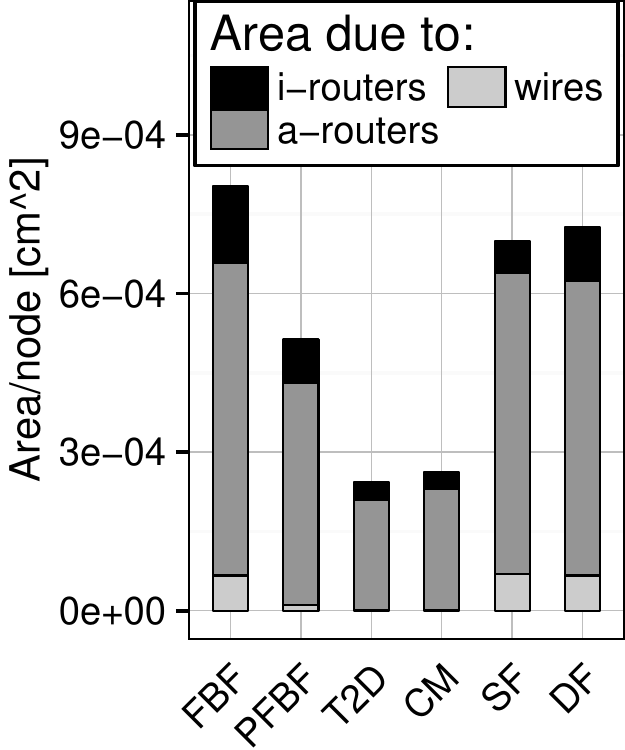}
  \label{fig:motiv_area}
 }
 \subfloat[Normalized power]{
  \includegraphics[width=0.225\textwidth]{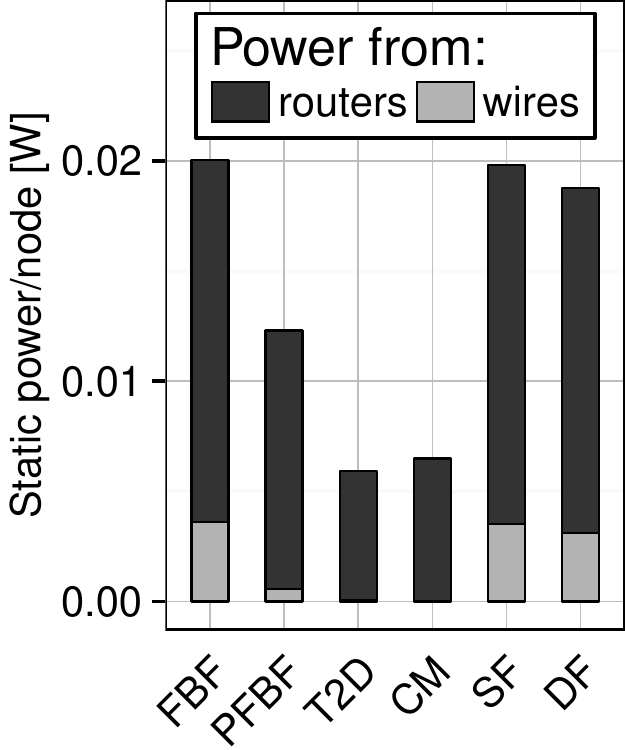}
  \label{fig:motiv_static_power}
 }
\caption{Disadvantages of Slim Fly and Dragonfly used straightforwardly as
NoCs (analyses in Figures~\ref{fig:motiv_area}--\ref{fig:motiv_static_power}
use 200 cores). \texttt{i-routers} and \texttt{a-routers} are router
areas in intermediate and active layers.}
\label{fig:motivation_analysis}
\end{figure*}


\subsection{Slim Fly and Dragonfly for On-Chip Networks}
\label{sec:motivation}

%

%
We first investigate whether \texttt{SF} or \texttt{DF} can be used
straightforwardly as NoCs (see Section~\ref{sec:evaluation} for our
methodology). We focus on \texttt{DF} and \texttt{SF} as they are the most
effective topologies with diameter three and two, respectively~\cite{dally08,
slim-fly}. They are also \emph{direct} topologies (each router is attached to
equally many cores) and thus easier to manufacture as NoCs. Figure~\ref{fig:motivation_analysis}
compares \texttt{SF} and \text{DF} to a torus (\texttt{T2D}), 
a concentrated mesh (\texttt{CM})~\cite{balfour2006design}, and two Flattened Butterflies:
a very high-radix full-bandwidth Flattened Butterfly topology (\texttt{FBF}) and
an alternative design that has the same bandwidth as the \texttt{SF} topology (\texttt{PFBF}).
%
%


%

  Based on Figure~\ref{fig:motivation_analysis}, we provide two observations. 
First, compared to \texttt{PFBF}, \texttt{SF}
requires $\approx$38\% longer wire length, consumes $>$30\% more area and power,
has $\approx$10\% higher latency (not shown), and provides $\approx$35\% lower throughput (not shown). 
Second, we find that \texttt{DF} used on-chip comes with similar overheads.
%
The reasons are as stated in~\mbox{\cref{sec:intro}}: both \texttt{SF} and \texttt{DF}
%
%
do not optimize for NoC
constraints, use layouts and cost models optimized for rack-level systems, and
minimize the use of resources such as bandwidth that are plentiful in NoC
settings.



\section{Slim NoC Design and Layouts}
\label{on-chip-slim-fly}

\commt{(§4) In this section, the most important changes are: (1) move the detailed
mathematical description of constructing Slim NoC to the technical report~\cite{slim-noc-report} and provide a
more intuitive discussion, (2) extend the description of all the introduced
concepts in models and layouts so that the mathematical elements remain for completeness and reproducibility but are not
required to understand the key concepts and serve only as an addition for the
more interested readers.}

We first describe the core ideas in \texttt{SN} and how to build it
(\cref{sec:key_ideas}). Then, we present generic \texttt{SN} models for router
placement, sizes of router buffers, and the total cost of a network with a
given layout (\cref{sec:layoutModel}). We then describe and analyze
cost-effective \texttt{SN} layouts (\cref{sec:sf_layouts}). Finally, we provide detailed
mathematical formulations of generating the underlying \texttt{SN} graphs.

\subsection{Key Ideas in Slim NoC Construction and Models}
\label{sec:key_ideas}

\change{(§4.1) The previous contents of this section are moved completely to
our technical report~\cite{slim-noc-report}. Instead, we provide a brief summary of the key research
ideas related to constructing Slim NoC and a summary of all the construction
steps detailed in §4.2.}

\change{(§4.1) We provide a new table that lists all feasible Slim NoC configurations with
detailed parameter values and comments on which one is easier to
manufacture, etc. This improves readability and reproducibility.}

Our key idea related to constructing \texttt{SN} is to use \emph{non-prime finite
fields}~\cite{slim-fly, strang1993introduction} to generate the underlying Slim NoC
graphs.
%
%
Specifically, we discover that graphs based on such fields fit various NoC
constraints (e.g., die dimensions or numbers of nodes) or reduce wiring
complexity. We analyze graphs resulting from non-prime finite fields that
enable Slim NoCs with at most 1300 nodes and summarize
them in Table~\ref{tab:slimnoc_configurations}.
The bold and shaded configurations in this table are the most desirable as
their number of nodes is a power of two (marked with bold font) or they have equally many groups of routers
on each side of a die (marked with grey shades).

\hl{
The underlying graph of connections in Slim NoC has the same structure based on
groups and subgroups as the graph in Slim Fly
(see~\mbox{\cref{sec:background_slim-fly}})}. \hl{To
construct Slim NoC, one first selects (or constructs) a graph that comes with
the most desirable configuration (see Table~\ref{tab:slimnoc_configurations}).
Second, one picks the most advantageous layout based either on the provided
analysis (\mbox{\cref{sec:sf_layouts}}), or one of the proposed particular Slim NoC
designs (\mbox{\cref{sec:fixed_sf_networks}}), or derives one's own layout using the
provided placement, buffer, and cost models (\mbox{\cref{sec:layoutModel}}).
}

\begin{table}[!h]
\centering
\footnotesize
\sf
\begin{tabular}{@{}lllllllll@{}}
\toprule
%
%
 & \makecell[c]{\textbf{Network}\\\textbf{radix} $k'$} & \makecell[c]{\textbf{Concen-}\\ \textbf{tration $p$}} & \makecell[c]{\textbf{$p=\left\lceil\frac{k'}{2}\right\rceil$} $^*$} & \makecell[c]{$p / \left\lceil \frac{k'}{2} \right\rceil$ $^{**}$} & \makecell[c]{\textbf{Network}\\\textbf{size} $N$} & \makecell[c]{\textbf{Router}\\\textbf{count} $N_r$} & \makecell[c]{\textbf{Input}\\\textbf{param.} $q$} \\
\midrule
\multirow{12}{*}{\begin{turn}{90}\makecell{\textbf{Non-prime} \textbf{finite fields}}\end{turn}} 
 & 6 & 3 & 2 & 66\% & \cellcolor{black!45} \textbf{64} & 32 & 4 \\ 
 & 6 & 3 & 3 & 100\% & \cellcolor{black!30} 96 & 32 & 4 \\ 
 & 6 & 3 & 4 & 133\% & \cellcolor{black!30} \textbf{128} & 32 & 4 \\ 
 & 12 & 6 & 4 & 66\% & \textbf{512} & 128 & 8 \\ 
 & 12 & 6 & 5 & 83\% & 640 & 128 & 8 \\ 
 & 12 & 6 & 6 & 100\% & 768 & 128 & 8 \\ 
 & 12 & 6 & 7 & 116\% & 896 & 128 & 8 \\ 
 & 12 & 6 & 8 & 133\% & \textbf{1024} & 128 & 8 \\ 
 & 13 & 7 & 5 & 71\% & \cellcolor{black!30} 810 & 162 & 9 \\
 & 13 & 7 & 6 & 85\% & \cellcolor{black!30} 972 & 162 & 9 \\
 & 13 & 7 & 7 & 100\% & \cellcolor{black!30} 1134 & 162 & 9 \\
 & 13 & 7 & 8 & 114\% & \cellcolor{black!50} 1296 & 162 & 9 \\
 \midrule
\multirow{12}{*}{\begin{turn}{90}\makecell{Prime finite fields}\end{turn}} 
 & 3 & 2 & 2 & 100\% & \textbf{16} & 8 & 2 \\ 
 & 5 & 3 & 2 & 66\% & 36 & 18 & 3 \\
 & 5 & 3 & 3 & 100\% & 54 & 18 & 3 \\
 & 5 & 3 & 4 & 133\% & 72 & 18 & 3 \\
 & 7 & 4 & 3 & 75\% & 150 & 50 & 5 \\
 & 7 & 4 & 4 & 100\% & 200 & 50 & 5 \\
 & 7 & 4 & 5 & 120\% & 250 & 50 & 5 \\
 & 11 & 6 & 4 & 66\% & 392 & 98 & 7 \\
 & 11 & 6 & 5 & 83\% & 490 & 98 & 7 \\
 & 11 & 6 & 6 & 100\% & 588 & 98 & 7 \\
 & 11 & 6 & 7 & 116\% & 686 & 98 & 7 \\
 & 11 & 6 & 8 & 133\% & 784 & 98 & 7 \\
\bottomrule
\end{tabular}
%
\caption{\hlcyan{The configurations of Slim NoC where the network size $N \le
1300$ nodes. Bold font indicates that in a particular configuration
$N$ is a power of two. 
%
%
Grey shade indicates that there are equally many groups on all die
sides. Dark grey shade means that, in addition to that, the number of nodes in a given
configuration is a square of some integer number.
$^*$The ideal concentration, 
$^{**}$over- or undersubscription.
}}
\label{tab:slimnoc_configurations}
\end{table}

\begin{figure*}[bp]
\centering
 \subfloat[(\cref{sec:layoutModel}) \hl{The placement model with two example wires}.]{
  \includegraphics[width=0.46\textwidth]{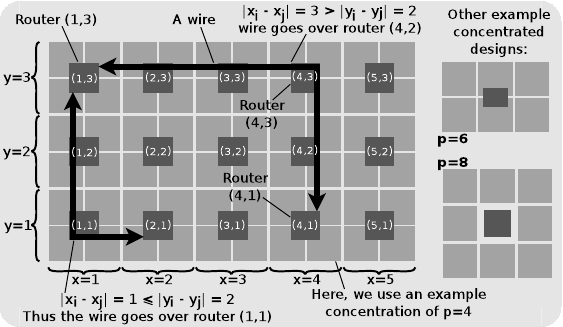}
  \label{fig:layout_model}
 }
  \subfloat[(\cref{sec:layoutModel}) Different physical \texttt{SN} layouts.]{
  \includegraphics[width=0.44\textwidth]{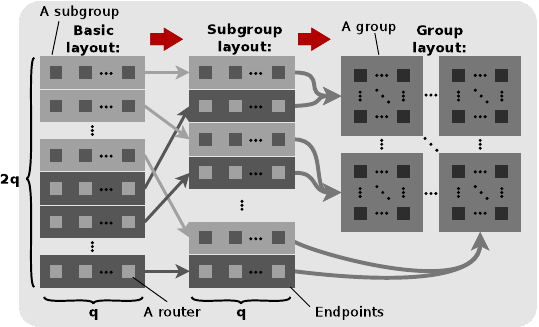}
  \label{fig:onchip-layouts}
 }
\caption{The details of the \texttt{SN} layout.}
\label{fig:placement_example}
\end{figure*}



\subsection{Models}
\label{sec:layoutModel}


\texttt{SN} reduces diameter to two for low latency. It also 
minimizes radix $k$ to limit area and power consumption. Yet, when
deploying the network on a chip, this may require long multi-cycle links that
span a large physical distance. Thus, one needs larger buffers to fully
utilize the wire bandwidth, overshadowing the advantages of minimized radix
$k$. To alleviate this, we develop new \textbf{placement}, \textbf{buffer}, and
\textbf{cost} models for \texttt{SN}
(\mbox{\cref{sec:placement_model}}--\mbox{\cref{sec:cost_model}}) and use them
to analyze and compare cost-effective layouts that ultimately minimize
the average Manhattan distance and the total buffer area.

%


\subsubsection{Placement Model}
\label{sec:placement_model}

\change{(§4.2.1) A completely rewritten section. The first main change: the section now
extensively explains every part of the model carefully and intuitively, with detailed
descriptions. The second main change: the text now is still accompanied by
equations, but the goal is to make the text self-contained and the discussions of the equations more intuitive. Even if the
equations are removed, the readers should still understand the most relevant model
elements.}

\change{(§4.2.1) We have a new name for the layout model, which is now called the
placement model. We found that placement model is more appropriate as this
model is about how to place routers on a 2D grid and how to place wires between
them.}

When building \texttt{SN}, we place routers on a chip seen as a 2D grid. To
analyze different placements, we introduce a model. The model must
assign routers their coordinates on a 2D grid, place wires between connected
routers, and be easy to use. The model consists of four parts:
assigning labels, indices, coordinates, and placing wires.
%


\change{(§4.2.1) The new structure of this section is now based on four parts
that correspond to the four model parts related to: assigning routers labels,
indices, coordinates, and placing wires between routers. We explain why each
part is important and how it is related to other parts.}

\macb{Labels}
\hlcyan{
We first name each router with a \emph{label} that uniquely encodes the
router position in the \texttt{SN} ``subgroup view'' (see the leftmost
picture in Figure~\ref{fig:sf_vs_df_intuition}). In this view, any router can
be identified using three numbers: (1) type of its subgroup (corresponds to
light or dark grey), (2) ID of its subgroup (top-most to bottom-most),
and (3) its position in the subgroup (leftmost to rightmost).
Labels are based on the subgroup view as its regular structure enables straightforward
visualization and identification of routers in \emph{any} \texttt{SN}.
Figure~\ref{fig:sf_numbering} shows the labeling. A router is
labeled as $[G|a,b]$. These symbols encode the subgroup type $\left(G \in
\{0, 1\}\right)$, the subgroup ID $\left(a \in \{1, ..., q\}\right)$, and the
position in a subgroup $\left(b \in \{1, ..., q\}\right)$.
%
%
}


%



\macb{Indices}
\hlcyan{Second, we translate labels into \emph{indices} such that each router has a
unique index~$i \in \{1 ... N_r\}$ based on its label. A
formula that ensures uniqueness is $i = G q^2 + (a-1) q + b$.
Figure~\ref{fig:sf_numbering} shows details.
%
%
We use indices derived from labels because, while labels are straightforward to
construct and use, indices facilitate reasoning about router 
coordinates on a 2D grid and wire placement.}

%
\macb{Coordinates}
\hlcyan{Indices and labels are used to assign the actual
\emph{coordinates} on a 2D grid; see Figure~\ref{fig:layout_model}. A router $i
\in \{1 ... N_r\}$ is assigned coordinates $(x_i,y_i)$. These coordinates
become concrete numbers based on the labels in each layout.
More details are in~\mbox{\cref{sec:sf_layouts}}.
We assume that routers form a rectangle and $1 \leq x_i \leq X$, $1 \leq y_i
\leq Y$.
%
}

%
\macb{Wires}
\hlcyan{For two connected routers~$i$ and $j$, we place the connecting link using the
shortest path (i.e., using the Manhattan distance). If the routers lie on the
same row or column of the grid, there is only one such path. Otherwise, there
are two such paths. We break ties by placing the first wire part
(originating at router~$i$) vertically (along the Y axis) or horizontally
(along the X axis) depending on whether the vertical or horizontal distance is
smaller.  Formally, we place a wire along the points $(x_i,y_i)$, $(x_i,y_j)$,
and $(x_j,y_j)$ (if $|x_i-x_j| > |y_i-y_j|$), or $(x_i,y_i)$, $(x_j,y_i)$, and
$(x_j,y_j)$ (if $|x_i-x_j| \leq |y_i-y_j|$), spreading wires over routers in a balanced way.
%
}

\change{(§4.2.1, ``Wires'') This paragraph was a separate section §4.2.5 (called ``Constraints'')
in the old submission. It is merged with §4.2.1 because all these
``constraints'' were only about the locations of wires between routers, which
belongs to this subsection.}

\change{(§4.2.1, ``Wires'') We extensively modified the description of the following wiring
equations, explaining each of the functions $\Phi, \phi, \Psi, \psi$,
their mutual relationships, and how they are used to determine the
placement of wires.}


\hlcyan{
The placed wires must adhere to certain placement constraints. We formally
describe these constraints for completeness and reproducibility (readers who are
\emph{not} interested in these formal constraints can proceed
to~\mbox{\cref{sec:buffer_model}}).
Specifically, there is a maximum number of wires $W$ that can be placed over a
router (and attached nodes).
To count wires that traverse a router (and its attached nodes) with given
coordinates, we use functions $\Phi$, $\Psi$, $\phi$, and $\psi$. First,
for any two routers with indices $i$ and $j$, $\Phi(i,j)$ and $\Psi(i,j)$
determine if the distance between $i$ and $j$ is larger along the X or Y
axis, respectively:
}

\begin{alignat}{1}
\Phi(i,j) = 1\ \text{if}\ |x_i-x_j| > |y_i-y_j|,\ \text{and}\ 0\ \text{otherwise} \\
\Psi(i,j) = 1\ \text{if}\ |x_i-x_j| \leq |y_i-y_j|,\ \text{and}\ 0\ \text{otherwise.}
\end{alignat}
\normalsize

\hlcyan{
Second, given a router pair $i$ and $j$, $\phi_{ij}(k,l)$ and $\psi_{ij}(k,l)$
determine if a router with coordinates $(k,l)$ is located on one of the two
shortest Manhattan paths between $i$ and $j$ ($\phi$ is responsible for the
``bottom-left'' path as seen in a 2D grid while $\psi$ is responsible for the
``top-right'' part)
}

\begin{alignat}{1}
 \phi_{ij}(k,l) = \left\{ 
  \begin{array}{l}
    1,\ \text{if } k = x_i\ \wedge\ \text{min}\{y_i,y_j\} \leq l \leq \text{max}\{y_i,y_j\} \\
    1,\ \text{if } l = y_j\ \wedge\ \text{min}\{x_i,x_j\} \leq k \leq \text{max}\{x_i,x_j\}\ \\
    0,\ \text{otherwise}
  \end{array}\nonumber \right.
  \end{alignat}
\normalsize

\begin{alignat}{1}
 \psi_{ij}(k,l) = \left\{ 
  \begin{array}{l}
    1,\ \text{if } k = x_j\ \wedge\ \text{min}\{y_i,y_j\} \leq l \leq \text{max}\{y_i,y_j\} \\
    1,\ \text{if } l = y_i\ \wedge\ \text{min}\{x_i,x_j\} \leq k \leq \text{max}\{x_i,x_j\} \\
    0,\ \text{otherwise.}
  \end{array}\nonumber \right.
  \end{alignat}
\normalsize

To derive the total count of wires crossing a router with coordinates $(k,l)$,
we iterate over all pairs of routers and 
use $\Phi$, $\Psi$, $\phi$, and $\psi$ to determine and count
wires that cross $(k,l)$.  For a single pair of routers $i$ and $j$, the expression
$\phi_{ij}(k,l) \Phi(i,j) + \psi_{ij}(k,l) \Psi(i,j)$ indicates whether the
Manhattan path between $i$ and $j$ crosses $(k,l)$ (the first or the second
product equals $1$ if the path is ``bottom-left'' or ``top-right'',
respectively).
Next, we multiply the sum of these products with $\varepsilon_{ij}$: this term
determines if routers $i$ and $j$ are connected with a link ($\varepsilon_{ij}
= 1$) or not ($\varepsilon_{ij} = 0$).
Finally, for each $(k,l)$, we verify whether its associated wire count is lower
than $W$, the maximum value dictated by the technology constraints:

\begin{alignat}{1}
    \sum_{i=1}^{N_r} \sum_{j=1}^{N_r} \varepsilon_{ij} [\phi_{ij}(k,l)
\Phi(i,j) + \psi_{ij}(k,l) \Psi(i,j)] \leq W \label{eq:constr_2} \end{alignat}
\normalsize

\ 



Consider two connected routers $A$ and $B$.  If $|x_A-x_B| > |y_A-y_B|$ then
the wire is placed over router/\hl{nodes} with coordinates $(x_A,y_B)$;
$\phi_{AB}(k,l) \Phi(A,B) = 1$ for any $(k,l)$ contained within line segments
connecting points $(x_A,y_A)$, $(x_A,y_B)$, and $(x_B,y_B)$.  If $|x_A-x_B|
\leq |y_A-y_B|$ then the wire is placed over router/\hl{nodes} with
coordinates $(x_B,y_A)$; $\psi_{AB}(k,l) \Psi(A,B) = 1$ for any $(k,l)$
contained within line segments connecting points $(x_A,y_A)$, $(x_B,y_A)$,
and $(x_B,y_B)$.

\begin{figure*}[bp]
%
\centering
 \subfloat[\hlgreen{Average wire length $M$}.]{
  \includegraphics[width=0.22\textwidth]{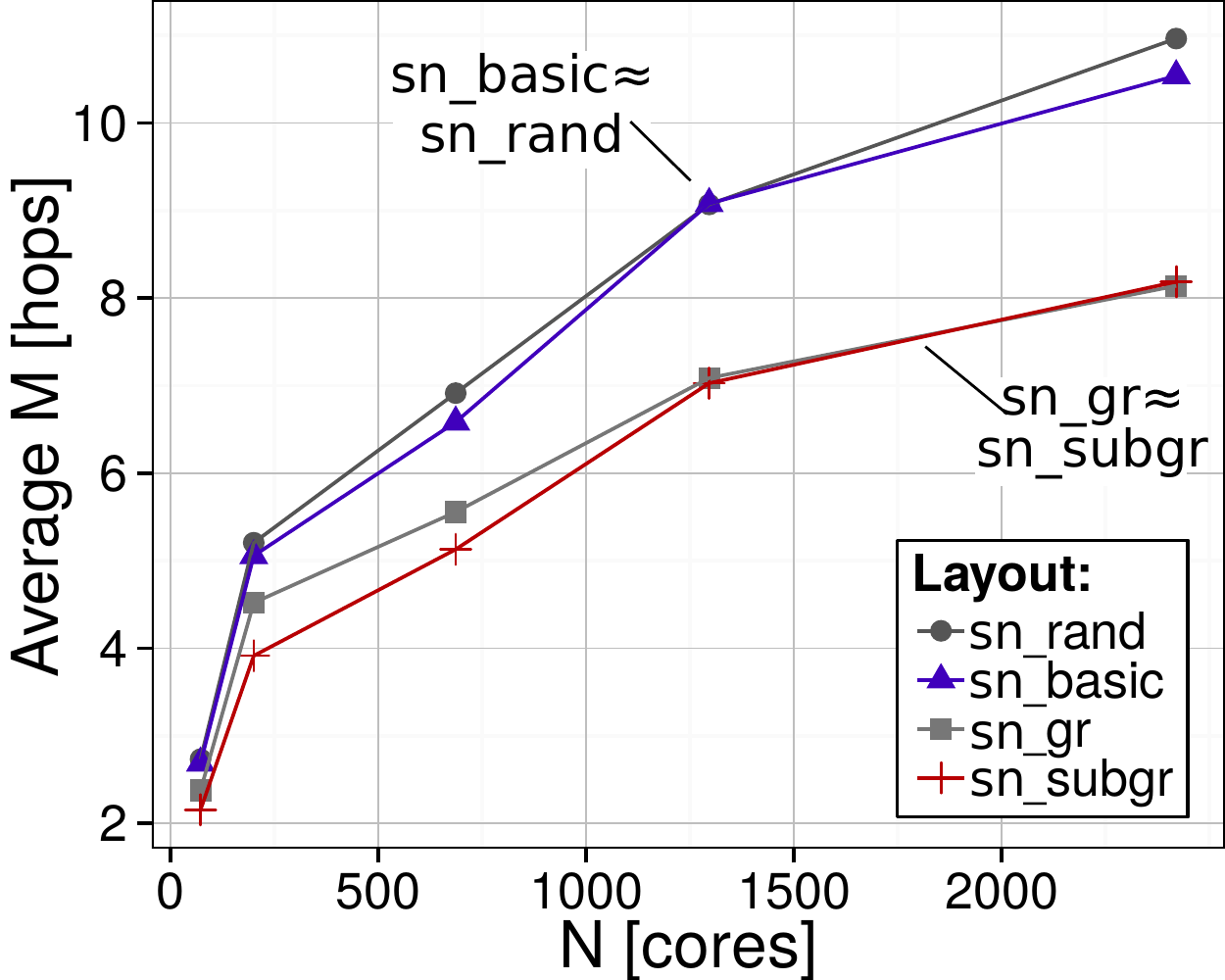}
  \label{fig:M-analysis-sf}
 }\hfill
  \subfloat[\hlgreen{Total size of all buffers in one router (no SMART)}.]{
  \includegraphics[width=0.22\textwidth]{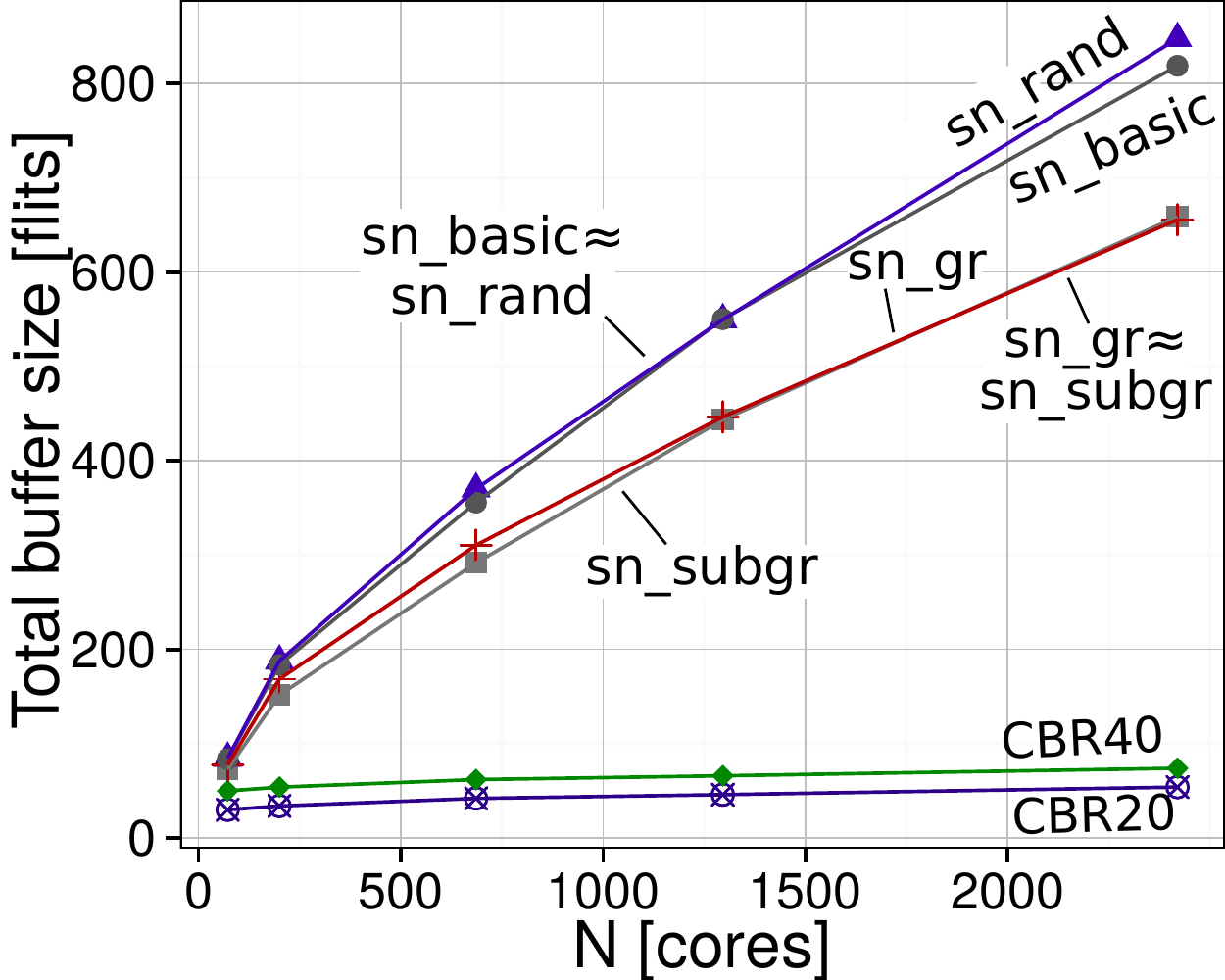}
  \label{fig:B-analysis-sf}
 }\hfill
  \subfloat[\hlgreen{Total size of all buffers in one router (with SMART)}.]{
  \includegraphics[width=0.22\textwidth]{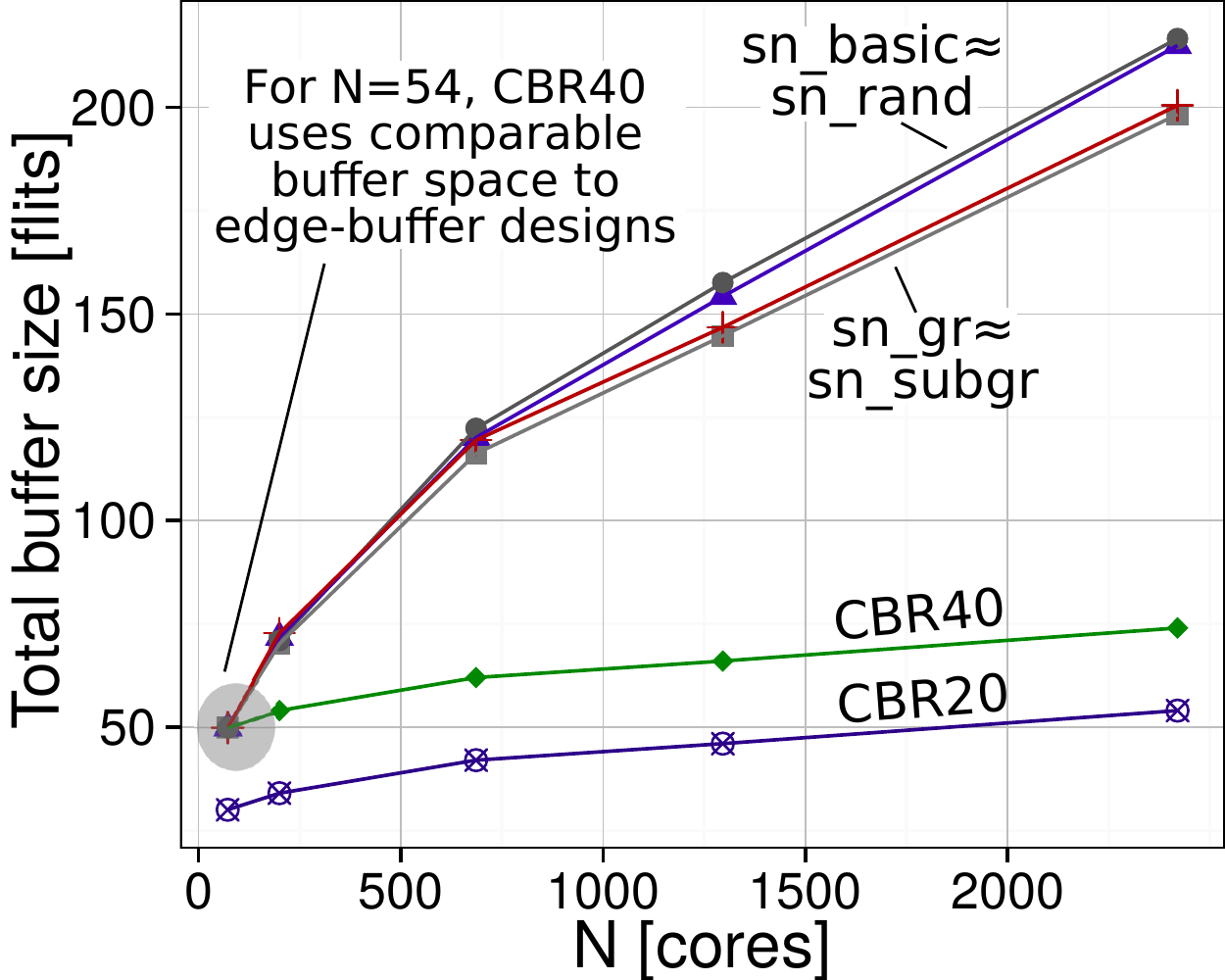}
  \label{fig:B-smart-analysis-sf}
 }\hfill
  \subfloat[\hlgreen{The maximum number of wires placed over a router $W$}.]{
  \includegraphics[width=0.22\textwidth]{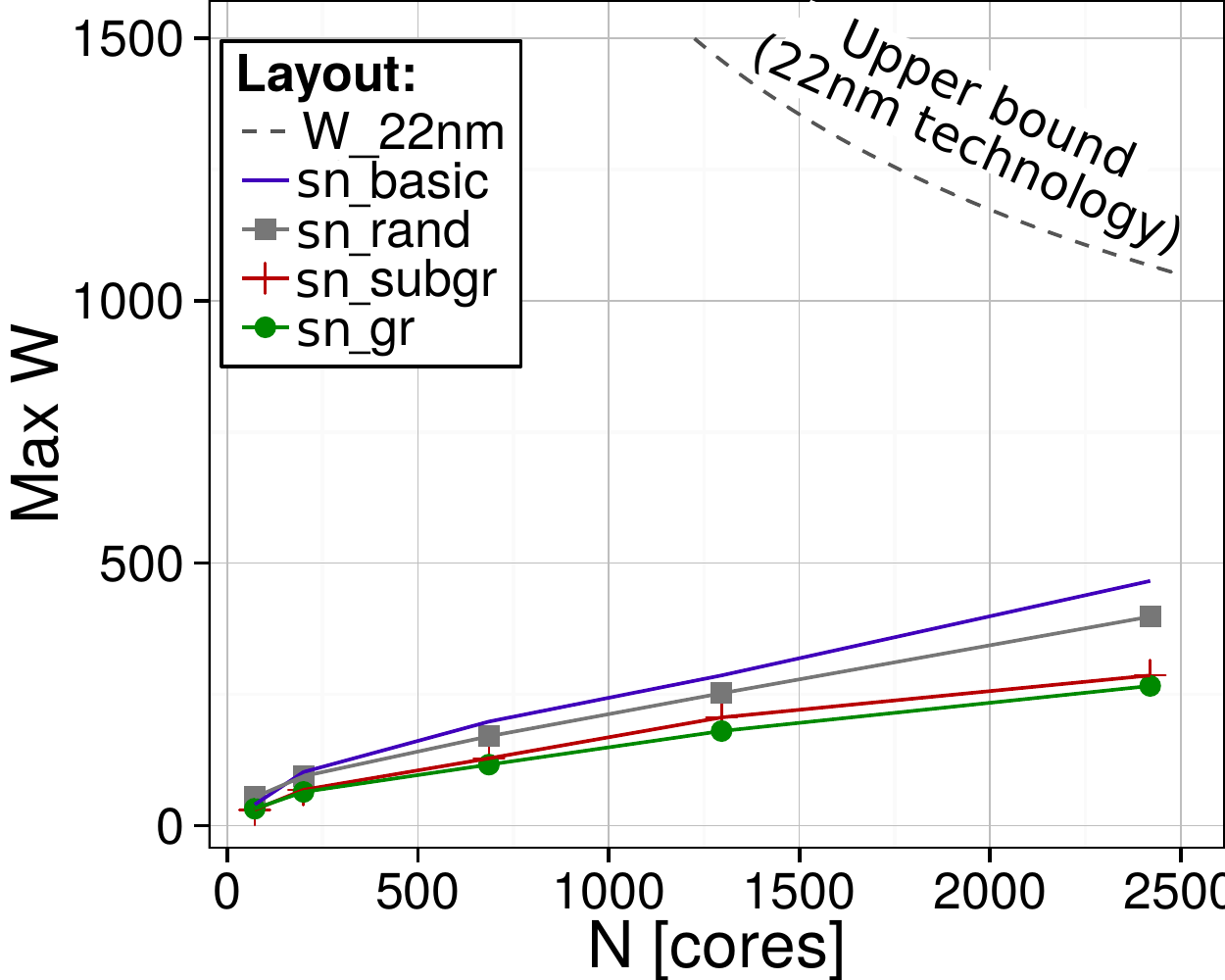}
  \label{fig:W-analysis-sf}
 }
%
%
\caption{(\cref{sec:sf_layouts}) \hlgreen{The analysis of link lengths and buffer sizes
in \texttt{SN} with different layouts
((a)--(c)). The illustration of technological constraints from
Eq.~(\ref{eq:constr_2}) (d).}}
%
\label{fig:wire-buffer-analysis-sf}
\end{figure*}


\subsubsection{Buffer Size Model}
\label{sec:buffer_model}


\change{(§4.2.2) We explained the provided equations intuitively.}

\change{(§4.2.2) We enhanced the clarity of symbols by removing the $\Delta$
symbol for a total buffer size and instead using two symbols, $\Delta_{eb}$ and $\Delta_{cbr}$,
for illustrating the buffer size for network designs based on, respectively,
edge (input) buffers and central buffers.}

Next, we formally model the size of buffers to 
provide a tool that 
enables comparing different Slim NoC layouts in how they reduce
the total buffer size.

\change{Additional explanations on the illustrated equations.}


%
%


\macb{Edge Buffers}
We model the size of an edge buffer integrated with a router $i$ and connected
to a wire leading to router~$j$ as $\delta_{ij} = \left( T_{ij} b |VC| \right)
/ L$. This size is proportional to the round trip time (RTT) between
$i$ and $j$ ($T_{ij}$), link bandwidth ($b$), and virtual channel count per
physical link ($|VC|$). It is inversely proportional to the flit size ($L$).
The RTT is $T_{ij} = 2 \left\lceil (|x_i-x_j| + |y_i-y_j|)/{H} \right\rceil +
3$ and is proportional to the Manhattan distance between $i$ and $j$.  $H$ is
the number of hops traversed in one link cycle (a hop is a part of a wire
between routers placed next to each other (vertically or horizontally) on a 2D
grid). $H$ is $1$ without wire enhancements (such as
SMART~\cite{Chen:2013:SSR:2485288.2485371}). We add two cycles for
router processing and one cycle for serialization latency.
%

%
%
%

\sloppypar{
\macb{Lower Latency in Long Wires with SMART Links}
SMART~\cite{Chen:2013:SSR:2485288.2485371} is a technique that builds on
driving links asynchronously and placing repeaters carefully to enable
single-cycle latency in wires up to 16mm in length at 1GHz at 45nm.
%
%
For wires
that cannot be made single-cycle this way, we assume that
SMART can be combined with EB links~\cite{4798250} to provide multi-cycle wires
and full link utilization.
We assume SMART has no adverse effect on \texttt{SN}'s error rates as it
achieves bit error rates $< 10^{-9}$, similarly to links with equivalent
full-swing repeaters~\cite{Chen:2013:SSR:2485288.2485371}.
}


With SMART links, the value of $H$
depends on the technology node and the operational frequency (typically, 8-11
at 1GHz in 45nm)~\cite{Chen:2013:SSR:2485288.2485371}. RTT still grows
linearly with wire length; SMART links simply limit the growth rate by a factor of
$H$ because a packet can traverse a larger distance on a chip (measured in the distances between neighboring routers) in one cycle.
Note that this model may result in edge buffers at different routers having different
sizes. To facilitate router manufacturing, one can also use edge buffers of
identical sizes. These sizes can be equal to: (1) the minimum edge buffer size in the whole network (reducing $\Delta_{eb}$ but also
potentially lowering performance), (2) the maximal edge buffer size in the whole network (increasing $\Delta_{eb}$ but potentially improving 
throughput), and (3) any tradeoff value between these two.

\macb{Central Buffers}
We denote the size of a CB as $\delta_{cb}$, this
number is a selected constant independent of $|VC|$, $b$, $L$, or $T_{ij}$.
CB size is empirically determined by an \texttt{SN} designer.


\subsubsection{Cost Model}
\label{sec:cost_model}

\change{(§4.2.3) We explained the provided equations intuitively.
In addition, we provided an intuitive formulation for each
buffer size model that does not use any symbols, see Eq.(4), (5), and (6).}

We now use the layout and buffer models to design the \texttt{SN} cost model. We
reduce the average router-router wire length ($M$)
and the sum of all buffer sizes in routers ($\Delta_{eb}$ or $\Delta_{cbr}$).
\\

\macb{Minimizing Wire Length}
The average wire length $M$ is 

%
 \begin{alignat}{1}
 M = \frac{\text{Sum of distances}}{\text{Number of links}} = \frac{\sum_{i=1}^{N_r} \sum_{j=1}^{N_r} \varepsilon_{ij}(|x_i - x_j| + |y_i
 - y_j|)}{\sum_{i=1}^{N_r} \sum_{j=1}^{N_r} \varepsilon_{ij}} . \label{eq:M}
 \end{alignat}
%
%
\normalsize

\noindent
To obtain $M$, we divide the sum of the Manhattan distances between all
connected routers (the nominator in Eq.~(\ref{eq:M})) by the number of connected
router pairs (the denominator in Eq.~(\ref{eq:M})). In Eq.~(\ref{eq:M}), we
iterate over all possible pairs of routers along both dimensions, and
$\varepsilon_{ij}$ determines if routers $i$ and $j$ are connected
($\varepsilon_{ij} = 1$) or not ($\varepsilon_{ij} = 0$).
%


\macb{Minimizing Sum of Buffer Sizes}
\hl{
One can also directly minimize the total sum of buffer sizes ($\Delta_{eb}$ for an \texttt{SN} with edge buffers and $\Delta_{cb}$ for an \texttt{SN} with central buffers). 
%
%
To derive the size of edge buffers ($\Delta_{eb}$), we sum all the terms $\delta_{ij}$ 
}

\begin{alignat}{1}
\Delta_{eb} = \sum_{\parbox{4em}{\centering\scriptsize All router\\pairs $i,j$}} \parbox{7.5em}{\centering\scriptsize If $i,j$ are connected ($\varepsilon_{ij}=1$), add the size of a buffer from $i$ to $j$\\($\delta_{ij} = T_{ij} b |VC| / L$)} = \sum_{i=1}^{N_r} \sum_{j=1}^{N_r} \varepsilon_{ij} \delta_{ij} 
\end{alignat}
\normalsize

We calculate the size of the central buffers by a combining the size
of the central buffer itself ($\delta_{cb}$) with the size of the staging 
I/O buffer ($|VC|$ per port).
The final sum is independent of wire latencies and
the use of SMART links. 


%
\begin{alignat}{1}
\Delta_{cb} = \sum_{\parbox{4em}{\centering\scriptsize All routers}} \parbox{9em}{\centering\scriptsize Size of a central buffer ($\delta_{cb}$)\\+ Size of I/O staging buffers ($2k' |VC|$)} = N_r (\delta_{cb} + 2k' |VC|) 
\end{alignat}
\normalsize

\begin{figure*}[t]
\centering
   \includegraphics[width=0.65\textwidth]{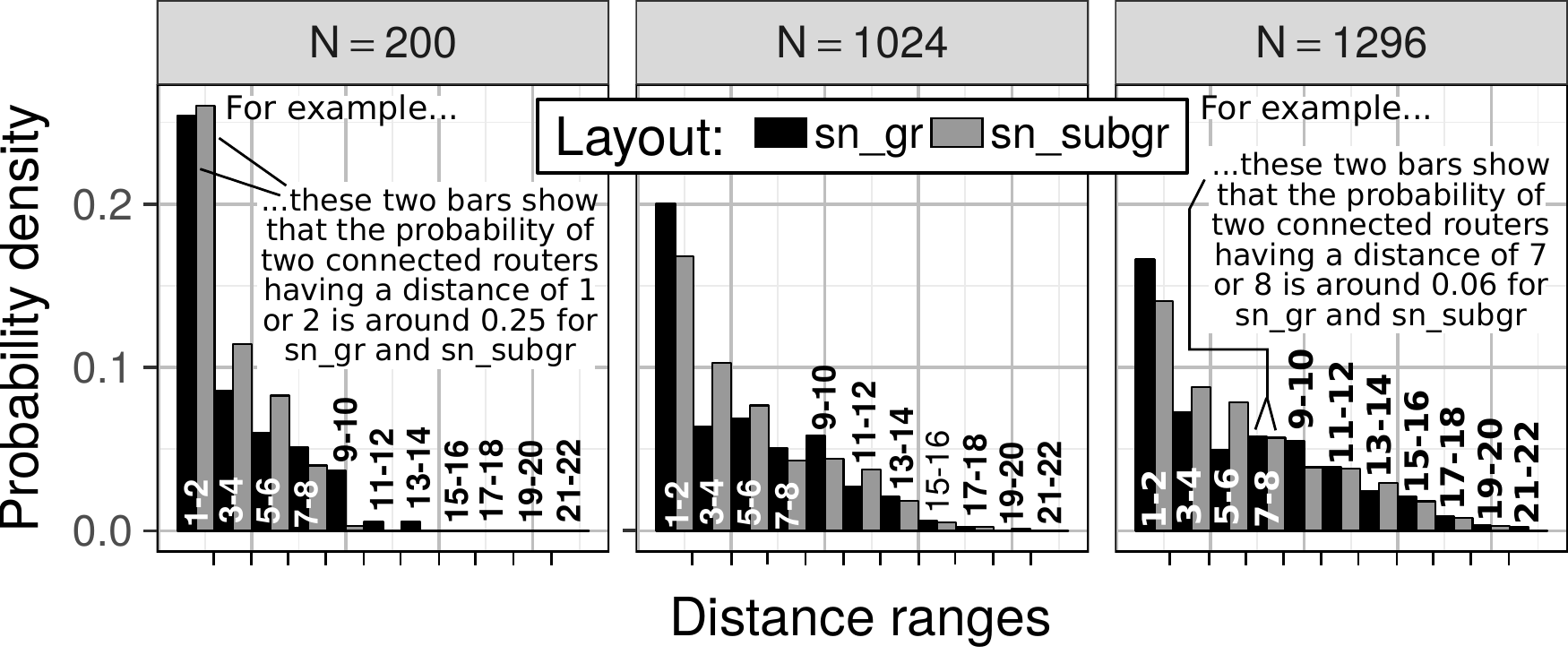}
 \caption{(\cref{sec:sf_layouts}) Distribution of link distances in \texttt{SN}s. A bar 
 associated with a distance range $X$
 illustrates the probability that, for a given layout, two routers are
 connected with a link that has the distance falling within $X$. 
 Bars of different colors are placed pairwise so that it is easier to compare the subgroup and group layouts.}
\label{fig:wire-dist-analysis-sf}
\end{figure*}


\subsection{Cost-Effective Slim NoC Layouts}
\label{sec:sf_layouts}

\change{(§4.3) We improved clarity and ensured better consistency between the
naming in this section and the terminology used in Figure 4.}

%
We now use the placement, buffer, and cost models from~\mbox{\cref{sec:layoutModel}}
to develop and analyze layouts that minimize
the average wire length and the sum of all buffers sizes (see
Figure~\ref{fig:onchip-layouts}). For each layout, we provide detailed
coordinates as a function of router labels $[G|a,b]$ (defined
in Table~\ref{tab:symbols} and in~\mbox{\cref{sec:placement_model}}, paragraph ``Labels'').
%
%
We start from the \emph{basic layout} (\texttt{sn\_basic}): subgroups with
identical intra-subgroup connections are grouped together and a router
$[G|a,b]$ has coordinates $(b,a+Gq)$. As such subgroups are not directly
connected, this layout may lengthen inter-subgroup links. To avoid this, the
\emph{subgroup layout} (\texttt{sn\_subgr}) mixes
subgroups pairwise to shorten wires between subgroups. In this layout, a router $[G|a,b]$ has coordinates
$(b,2a-(1-G))$. Both \texttt{sn\_basic} and \texttt{sn\_subgr} have a
rectangular shape ($q \times 2q$ routers) for easy manufacturing.
Finally, to reduce the wiring complexity, we use the \emph{group layout}
(\texttt{sn\_gr}) where subgroups of different types are merged pairwise and 
the resulting groups are placed in a shape as close to a square as possible.
%
%
In \texttt{sn\_gr}, there are $q$ groups, each group has identical intra-group connections, and
$2(q-1)$ wires connect every two groups. 
The router coordinates are as follows:

\footnotesize
\small
\begin{alignat}{1}
x &= \left[(a-1) \left\lceil \sqrt{2q} \right\rceil\right] \Mod \left(\left\lceil \sqrt{2q} \right\rceil \left\lceil \sqrt{q} \right\rceil\right) + (b+Gq)\Mod \left\lceil \sqrt{2q} \right\rceil \nonumber\\
y &=  \left\lfloor \frac{a-1}{\left\lceil \sqrt{q} \right\rceil} \right\rfloor \left\lceil \frac{2q}{\left\lceil \sqrt{2q} \right\rceil} \right\rceil + \left\lceil \frac{b+Gq}{\left\lceil \sqrt{2q} \right\rceil} \right\rceil \nonumber
\end{alignat}
\normalsize

\subsubsection{Evaluating Average Wire Length $M$ and Total Buffer Size $\Delta_{eb}$, $\Delta_{cb}$}
\label{sec:eval_m_delta}

%
%
%
%
We evaluate each layout by calculating $M$, $\Delta_{eb}$, and $\Delta_{cb}$; the results are
shown in Figure~\ref{fig:wire-buffer-analysis-sf}.
We also compare to a layout where routers are placed \emph{randomly} 
in $q \times 2q$ slots (\texttt{sn\_rand}).
Figure~\ref{fig:M-analysis-sf} shows $M$. Both
\texttt{sn\_subgr} and \texttt{sn\_gr} reduce the average wire length by $\approx$25\% 
compared to \texttt{sn\_rand} and \texttt{sn\_basic}. This reduces $\Delta_{eb}$
as illustrated in Figure~\ref{fig:B-analysis-sf};
\texttt{sn\_gr} reduces $\Delta_{eb}$ by $\approx$18\%. 

\macb{Investigating Details}
\hlgreen{
Figure~\ref{fig:wire-dist-analysis-sf} shows the distribution of wire distances in
\texttt{SN}s with $N \in \{200, 1024, 1296\}$ 
(these \texttt{SN}s are described in detail in~\mbox{\cref{sec:fixed_sf_networks}})
for two best layouts:
\texttt{sn\_subgr}, \texttt{sn\_gr}. 
%
%
The distributions for $N \in \{1024, 1296\}$ are similar. We find that 
\texttt{sn\_gr} uses the largest number of the longest
links for $N=200$, 
%
%
while \texttt{sn\_subgr} uses fewer links traversing the whole die. We use
this for designing example ready-to-use Slim NoCs 
(\mbox{\cref{sec:fixed_sf_networks}}) that reduce the average wire length the most. 
We analyze other \texttt{SN}s  for $1 \leq q \leq 37$: 
Both \texttt{sn\_gr} and \texttt{sn\_subgr} consistently reduce the number
of the longest wires compared to \texttt{sn\_rand} and \texttt{sn\_basic},
lowering $M$ and $\Delta_{eb}$ or $\Delta_{cb}$.
}

\begin{figure*}[hbtp]
\centering
 \subfloat[The layout of \texttt{SN-S}, a small-scale \texttt{SN} with $N=200, N_r = 50$.]
{
  \includegraphics[width=0.7\textwidth]{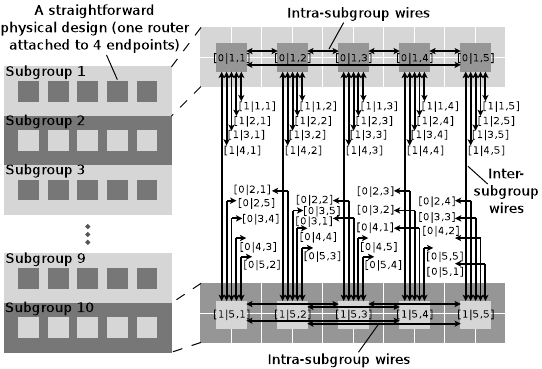}
  \label{fig:sf-s-concrete}
 }\\
  \subfloat[The layout of \texttt{SN-L}, a large-scale \texttt{SN} with $N=1296, N_r = 162$.]
{  \includegraphics[width=0.67\textwidth]{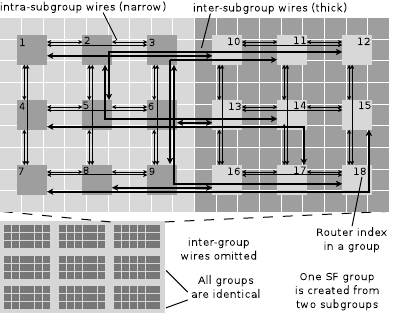}
  \label{fig:sf-l-concrete}
 }
\caption{(\cref{sec:fixed_sf_networks}) \hlgreen{Example SNs. For clarity, we only show wires
connected to two subgroups (in \texttt{SN-S}) and omit inter-group
wires (in \texttt{SN-L})}.}
\label{fig:concrete-examples}
\end{figure*}

\macb{Adding SMART}
%
With SMART~\cite{Chen:2013:SSR:2485288.2485371},
our \texttt{sn\_subgr}
and \texttt{sn\_gr} layouts reduce $\Delta_{eb}$ by $\approx$10\% compared to
\texttt{sn\_basic} (Figure~\ref{fig:B-smart-analysis-sf}).

\macb{Adding CBRs}
%
We consider small CBs ($\delta_{cb} = 20$) and large CBs ($\delta_{cb} = 40$).
As shown in Figures~\ref{fig:B-analysis-sf}--\ref{fig:B-smart-analysis-sf}, CBs result in the lowest total buffer size because
the CB size is independent of network radix ($k'$) and the round-trip time between routers ($T_{ij}$).
\subsubsection{Verifying Constraints}


We verify that \texttt{SN} layouts satisfy Eq.~(\ref{eq:constr_2}), i.e., 
the technological wiring constraints.
%
%
We assume intermediate metal layers~\cite{balfour2006design},
wiring densities of 3.5k/7k/14k wires/mm, and processing core areas of
4mm$^2$ / 1mm$^2$ / 0.25mm$^2$ at 45nm / 22nm / 11nm~\cite{6831831}.
$W$ is the product of the wiring density and the length of a side of a core.
We assume that we use a single metal layer (the worst-case scenario); all
the remaining available layers can be used by caches and other core
components.
We present the 45nm analysis in Figure~\ref{fig:W-analysis-sf} (other technology nodes are
similar). No layout violates Eq.~(\ref{eq:constr_2}).
%
%
We conclude that \texttt{SN} layouts offer advantageous wire length distributions that satisfy the
considered technology constraints, enabling feasible manufacturing.
%

\subsubsection{Theoretical Analysis}


To formalize \texttt{SN} layouts, we present a brief theoretical analysis.

\goal{Calculate asymptotic upper bounds}
\textbf{Bounds: }
The maximum Manhattan distance between two routers is $q-1$
or $2q-1$ if they
are in the same or different subgroups.
With this, {\small $M \leq q-1+2q^4/N = \sqrt[3]{2N/3} +
(\sqrt[3]{4/9})^2 \sqrt[3]{N}$}.
Similarly, {\small $\Delta \leq 6Nb(2\sqrt[3]{2N/3}+1)/L$}.
%
%
Thus, {\small $M \in O(\sqrt[3]{N})$ and $\Delta \in O(N \sqrt[3]{N})$}.
%
%
%
%
\goal{Calculate asymptotic lower bounds and conclude}
Next, we denote the sum of the lengths of all the intra-group and the inter-group
cables as  {\small $\mathfrak{L}$} and {\small $\mathfrak{G}$}, respectively.
One can show that {\small $\mathfrak{L} \geq {3 N_r q}/{8}$}.
%
%
Then, each router has connections to $q$ other subgroups. The inter-group wires
with minimum distances originate in two subgroups containing routers with
coordinates $(q,\cdot)$ and $(q+1,\cdot)$ that are in the center of the die.
For both subgroups, the minimum possible distances to the connected subgroups
are: {\small $1, 1, 3, 3, ..., q-2, q-2, q$}; using it one can show that {\small $\mathfrak{G} \in
\Omega(q^4)$}.
We have {\small $M = 2 {\mathfrak{L}+\mathfrak{G}}/{N_r k'} \in
\Omega(\sqrt[3]{N})$}.
Similarly, {\small $\Delta \in \Omega(N \sqrt[3]{N})$}. Thus: 
{\small $M \in \Theta(\sqrt[3]{N}),\ \Delta \in \Theta(N \sqrt[3]{N})$}.

\begin{thm}
In an SF with $u = 0$ and the subgroup layout, $M = \Theta(\sqrt[3]{N})$ and
$\Delta = \Theta(N \sqrt[3]{N})$.
\end{thm}

\vspace{-1em}
\begin{alignat}{2}
M &\in \Theta(\sqrt[3]{N}),\ \Delta &\in \Theta(N \sqrt[3]{N})
\end{alignat}
\normalsize

\subsection{Examples of Slim NoC Networks}
\label{sec:fixed_sf_networks}


%
%
%
%
%
We now illustrate example \texttt{SN}s that use layouts
from~\mbox{\cref{on-chip-slim-fly}} and can be used to manufacture future massively
parallel chips.

\begin{table*}[b]
 \setlength{\tabcolsep}{0.4pt} 
     \renewcommand{\arraystretch}{0.4} 
\centering
\fontfamily{zi4}\selectfont
{\begin{tabular}{@{}l|lllllllll@{}}
+ & 0 & 1 & 2 & u & v & w & x & y & z \\ \midrule
0 & 0 & 1 & 2 & u & v & w & x & y & z \\
1 & 1 & 2 & 0 & v & w & u & y & z & x \\
2 & 2 & 0 & 1 & w & u & v & z & x & y \\
u & u & v & w & x & y & z & 0 & 1 & 2 \\
v & v & w & u & y & z & x & 1 & 2 & 0 \\
w & w & u & v & z & x & y & 2 & 0 & 1 \\
x & x & y & z & 0 & 1 & 2 & u & v & w \\
y & y & z & x & 1 & 2 & 0 & v & w & u \\
z & z & x & y & 2 & 0 & 1 & w & u & v 
\end{tabular}}
\ 
{\begin{tabular}{@{}l|lllllllll@{}}
$\times$ & 0 & 1 & 2 & u & v & w & x & y & z \\ \midrule
0 & 0 & 0 & 0 & 0 & 0 & 0 & 0 & 0 & 0 \\
1 & 0 & 1 & 2 & u & v & w & x & y & z \\
2 & 0 & 2 & 1 & x & z & y & u & w & v \\
u & 0 & u & x & 2 & w & z & 1 & v & y \\
v & 0 & v & z & w & x & 1 & y & 2 & u \\
w & 0 & w & y & z & 1 & u & v & x & 2 \\
x & 0 & x & u & 1 & y & v & 2 & z & w \\
y & 0 & y & w & v & 2 & x & z & u & 1 \\
z & 0 & z & v & y & u & 2 & w & 1 & x
\end{tabular}}
\ 
{\begin{tabular}{@{}c|c@{}}
$el$ & $-el$ \\ \midrule
0      & 0       \\
1      & 2       \\
2      & 0       \\
u      & x       \\
v      & z       \\
w      & y       \\
x      & u       \\
y      & w       \\
z      & v      
\end{tabular}}
\
{\begin{tabular}{@{}l|llllllll@{}}
+ & 0 & 1 & u & v & w & x & y & z \\ \midrule
0 & 0 & 1 & u & v & w & x & y & z \\
1 & 1 & 0 & v & u & x & w & z & y \\
u & u & v & 0 & 1 & y & z & w & x \\
v & v & u & 1 & 0 & z & y & x & w \\
w & w & x & y & z & 0 & 1 & u & v \\
x & x & w & z & y & 1 & 0 & v & u \\
y & y & z & w & x & u & v & 0 & 1 \\
z & z & y & x & w & v & u & 1 & 0
\end{tabular}}
\ 
{\begin{tabular}{@{}l|llllllll@{}}
$\times$ & 0 & 1 & u & v & w & x & y & z \\ \midrule
0 & 0 & 0 & 0 & 0 & 0 & 0 & 0 & 0 \\
1 & 0 & 1 & u & v & w & x & y & z \\
u & 0 & u & w & y & x & z & 1 & v \\
v & 0 & v & y & x & 1 & u & z & w \\
w & 0 & w & x & 1 & z & v & u & y \\
x & 0 & x & z & u & v & y & w & 1 \\
y & 0 & y & 1 & z & u & w & v & x \\
z & 0 & z & v & w & y & 1 & x & u
\end{tabular}}
\ 
{\begin{tabular}{@{}c|c@{}}
$el$ & $-el$ \\ \midrule
0      & 0       \\
1      & 1       \\
u      & u       \\
v      & v       \\
w      & w       \\
x      & x       \\
y      & y       \\
z      & z      
\end{tabular}}

\caption{Addition, product, and inverse element tables for an \texttt{SN} based on $\mathbb{F}_9$
(three tables on the left) and on $\mathbb{F}_8$ (three tables on the right).}
\label{tab:add_mul_tables}
\end{table*}


\change{(§5.1) We use full names of the mentioned parameters for better readability.}

\macb{A Small Slim NoC for Near-Future Chips}
\label{sec:small_sf}
\hlgreen{
We first sketch an \texttt{SN} design in Figure~\ref{fig:sf-s-concrete} with $200$ nodes and $50$ routers (denoted as
\texttt{SN-S}), targeting the scale of the SW26010~\cite{fu2016sunway} manycores that are becoming more common~\cite{tile-mx100}. 
%
%
%
The input parameter $q=5$ is prime and \texttt{SN-S} is based on a simple finite
field $\{0, ..., 4\}$.
\texttt{SN-S} uses concentration $p=4$ and network radix $k'=7$, and it consists of 10 subgroups and five
groups. Thus, for a rectangular die ($10 \times 5$ routers), we
use the subgroup layout. This layout also minimizes the number of the longest links
traversing the whole chip, cf.~Figure~\ref{fig:wire-dist-analysis-sf}. 
}

\change{(§5.2) We use full names of the mentioned parameters for better readability.}

\macb{A Large Slim NoC for Future Manycores}
\label{sec:big_sf}
\hlgreen{
The next \texttt{SN} design (denoted as \texttt{SN-L}) addresses future massively parallel chips with
$>$1k cores. We use network radix $k'=13$ and concentration $p=8$ (one router with its nodes form a
square). As the input parameter $q=9=3^2$ is a prime power, we use a finite field $\mathbb{F}_9$ 
that cannot be a simple set $\{0, ..., 8\}$ but must be designed by hand
(details are at the end of this section).
%
%
\texttt{SN-L} has a regular structure with 1296 nodes and 162 routers
belonging to 9 identical groups ($18 \times 9$ routers).
Thus, we use the group layout for easy manufacturing ($3 \times 3$ groups) that
is illustrated for this particular design in 
Figure~\ref{fig:sf-l-concrete}.
}

\change{(§5.3) We use full names of the mentioned parameters for better readability.}

\macb{A Large Slim NoC with Power-of-Two $N$}
\hlgreen{
We also construct an \texttt{SN} with $1024$ nodes and router radix $12$. Its core
count matches the future Adapteva Epiphany~\cite{olofsson2016epiphany} chip.
This \texttt{SN} uses a subgroup layout. Similarly to \texttt{SN-L}, it is
based on a prime power $q = 8$. 
%
}

\subsection{Generating Slim NoCs: Details}
\label{sec:slim_fly_simplified_alg}

We now provide details on the general algorithm for developing SNs.
We first extend the algorithm by Besta and Hoefler~\cite{slim-fly}
(\cref{sec:slim_fly_simplified_alg}) so that its input is port counts $k'$ and
$p$, enabling \texttt{SN}s for a selected router design. Then, we
snow how to use non-prime finite fields with \texttt{SN} to
enable networks that better fit various NoC constraints such as
die dimensions or core counts (\cref{sec:non-prime-fields}). Finally, we discuss using the number of
\hl{nodes} $N$ as input to construct \texttt{SN}s for a give network size (\cref{sec:use-N}).
%

\subsubsection{Constructing \texttt{SN} for a Fixed Router Radix}
\label{sec:slim_fly_simplified_alg_fixed_k}

\ 
We first select router parameters
$k'$ and $p$ to generate an \texttt{SN} with
a desired router design.
Here, $k' = 2w+q$; $w \in \mathbb{N}$ and $q$ is a prime power described
in~\cref{sec:background_slim-fly} ($q=4w+u$; $u \in \{\pm1,0\}$).
%
%
%
Second, we construct a finite field $\mathbb{F}_q$ with $q$ elements.
Third, we find an element $\xi \in \mathbb{F}_q$ that \emph{generates}
$\mathbb{F}_q$: all non-zero elements of $\mathbb{F}_q$ can be written as
$\xi^{i}$ ($i \in \mathbb{N}$). There is no universal algorithm for finding
$\xi$ but a simple exhaustive search can be used. Next, we use $\xi$ to
construct two \emph{generator sets} $X$ and $X'$; for $u=1$ we have: $X=\{1,
\xi^2,..., \xi^{q-3}\}$ and $X'=\{\xi, \xi^3,..., \xi^{q-2}\}$.
Finally, intra-subgroup and inter-subgroup router connections are established by the following
equations (the notation $[G|a,b] \rightleftharpoons [G'|a',b']$ indicates that
routers $[G|a,b]$ and $[G'|a',b']$ are connected), respectively:
%

%

\begin{gather}
\Big([0|a,b] \rightleftharpoons [0|a,b']\Big) \Leftrightarrow \Big((b+(-b')) \in X\Big) \\
\Big([1|m,c] \rightleftharpoons [1|m,c']\Big) \Leftrightarrow \Big((c+(-c')) \in X'\Big)
\end{gather}

\begin{gather}
\Big([0|a,b] \rightleftharpoons [1|m,c] \Big) \Leftrightarrow \Big(b = (m \cdot a + c)\Big)
\end{gather}

%


\subsubsection{Using Non-Prime Finite Fields}
\label{sec:non-prime-fields}

\ 
Fields $\mathbb{F}_q$ based on a prime $q$ are simple:
$\mathbb{F}_q = \{0, ..., q-1\}$; they are used in the original \texttt{SF}. 
Yet, we discover that graphs based on non-prime $q$ 
(non-prime finite fields) fit with various NoC
constraints (such as die dimensions) or reduce wiring complexity. We analyzed
all the graphs resulting from non-prime $q$ with $N < 2,000$ and we generated
and evaluated the ones that are appropriate candidates for \texttt{SN}. For
example, $q = 9$ results in a graph with 9 identical groups, facilitating
manufacturing.

We build the associated fields by hand. For this, we first
define $\mathbb{F}_q$ to be a set of arbitrary $q$ elements.  Then, we
construct \emph{addition}, \emph{product}, and \emph{inverse element} tables:
they define the operations on $\mathbb{F}_q$ (adding two elements, multiplying
two elements, and taking the reverse of an element; all done modulo $q$).  Such
tables can easily be derived using an exhaustive search.  Finally, connections
between routers are defined by equations
from~\cref{sec:slim_fly_simplified_alg}; all the operations in these equations
are defined by the respective operation tables.

As an example, for $p=8$ and $k'=13$ we have $q=9$. As $q$ is a prime
\emph{power}, there is no finite field $\mathbb{F}_9 =
\{0,...,8\}$~\cite{lidl1997finite}. Instead, we define $\mathbb{F}_9 =
\{$\texttt{0,1,2,}\texttt{u,v,w} \texttt{x,y,z}$\}$; the symbols are chosen
arbitrarily. Then, we construct {addition}, {product}, and {inverse element}
tables that define the operations on $\mathbb{F}_9$; see
Table~\ref{tab:add_mul_tables}. Next, we calculate $\xi$.  There are 4 such
(equivalent) elements: \texttt{v,w,y,z}. We then construct
$X=\{$\texttt{1,x,2,u}$\}$ and $X' = \{$\texttt{v,y,z,w}$\}$ and use them to
connect routers.
 
\subsubsection{Constructing \texttt{SN} for a Fixed Network Size} \label{sec:use-N}

\ 
Alternatively, one can also select network size.
One first verifies if the selected $N$ is feasible: $N$ must be
equal to $N_r \cdot p$. Thus, in step~1, a $p$ is selected such that $N = p
\cdot N_r$ and $N_r = 2 q^2$, where $q$ is a prime power as specified
in~\cref{sec:background_slim-fly}
{($N \ne N_r \cdot p$ is also feasible by removing
some \hl{nodes} from selected tiles; a strategy used in, e.g, fat trees)}.
{Then, in step~2, one verifies
the $\kappa$ parameter that determines the tradeoff between node density
and contention (see paragraph~3 in~}\cref{sec:background_slim-fly}){. If
$\kappa$ is acceptable, \texttt{SN} can be developed as described
in~}\cref{sec:slim_fly_simplified_alg}.
%


\section{Slim NoC Microarchitecture}
\label{sec:cb_sf}

\commt{(§6) 
We enhanced the writing style for more clarity by, for example, using fewer
acronyms or simplifying sentences.}

\texttt{SN} provides the lowest radix $k'$ for the diameter of two, minimizing buffer area
and power consumption while providing low latency.\footnote{We use
\texttt{SF} as the basis of our new NoC design (as \texttt{SF} consistently
outperforms \texttt{DF} as shown in Figure~\ref{fig:motivation_analysis}).
While we select a variant of \texttt{SF} with diameter two as the main design
in this work~\cite{slim-fly}, most of our schemes are generic and can be
applied to any \texttt{SF} and \texttt{DF} topology.} In this section, we further reduce
buffer space by extending \texttt{SN} with Central Buffers (CBs)~\cite{6558397}
and optimizing it with ElastiStore
(ES)~\cite{Seitanidis:2014:EEB:2616606.2616900}. We first provide
background information on CBs and ES (\mbox{\cref{sec:background-arch}}).
Then, we show how to combine CBs with virtual channels (VCs) to enable
deadlock-freedom (\mbox{\cref{sec:micro-es}}), describe our deadlock-freedom
mechanism (\mbox{\cref{sec:deadlcok}}), and explain how to maintain full utilization of links when using routers with central buffers
(\mbox{\cref{sec:full_link}}).

%




\subsection{Techniques to Improve Slim NoC Performance}
\label{sec:background-arch}

%
  In order to enhance Slim NoC for high performance and
low energy consumption, we use two additional mechanisms: \emph{Elastic Links} 
and \emph{Central Buffer Routers}.
%


\macb{Elastic Links: Lower Area and Power Consumption}
\label{sec:e_links_background}
%
%
%
%
To reduce the area and power consumption of a NoC, Elastic Buffer (EB) Links~\cite{4798250, Seitanidis:2014:EEB:2616606.2616900}
remove input buffers and repeaters within the link pipelines and replace
them with master-slave latches.
%
%
To prevent deadlocks in Slim NoC, we use ElastiStore (ES), which is an
extension of EB links~\cite{Seitanidis:2014:EEB:2616606.2616900}. We present
design details in \mbox{\cref{sec:micro-es}}.

\macb{Central Buffer Routers: Less Area}
\label{sec:onchip_background}
%
  To further reduce area, we use Central Buffer Routers (CBRs)~\cite{6558397}. In a CBR, multi-flit edge (input) buffers
are replaced with single-flit input staging buffers and a central buffer (CB)
shared by \emph{all} ports. At low loads, the CB is bypassed, providing a two-cycle
router latency. At high loads, in case of a conflict at the output port, a flit
passes via the CB, taking four cycles.
A CBR employs 3 allocation and 3 traversal stages, which increase the area
and power in arbiters and allocators. 
%
%
Yet, it significantly reduces buffer space and thus overall router area
and power consumption~\cite{6558397}. 
%
%
%

\subsection{Combining Virtual Channels with CBs}
\label{sec:micro-es}

\begin{figure}[h!]
\centering
 \subfloat{
  \includegraphics[width=0.46\textwidth]{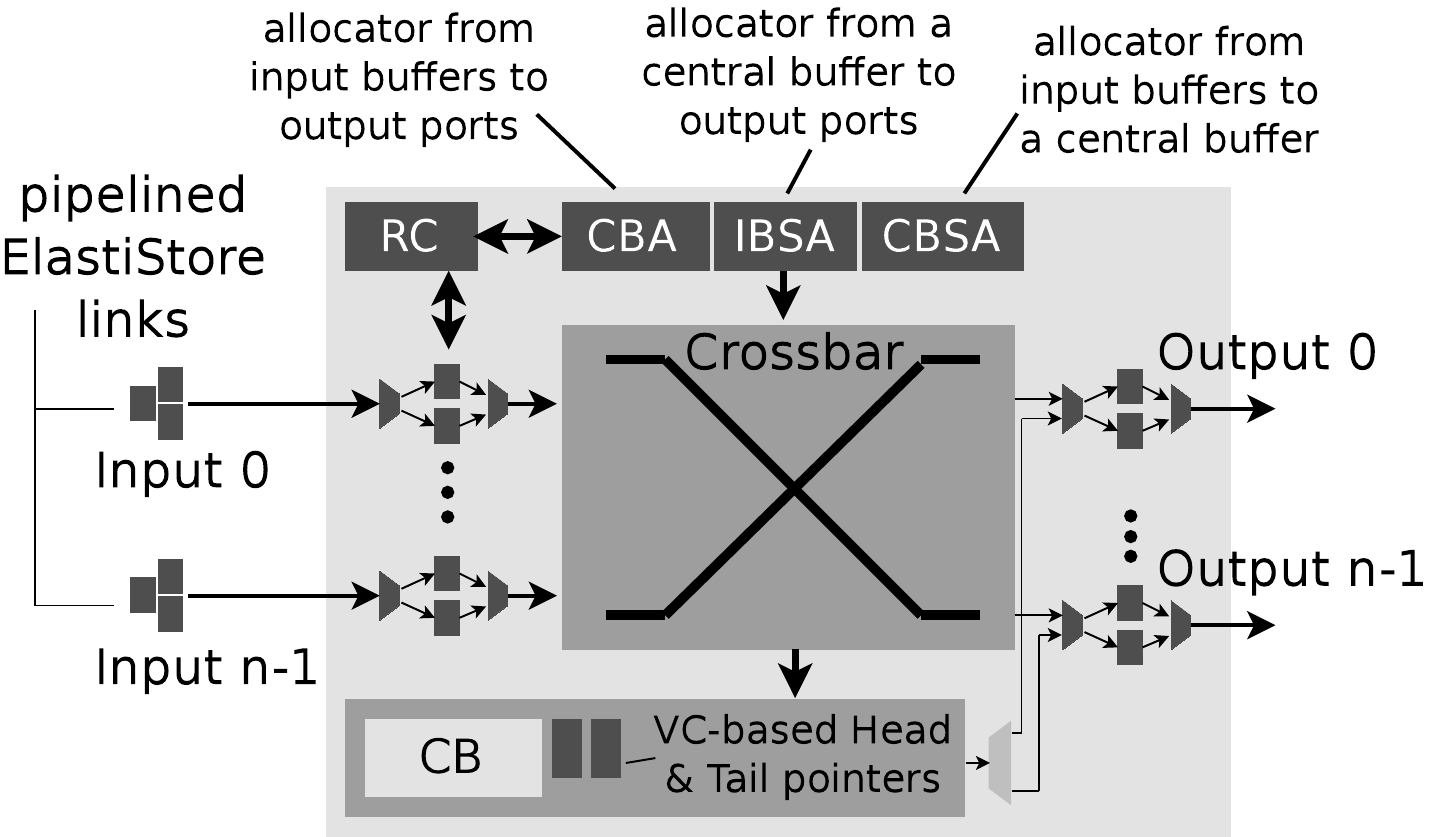}
 }
  \caption{Modifications to the CB router in~\cref{sec:cb_sf}.}
 %
%
\label{fig:cb_pfbf}
\end{figure}

%
Original CBs do not support VCs. To alleviate this, we use ElastiStore (ES)
links~\cite{Seitanidis:2014:EEB:2616606.2616900} that enable multiple VCs in
the EB channels. The design is presented in Figure~\ref{fig:cb_pfbf}.
ElastiStore links use a separate pipeline buffer and associated control logic
for each VC. The per-VC ready-valid handshake signals independently handle the
flits of each VC, removing their mutual dependence in the pipelined link. We
only keep a slave latch per VC and share the master latch between all VCs.
This reduces the overall area and power due to ElastiStore links. The
resulting performance loss is minimal and reaches $\frac{1}{|VC|}$ only when
all VCs except one are blocked in the pipeline.
%
%
Other modifications (shown in dark grey) include using per-VC (instead of
per-port) I/O staging buffers and CB head/tail pointers to keep VCs
independent. The crossbar radix is $k' (k'+1)$, like in the original
CBR. For this, we use a small mux/demux before and
after the crossbar inputs and outputs. We maintain single input and output for
the CB, which only negligibly impacts performance~\cite{6558397}.


%



\subsection{Ensuring Deadlock Freedom with CBRs}
\label{sec:deadlcok}

\change{(§6.1) We explain the bypass
in central buffer routers for more clarity.}

\texttt{SN} with $D=2$ uses two VCs to avoid deadlocks: VC0 for the first and
VC1 for the second hop (assuming paths of lengths up to 2).
%
%
Here, the only dependency is that of VC0 on VC1, which is not enough to form
cycles.  We now extend this scheme to the CBR design. To ensure
deadlock freedom, two conditions must be met.
First, 
CB allocation for a packet must be atomic: it cannot happen that some
flits have entered the CB and the rest are 
stalled in the links waiting for the CB.
%
%
Second, head flits of all the packets in different ports and VCs should always be
able to compete for the allocation of output ports and VCs. However, since we focus 
on deterministic routing algorithms, this condition is not required for single deadlock-free deterministic paths. 


%

\hlgreen{
We satisfy the first condition by reserving the space required for a complete packet during the
CB allocation stage. Thus, once a packet takes the CB path, it is guaranteed to move
completely into the CB
(note that a packet may bypass the CB via the 
low-load CB bypass path; see~\mbox{\cref{sec:onchip_background}}).
A packet in the CB is always treated as a part of the
output buffer of the corresponding port and VC.
Thus, if the baseline routing is deadlock-free, it remains
deadlock-free with the CBR design.
%
%
%
}
{Finally, we use SMART links orthogonally to ElastiStore, only to reduce link latency,
ensuring no deadlocks. We avoid livelocks with deterministic paths.}

\subsection{Maintaining Full Link Utilization with CBRs}
\label{sec:full_link}

\goal{Say how we efficiently use bandwidth without edge buffers}
Large edge buffers enable full utilization of the bandwidth of long wires.
For CBR, we obtain the same effect with elastic links
(EBs)~\cite{Seitanidis:2014:EEB:2616606.2616900}.
Varying the central buffer size reduces head of line blocking.
%


\section{EVALUATION}
\label{sec:evaluation}

%
%
%
We evaluate \texttt{SN} versus other topologies in terms of latency,
throughput, $\Delta_{cb}$, $\Delta_{eb}$, area, and power consumption.

\subsection{Experimental Methodology}
\label{sec:targets_}

\change{(§7.1) We clarify the definition and functionalities of the edge buffer. We provide
the description of the minimum static routing and provide the citation. We clarify that the
edge buffer is used for the input ports in our router architecture. We provide the name
of the tool (DSENT) we use for area and power evaluation along with its citation.}

\change{(§7.2) We merged Section 7.2 with Section 7.1, as the simulation setup is a part of our
simulation methodology.}

\begin{figure}[h!]
\centering
\includegraphics[width=0.45\textwidth]{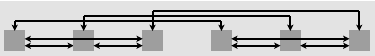}
\caption{An example partitioned FBF (1D). Nodes are not shown.}
\label{fig:pfbf}
\end{figure}

\macb{Considered Topologies}
\hl{
We compare \texttt{SN} to both low- and high-radix baselines, summarized in
Table~\ref{table:net_cfg}: (1)
tori~\cite{Alverson:2010:GSI:1901617.1902283} (\texttt{T2D}), (2) concentrated
meshes~\cite{balfour2006design} (\texttt{CM}), (3) Flattened
Butterflies~\cite{dally07} (\texttt{FBF}). We also consider \texttt{SF} and
\texttt{DF}; their results are consistently worse than others and are excluded
for brevity.} \hl{Note that, for a fixed $N$ and for $D=2$, $k$ and bisection
bandwidth of \texttt{FBF} are much higher than those of \texttt{SN}. Thus, for a
fair comparison, we develop a \emph{partitioned} \texttt{FBF} (\texttt{PFBF})
with fewer links to match \texttt{SN}'s $k$ and bisection bandwidth; see
%
%
Figure~\ref{fig:pfbf}.} \hl{We partition an original \texttt{FBF} into
smaller identical \texttt{FBF}s connected to one another by one port per node
in each dimension. \texttt{PFBF} has $D=4$ while the Manhattan distance between
any two routers remains the same as in \texttt{FBF}. Finally, even though we 
focus on direct topologies, we also briefly compare to indirect hierarchical
networks~\cite{abeyratne2013scaling}.
}

\begin{table}[h!]
\centering
\footnotesize
\sf
\setlength{\tabcolsep}{1.2pt} 
\renewcommand{\arraystretch}{0.9} 
\begin{tabular}{@{}l||c||c|c|c|c|c|c||c|c|c|c|c|c@{}}
\toprule
\multirow{2}{*}{Network} & \multirow{2}{*}{$D$} & \multicolumn{6}{c||}{$N \in \{192,200\}$}                                                     & \multicolumn{6}{c}{$N = 1296$}                                                     \\ 
                      &  & Sym. & $p$ & $k'$ & $k$ & Routers & $N$                                               & Sym. & $p$ & $k'$ & $k$ & Routers  & $N$                                              \\ \midrule
\multirow{2}{*}{\ttfamily{T2D}}      & \multirow{2}{*}{$\left\lceil \frac{1}{2} \sqrt{N_r} \right\rceil$} & t2d3 & 3 & 4 & 7 & 8x8   & 192                                                      & t2d9 &9  &4  &13 & 12x12       & 1296                                          \\
                                 &    & t2d4 & 4 & 4 & 8 & 10x5  & 200                                                      & t2d8 &8  &4  &12 & 18x9        & 1296                                                \\
\multirow{2}{*}{\ttfamily{CM}}  & \multirow{2}{*}{$\left\lceil\sqrt{N_r} - 2\right\rceil$}  & cm3  & 3 & 4 & 7 & 8x8   & 192                                                      & cm9  &9  &4  &13 & 12x12       & 1296                                                \\
                                   &  & cm4  & 4 & 4 & 8 & 10x5 & 200                                                       & cm8  &8  &4  &12 & 18x9         & 1296                                               \\
\multirow{2}{*}{\ttfamily{FBF}}   & \multirow{2}{*}{2}  & fbf3 & 3 & 14&17 & 8x8   & 192                                                      & fbf9 &9  &22 &31 & 12x12        & 1296                                               \\
                                   &  & fbf4 & 4 & 13&17 & 10x5  & 200                                                      & fbf8 &8  &25 &33 & 18x9         & 1296                                               \\
\multirow{2}{*}{\ttfamily{PFBF}}  & \multirow{2}{*}{4}   & pfbf3& 3 & 8 &11 & \begin{tabular}[c]{@{}c@{}}4 \texttt{FBF}s\\ (4x4 each)\end{tabular} & 192 & pfbf9&9  &12 &21 & \begin{tabular}[c]{@{}c@{}}4 \texttt{FBF}s\\ (6x6 each)\end{tabular} & 1296\\
                                 &    & pfbf4& 4 & 9 &13 & \begin{tabular}[c]{@{}c@{}}2 \texttt{FBF}s\\ (5x5 each)\end{tabular} & 200 & pfbf8&8  &17 &25 & \begin{tabular}[c]{@{}c@{}}2 \texttt{FBF}s\\ (9x9 each)\end{tabular} & 1296\\
\ttfamily{SN}                   & 2     & sn\_*& 4 & 7 &11 & 10x5 & 200                                                       & sn\_*&8  & 13&21 &  18x9       & 1296                                               \\ \bottomrule
\end{tabular}
\caption{Considered configurations for two example class sizes.}
\label{table:net_cfg}
\end{table}

\change{(§7.2, ``Layouts and Sizes'') We make the text more precise. For example,
We avoid imprecise statements (``Other designs are similar in performance'')
to avoid an extensive discussion on what ``similar'' means, and instead - when
necessary - provide the exact numbers.}

\macb{Layouts, Sizes}
\hl{
We compare all the \texttt{SN} layouts from~\mbox{\cref{sec:sf_layouts}} for
two size classes: $N \in \{192,200\}$ and $N=1296$.
We use both square networks as comparison points that are close in size to
\texttt{SN} ($N=192$) and rectangular ones with identical $N$ ($N=200$); 
see Table~\ref{table:net_cfg}.
}

\macb{Cycle Times}
\hlgreen{
 We use router clock cycle times \emph{to account
for various crossbar sizes}: 0.5ns for \texttt{SN} and \texttt{PFBF}, 0.4ns for
  topologies with lower radix (\texttt{T2D}, \texttt{CM}), and 0.6ns for
  high-radix \texttt{FBF}. In specified cases, for analysis purposes, we also use cycle times that are constant across
  different topologies.
}

\change{(§7.2, ``Routing'') we extend the description of minimum static routing
and justify selecting this routing scheme. We also describe our selection of
adaptive routing schemes in our preliminary analysis of the performance of Slim NoC
under adaptive routing.}
%

\macb{Routing}
%
We focus on static minimum routing where paths between routers are
calculated using Dijkstra's Single Source Shortest Path algorithm~\cite{skiena1990dijkstra}.
This is because we aim to design an \emph{energy-efficient} topology. 
Adaptive routing would increase overall router complexity and power consumption. 
%
%
%
%
Our choice is similar to what is done in many prior works that explored new
topologies~\cite{4798251, xu2009low, udipi2010towards, kilo-noc, hird-journal}.
Moreover, various works that do not introduce new topologies but 
conduct topology-related evaluations also follow
this strategy~\cite{aergia, das_micro09, kim2009low}.

Yet, it is clear that the routing algorithm can be customized or designed for each topology 
to maximize performance or energy efficiency and determining the best routing algorithm on 
a per-topology basis for a given metric is an open research problem. 
For example, \texttt{SN} can be extended to adaptive routing
using paradigms such as the Duato protocol~\cite{Dally:2003:PPI:995703},
UGAL~\cite{ugal2005scheme}, or up-down routing~\cite{Dally:2003:PPI:995703}.
We later (\mbox{\cref{sec:DISNUS}}) provide a discussion on adaptive routing in \texttt{SN}.
A full exploration of adaptive routing schemes in \texttt{SN} 
is left for future work.

%


\goal{Describe considered wire architectures}


\macb{Wire Architectures} 
\hlgreen{
We compare designs with and without SMART links. We use the latency 
from~\mbox{\cref{sec:buffer_model}} and set the number of hops
traversed in one link cycle as $H=9$ (with SMART links) and $H=1$
(no SMART links). We fix the packet size to 6 flits (except for
real benchmark traces, see below). All links are 128 bits wide.
}

\macb{Router Architectures}
\hlgreen{
  %
We use routers with central buffers or with edge buffers. An edge router has a standard
2-stage pipeline with two VCs~\cite{Peh:2001:DMS:580550.876446}. The CB router
delay is 2 cycles in the bypass path and 4 cycles in the buffered path. 
Buffer sizes in flits for routers with central buffers are: 1 (input buffer size per VC), 1
(output buffer size per VC), 20 (central buffer size), 20 (injection and
ejection queue size). The corresponding sizes for routers without central buffers are,
respectively, 5, 1, 0, 20.
}


\macb{Buffering Strategies}
\hlgreen{
\texttt{EB} and \texttt{CBR} prefixes indicate \emph{E}dge and \emph{C}entral
\emph{B}uffer \emph{R}outers. We use: \texttt{EB-Small} and \texttt{EB-Large}
(all edge buffers have the size of 5 and 15 flits), \texttt{EB-Var-S} and
\texttt{EB-Var-N} (edge buffers have minimal possible sizes for 100\% link
utilization with/without SMART links), \texttt{CBR-$x$} (CBs of size~$x$), and
\texttt{EL-Links} (only elastic links).
}


\goal{Describe the synthetic traffic used}

\macb{Synthetic Traffic}
\hlgreen{
We use 5 traffic patterns: random (\texttt{RND}, each source $s$
selects its destination $d$ with uniform random distribution), bit shuffle
(\texttt{SHF}, bits in destination ID are shifted by one position), bit reversal
(\texttt{REV}, bits in destination ID are reversed), and two adversarial patterns
(\texttt{ADV1} and \texttt{ADV2}; they maximize load on single- and
multi-link paths, respectively).
We omit the ADV2 results when they are similar to the ADV1 results.
}

%

\macb{Real Traffic}
\hlgreen{
We use PARSEC/SPLASH benchmark traces to evaluate various real workloads. We run three 
copies of 64-threaded versions of each benchmark on 192 cores \emph{to
model a multiprogrammed scenario}.
%
%
We obtain traces for 50M cycles (it corresponds to $\approx$5 billion
instructions for \texttt{SN-S}) with the Manifold
simulator~\cite{wang2014manifold}, using the DRAMSim2 main memory
model~\cite{Rosenfeld:2011:DCA:1999163.1999216}.
%
%
As threads are spawned one by one, we warm up simulations by waiting for
75\% of the cores to be executing. The traces are generated at L1's back side;
messages are read/write/coherence requests. Read requests and coherence
messages use 2 flits; write messages use 6 flits (\emph{we thus test variable
packet sizes}). A reply (6 flits) is generated from a destination for each
received read request. 
%
}


\macb{Performance Evaluation}
\hlgreen{
We use a cycle-accurate in-house simulator (described by Hassan and
Yalamanchili~\cite{6558397,wang2014manifold}). Simulations are run for 1M cycles.
For $N \in \{192, 200\}$ we use detailed topology models (each
router and link modeled explicitly). If $N=1296$, due to large memory
requirements ($>$40GB), we simplify the models by using average wire lengths and
hop counts. 
%
}

\macb{Area and Power Evaluation}
\hlgreen{
We estimate general power consumption using the DSENT tool~\cite{dsent}.
We break down area and static power (leakage) due to (1) router-router wires,
(2) router-node wires, and (3) actual routers (\texttt{RR-wires}, \texttt{RN-wires}, and
\texttt{routers}). We further break down area into global, intermediate, and active layers
(denoted as \texttt{RRg-wires}, \texttt{RRi-wires}, and \texttt{RRa-wires}; \texttt{RNg-wires}, \texttt{RNi-wires}, and
\texttt{RNa-wires}; \texttt{g-routers}, \texttt{i-routers}, and \texttt{a-routers}, respectively).
We break down dynamic power into buffers,
crossbars, and wires. 
}

\macb{Technologies and Voltages}
\hlgreen{
We use 45nm and 22nm technologies with 1.0V and 0.8V voltages.
}


\hlgreen{
%
We next present a representative subset of our results.
%
}

\subsection{Analysis of Performance (Latency and Throughput)}
\label{sec:syn_traffic}


We first examine the effects of \texttt{SN} layouts and
various buffering strategies on latency and throughput (\mbox{\cref{sec:perf_layouts}}); 
we next compare \texttt{SN} to other topologies
(\mbox{\cref{sec:other_topos_perf}}).

\subsubsection{Analysis of Layouts and Buffers}
\label{sec:perf_layouts}

\begin{figure}[h!]
\centering
 \subfloat[Synthetic traffic.]{
  \includegraphics[width=0.255\textwidth]{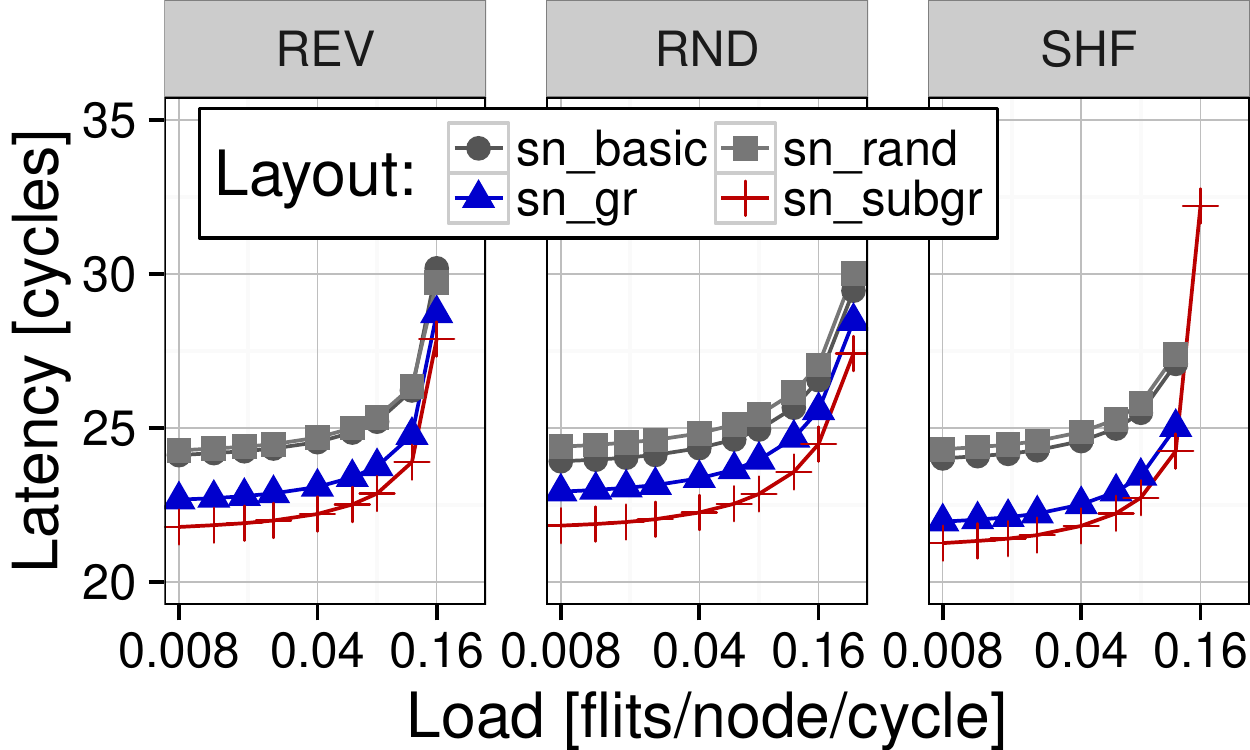}
  \label{fig:layouts-synthetic-perf-small}
 }
 \subfloat[PARSEC/SPLASH.]{
  \includegraphics[width=0.215\textwidth]{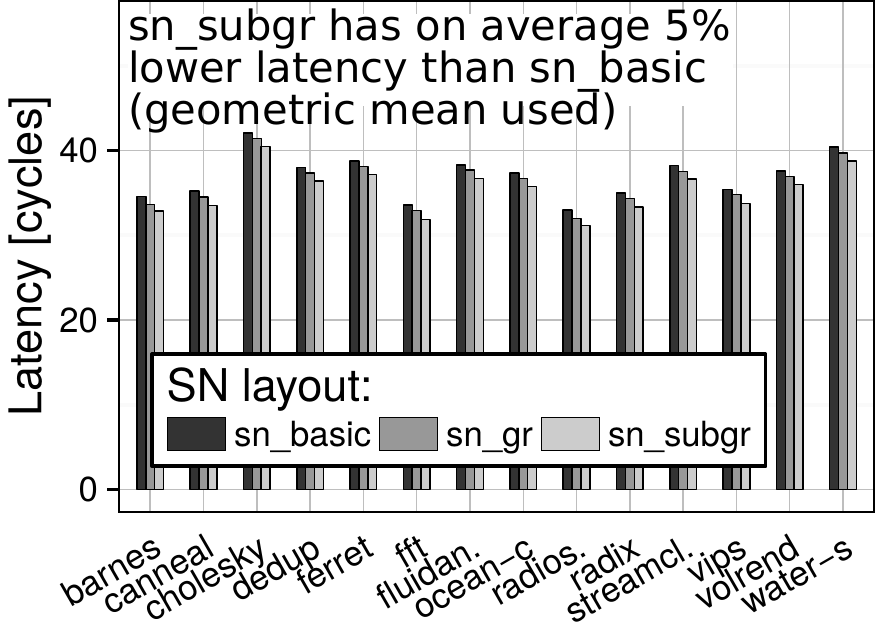}
  \label{fig:layouts-real-perf}
 }
\caption{(\cref{sec:perf_layouts}) Average packet latency with different \texttt{SN} layouts (without SMART links) for synthetic traffic and real applications, for $N=200$.} 
\label{fig:perf-layouts}
\end{figure}

Figure~\ref{fig:perf-layouts} shows how the layouts improve the performance of
\texttt{SN}. All traffic patterns follow similar trends; we focus on
\texttt{RND}.
Without SMART links, \texttt{sn\_basic} and
\texttt{sn\_rand} entail higher overheads than \texttt{sn\_subgr} and
\texttt{sn\_gr} due to longer wires.
As predicted theoretically (in~\cref{on-chip-slim-fly}), \texttt{sn\_subgr} and
\texttt{sn\_gr} are the best for respectively $N=200$ and $N=1296$ in terms of
latency and throughput.
%
%

\begin{figure}[h!]
\centering
\includegraphics[width=0.48\textwidth]{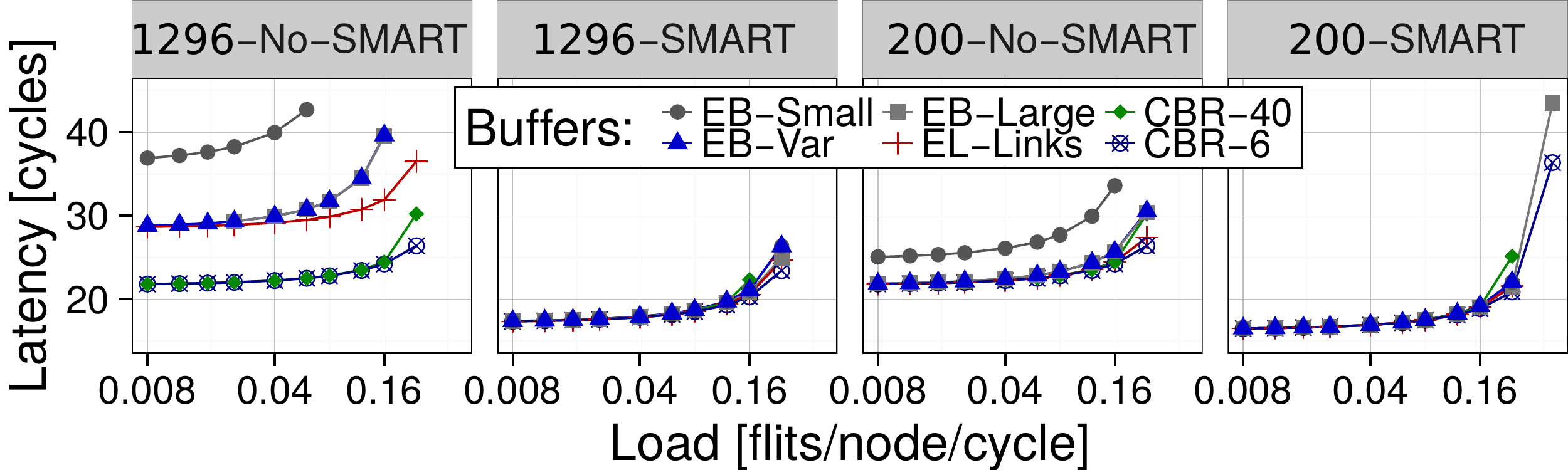}
%
\caption{(\cref{sec:perf_layouts}) Impact of buffers (edge
buffers, central buffers, no buffers); ``200''/``1296'' mean $N =
200$/$1296$; all variants explained in~\cref{sec:targets_}.} \label{fig:perf-buffers}
\end{figure}

\hl{
Figure~\ref{fig:perf-buffers} shows the average packet latency with \texttt{SN} using 
edge buffers. Without SMART links, small
edge buffers lead to higher latency due to high congestion and high overhead
of credit-based flow control.
\texttt{EL-Links} improve throughput but lead to head-of-line blocking. 
Both edge buffers and elastic links offer comparable
performance to that of central buffers for $N \in \{192, 200\}$.
}

\hlgreen{
We analyze two representative CB sizes (6 and 40 flits) in
Figure~\ref{fig:perf-buffers}; we also test sizes with 10, 20, 70, and 100
flits.
We observe that small \texttt{CB}s outperform (especially for $N > 1000$) both
edge buffers and \texttt{EL-Links} by removing head-of-line-blocking.
Large CBs (e.g., \texttt{CBR-40}) can contain more packets, increasing overall
latency.
%
%
}

\hlgreen{
We also derive the total buffer area for each buffering scheme. 
%
%
we show that \texttt{SN} gives the best tradeoff of radix (hence the crossbar size) for a given
diameter and thus \emph{it ensures the lowest total buffer size for networks with diameter two, for both edge and central buffer designs}.
}

\macb{Impact of SMART Links}
\hlgreen{
SMART links reduce the relative latency differences between Slim NoCs based on
different buffering schemes to
$\approx$1-3\% for most data points, and to up to $\approx$16\%
for high injection rates that approach the point of network saturation. 
SMART links accelerate \texttt{SN} by up to $\approx$35\%
for the \texttt{sn\_subgr} layout. 
}

\change{(§7.3.1, ``Conclusion'') Besides simplifying text, we explain
more extensively than in the previous submission the performance of edge
input buffers.}

\hl{
We conclude that: (1) group and subgroup layouts 
outperform default \texttt{SN} designs, (2) \texttt{SN} with edge buffers
can have similar latency and throughput to those of \texttt{SN} designs with elastic links or central buffers,
and (3) \texttt{SN} with small CBs has the best performance.
}

%

%


%
%
%


\subsubsection{SN versus Other Network Designs}
\label{sec:other_topos_perf}

\begin{figure}[!h]
\centering
\includegraphics[width=0.49\textwidth]{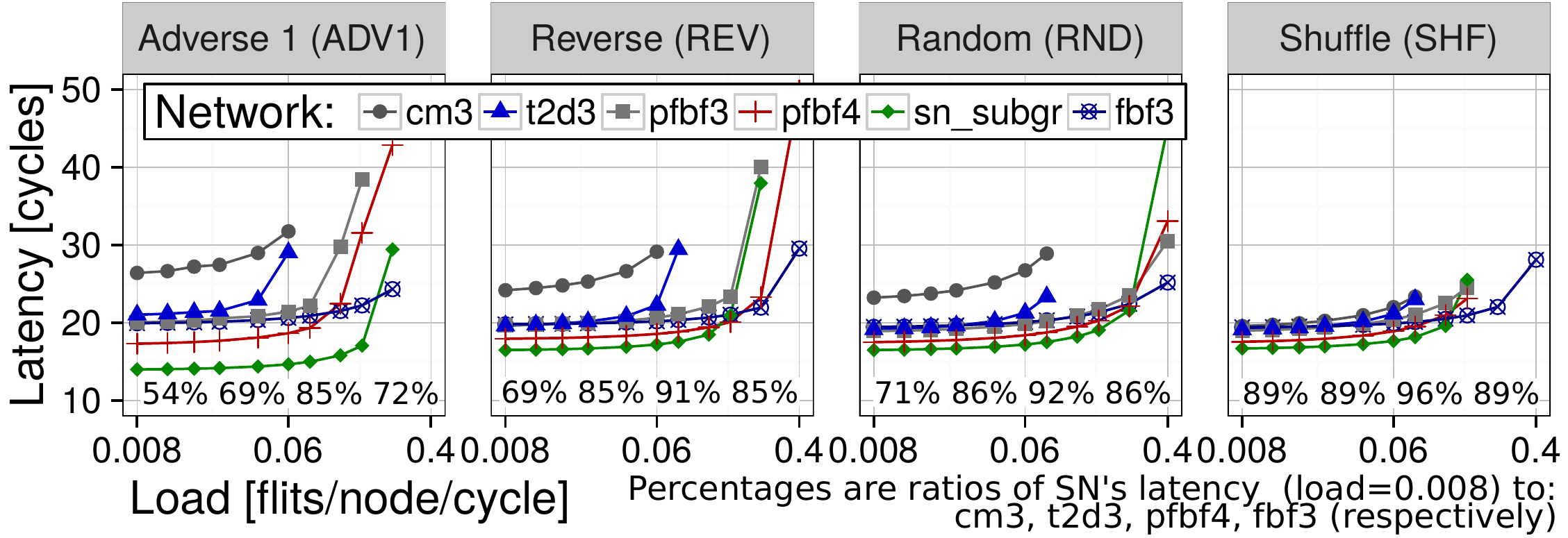}
\caption{(\cref{sec:other_topos_perf}) Performance of synthetic traffic with SMART links for small networks ($N \in \{192,200\}$) and different cycle times for different designs.}
\label{fig:small_smart_var}
\end{figure}

\begin{figure}[!h]
\centering
\includegraphics[width=0.49\textwidth]{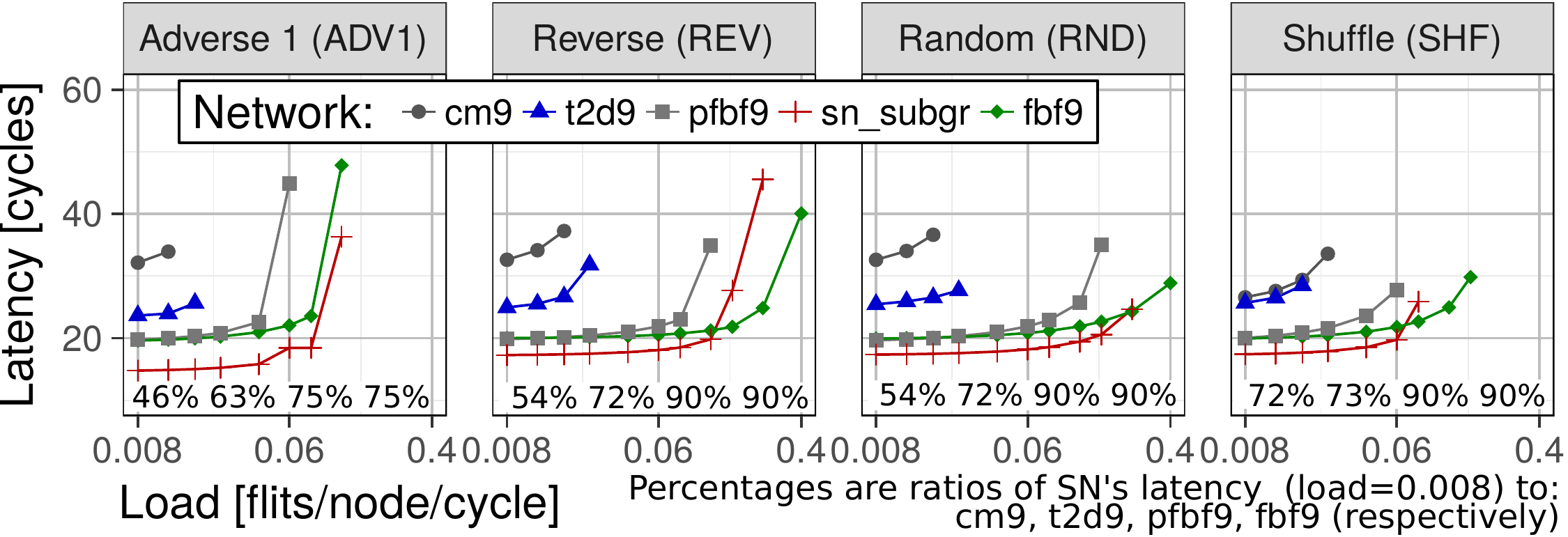}
\caption{(\cref{sec:other_topos_perf}) Performance of synthetic traffic with SMART links for large networks ($N = 1296$) and different cycle times for different designs.}
\label{fig:large_smart_var}
\end{figure}

\begin{figure}[!h]
\centering
  \includegraphics[width=0.49\textwidth]{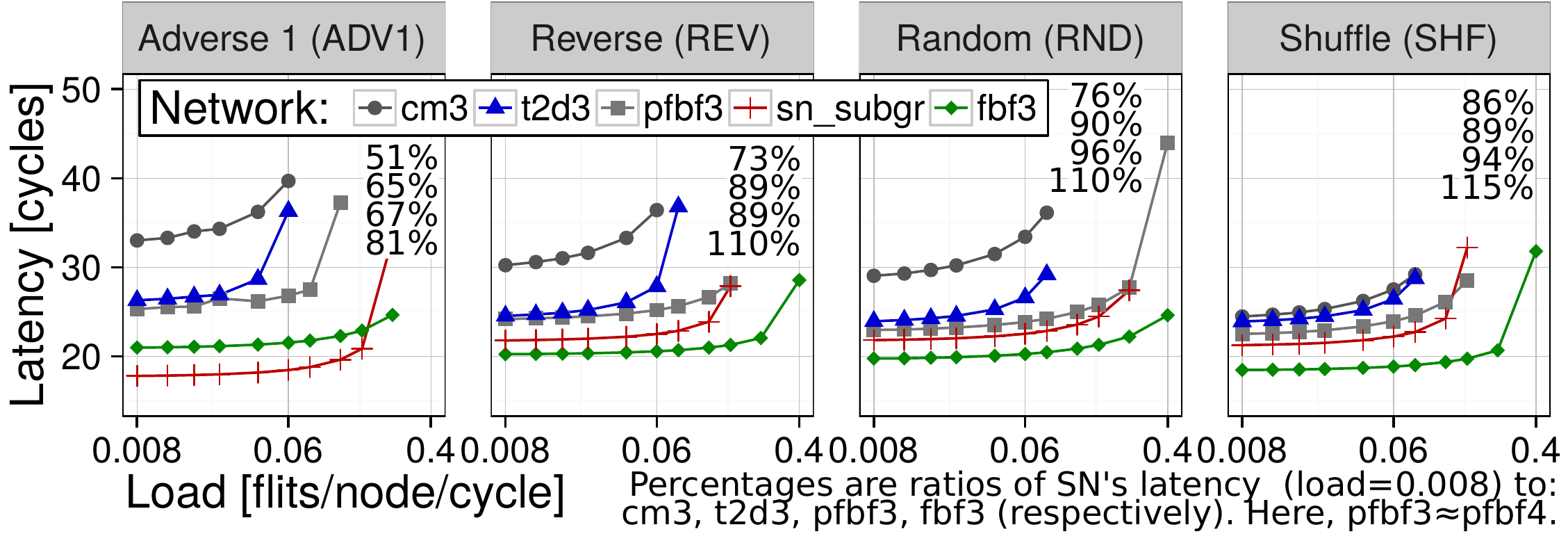}
%
\caption{(\cref{sec:other_topos_perf}) {Performance analysis, $N \in \{192,200\}$, no SMART links.}}
\label{fig:small_nosmart_perf}
\end{figure}

We show that \texttt{SN} outperforms other network designs from Table~\ref{table:net_cfg}.
%
%
The results are in 
Figures~\ref{fig:small_smart_var}--\ref{fig:large_smart_var} (SMART links) and~\ref{fig:small_nosmart_perf} (no SMART links).
As expected, \texttt{SN} \emph{always} outperforms \texttt{CM} and \texttt{T2D}.
For example, for RND and $N=1296$, \texttt{SN} improves average packet latency by
$\approx$45\% (over \texttt{T2D}) and $\approx$57\% (over \texttt{CM}), and
throughput by 10x. This is a direct consequence of \texttt{SN}'s asymptotically lower $D$
and higher bandwidth.
\texttt{SN}'s throughput is marginally lower than that of \texttt{PFBF}
in some cases (e.g., $N \in \{192,200\}$, REV) because of \texttt{PFBF}'s minimum Manhattan
paths.
Yet, in most cases \texttt{SN} has a higher throughput than \texttt{PFBF} (e.g., $>$60\% for
$N=1296$ and RND).
\texttt{SN}'s latency is always lower ($\approx$6-25\%) than that of
\texttt{PFBF} due to its lower $D$.
Finally, without SMART links, \texttt{SN}'s longer wires result in higher
latency than \texttt{FBF} in several traffic patterns ($\approx$\%26 for RND
and $N=1296$). In ADV1, \texttt{SN} outperforms \texttt{FBF} (by
$\approx$18\%). 
%
%
%
We later (\mbox{\cref{sec:real_traces_eval}}, \mbox{\cref{sec:DISNUS}}) show that
\texttt{SN} also offers a better power/performance tradeoff than
\texttt{FBF}. 

\macb{Impact of SMART Links}
\hlgreen{
SMART links do not impact the latency of \texttt{CM} and \texttt{T2D} as these
topologies mostly use single-cycle wires.  As expected, SMART links diminish the
differences in the performance of different networks with multi-cycle wires. 
%
%
%
}


\begin{figure}
\centering
 \subfloat[Area of various \texttt{SN} layouts.]{
  \includegraphics[width=0.13\textwidth]{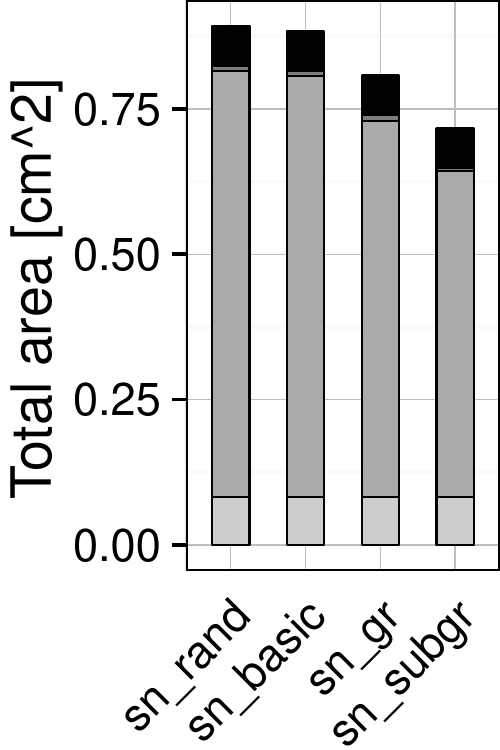}
  \label{fig:nsap1}
 }\hfill
 \subfloat[Area of various networks.]{
  \includegraphics[width=0.13\textwidth]{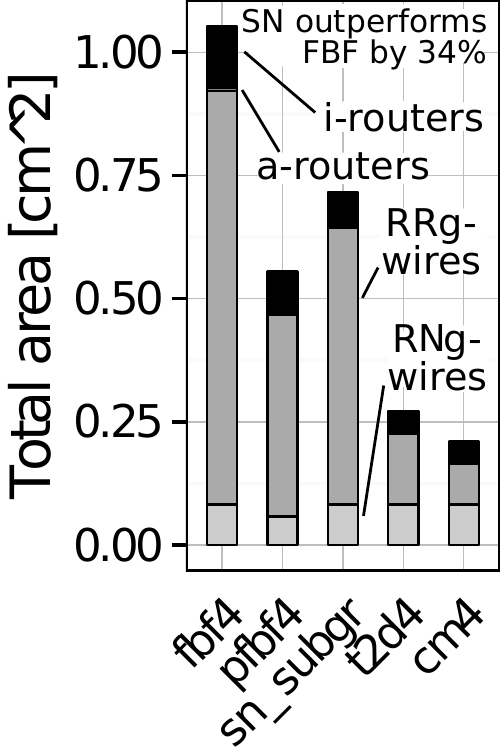}
  \label{fig:nsap2}
  }\hfill
 \subfloat[Static power consumption.]{
  \includegraphics[width=0.13\textwidth]{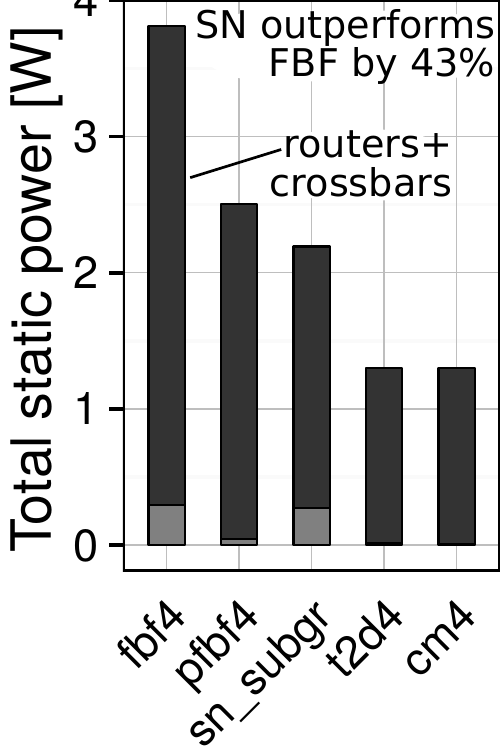}
  \label{fig:nsap3}
 }
\caption{Area/power analysis without SMART, $N=200$ (\cref{sec:area_pwr_analysis}).}
\label{fig:no_smart_area_power}
\end{figure}

\subsection{Analysis of Area and Power Consumption}
\label{sec:area_pwr_analysis}

\change{(§7.4)
As in other parts of §7, we mainly: (1) improve the writing style, (2)
dedensify the text, (3) make the discussion of the results more precise
(avoiding any vague and non-numerical adjectives).}

\change{(§7.4) We provide the citation to our technical report for the area and power consumption
for $N=1296$.}


We first briefly analyze area and power differences between various \texttt{SN} layouts. As predicted,
\texttt{sn\_subgr} outperforms others; for example, see
Figure~\ref{fig:nsap1}.



Figures~\ref{fig:nsap2}--\ref{fig:nsap3} present \texttt{SN}'s advantages
for $N \in \{192,200\}$ \emph{without SMART and central buffers. These are
  gains from the proposed layouts}.  \texttt{SN} significantly outperforms
  \texttt{FBF} in all the evaluated metrics, and \texttt{PFBF} in consumed
  power. \texttt{PFBF}'s area is smaller; we later use SMART to alleviate this.



\hl{
%
%
Similarly to $N \in \{192,200\}$, \texttt{SN} with $N=1296$ reduces area (by up to
$\approx$33\%) and power consumption (by up to $\approx$55\%) compared to \texttt{FBF}.
%
%
%
An exception is \texttt{pfbf9} as it improves upon \texttt{SN} in
both metrics (by $\approx$10-15\%).
Yet, \texttt{SN}'s higher throughput improves the power/performance
tradeoff by $\approx$24\% (more details in~\mbox{\cref{sec:real_traces_eval}}).
Thus, \texttt{SN} outperforms \texttt{FBF} (in area and power
consumption) and \texttt{PFBF},
\texttt{CM}, and \texttt{T2D} (in power/performance) in designs with
$N>1000$.
}

\begin{figure}[!h]
\centering
 \subfloat[Area.]{
  \includegraphics[width=0.155\textwidth]{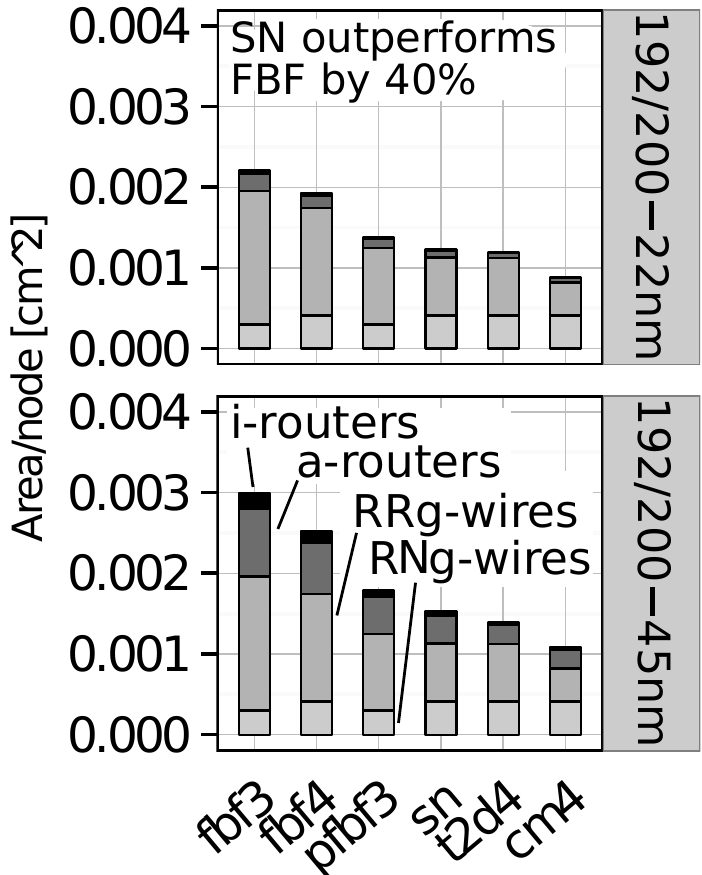}
  \label{fig:ap1}
 }
 \subfloat[Static power.]{
  \includegraphics[width=0.155\textwidth]{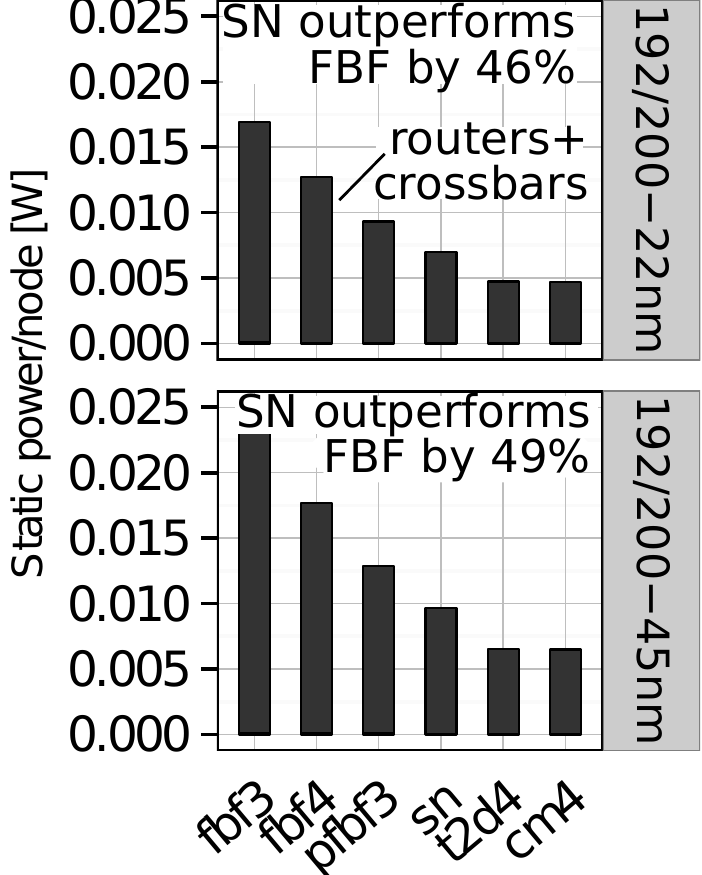}
  \label{fig:ap2}
 }
 \subfloat[Dynamic power.]{
  \includegraphics[width=0.155\textwidth]{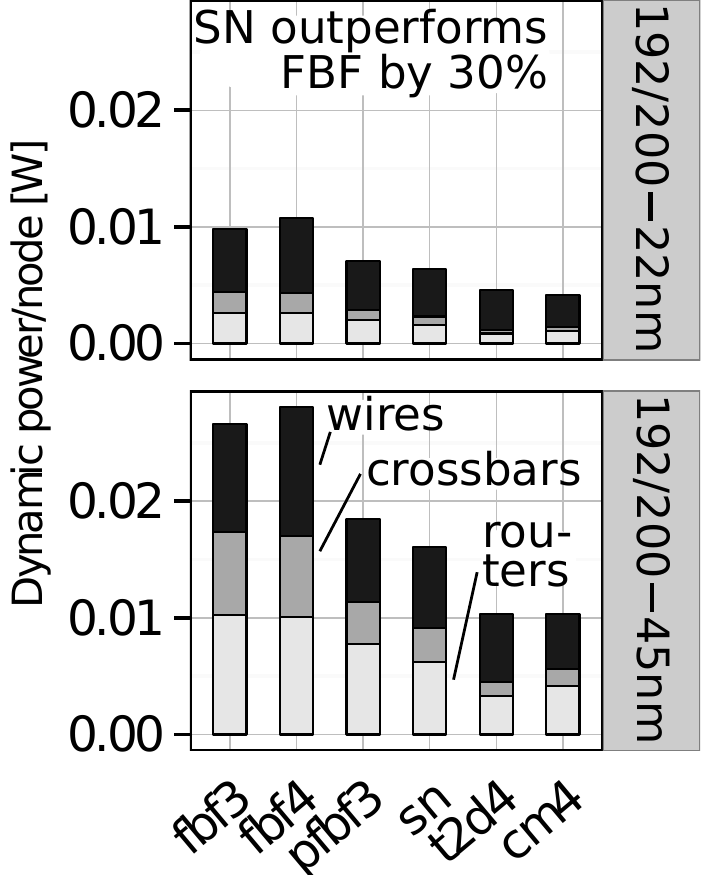}
  \label{fig:ap3}
 }
%
\caption{(\cref{sec:area_pwr_analysis}) Area/power analysis with SMART links, $N \in \{192,200\}$.}
\label{fig:area_power_analysis_small}
\end{figure}

\begin{figure}[!h]
\centering
 \subfloat[Area.]{
  \includegraphics[width=0.155\textwidth]{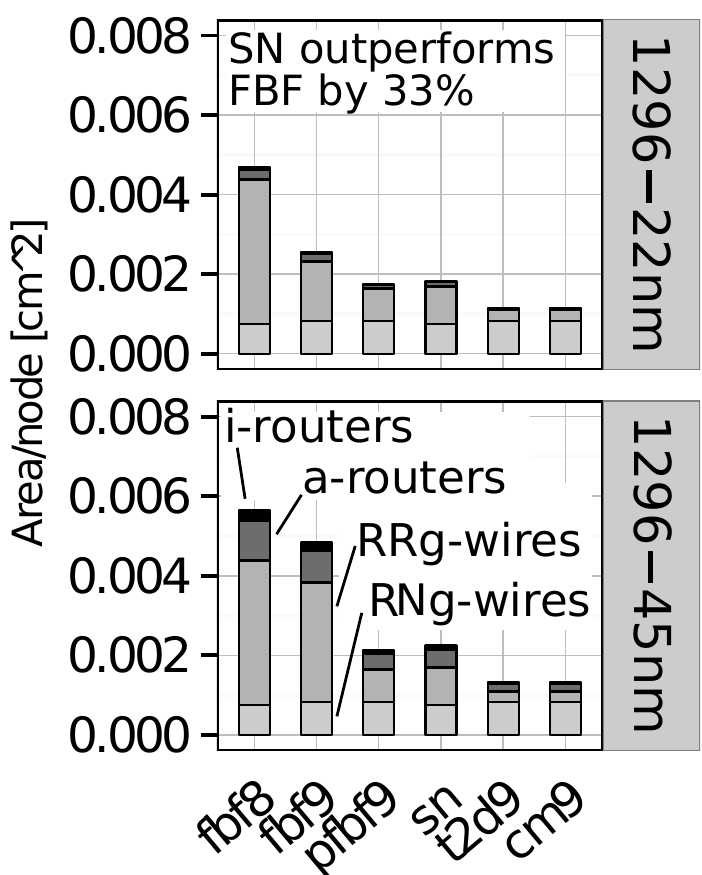}
  \label{fig:ap5}
 }
 \subfloat[Static power.]{
  \includegraphics[width=0.155\textwidth]{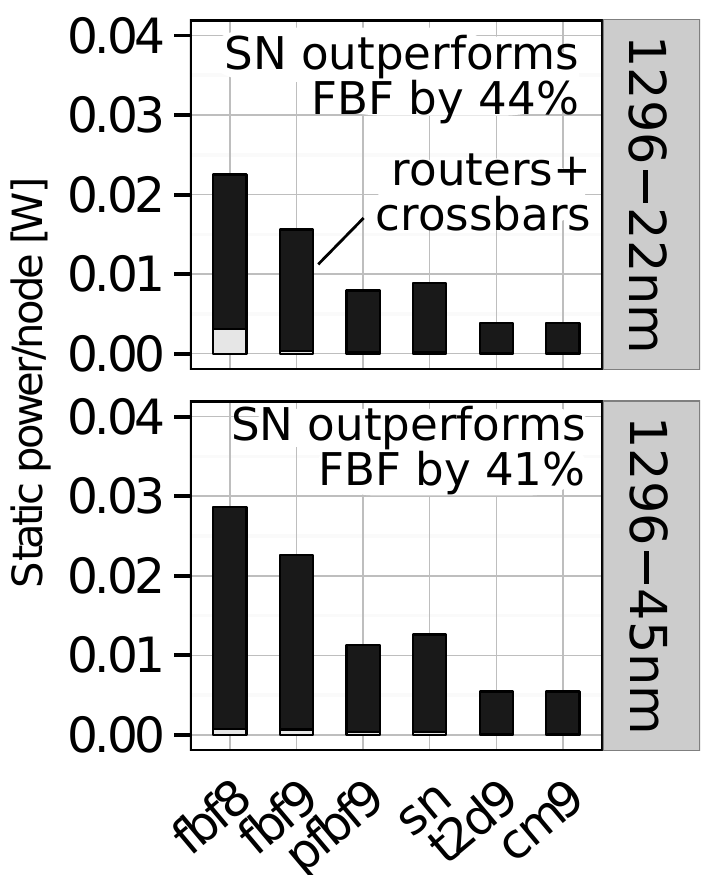}
  \label{fig:ap6}
 }
 \subfloat[Dynamic power.]{
  \includegraphics[width=0.155\textwidth]{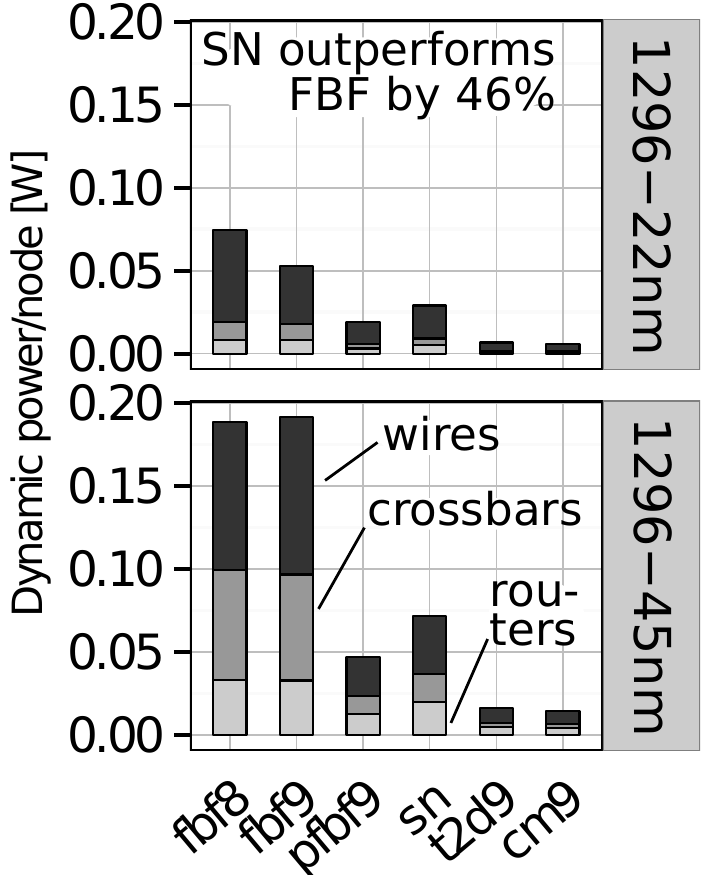}
  \label{fig:ap7}
 }
%
\caption{(\cref{sec:area_pwr_analysis}) Area/power analysis with SMART links, $N=1296$.}
\label{fig:area_power_analysis_large}
\end{figure}

\macb{Impact of SMART Links}
%
Figures~\ref{fig:area_power_analysis_small}--\ref{fig:area_power_analysis_large} shows the effect of SMART links on \texttt{SN}.
\texttt{SN} reduces area over
\texttt{FBF} ($\approx$40-50\%) and \texttt{PFBF} ($\approx$9\%)
as it ensures the lowest $k'$ for a given $D$, reducing
the area due to fewer buffers and ports as well as smaller
crossbars.
Low-radix networks deliver the lowest areas but they also entail a
worse power/performance tradeoff as shown
in~\mbox{\cref{sec:real_traces_eval}}.
%
%
%
Finally, static/dynamic power consumption follows similar trends as the area.
For example, \texttt{SN} reduces static power over both \texttt{FBF} ($\approx$45-60\%) and \texttt{PFBF}
($\approx$14-27\%),
%
%
%
as a consequence of providing the
lowest $k$ for a given $D$.

\subsection{Analysis of the Performance--Power Tradeoff}
\label{sec:real_traces_eval}

\change{(§7.5)
As in other parts of §7, we mainly: (1) improve the writing style, (2)
make the text less dense, (3) make the discussion of the results more precise
(avoiding any vague and non-numerical adjectives).}

\change{(§7.5) We provide the reference to Table (Table 4) in the throughput discussion and
Figure (Figure 15) in the energy--delay discussion.}


We demonstrate that \texttt{SN} provides the best tradeoff between performance and power
out of all the topologies.
%

\macb{Throughput/Power}
\hlgreen{
Table~\ref{tab:power-tradeoffs} shows
\texttt{SN}'s relative improvements over other topologies in 
the throughput delived per unit of consumed power. To calculate this
metric, we divide the number of flits delivered in a cycle by the power consumed
during this delivery.
\texttt{SN} outperforms all the designs; the lowest gain is over \texttt{FBF}
due to its high throughput ($\approx$5-12\%) and the highest over low-radix networks
($\ge$50\%).
Thus, Slim NoC achieves the sweetspot between power consumption
and performance (for random traffic).
}

\begin{table}[h!]
\centering
\footnotesize
\sf
 \setlength{\tabcolsep}{1.2pt} 
     \renewcommand{\arraystretch}{0.9} 
\begin{tabular}{l||lllll||lllll}
\toprule
 & \multicolumn{5}{c||}{$N \in \{192,200\}$}                     & \multicolumn{5}{c}{$N = 1296$}                        \\ 
       &  t2d4 & cm4 & pfbf3 & fbf3 & fbf4                                & t2d9 & cm9 & pfbf9 & fbf8 & fbf9                                \\\midrule
 45nm  & 96\% & 97\% & 17\% & 12\%  & 6\%                                & 155\% & 235\% & 38\% & 54\% & 52\%    \\
 22nm &  209\% & 199\% & 17\% & 14\% & 5\%                               & 182\% & 273\% & 43\% & 53\% & 8\% \\\bottomrule
\end{tabular}
\caption{(\cref{sec:real_traces_eval}) \texttt{SN}'s advantages in
throughput/power (the \texttt{RND} traffic).  The percentages are \texttt{SN}'s
relative improvements over other topologies in the throughput delived per unit
of consumed power. To calculate this metric, we divide the number of flits
delivered in a cycle by the power consumed during this delivery.
}
\label{tab:power-tradeoffs}
\end{table}

\begin{figure}[h!]
 \centering
 \includegraphics[width=0.9\linewidth]{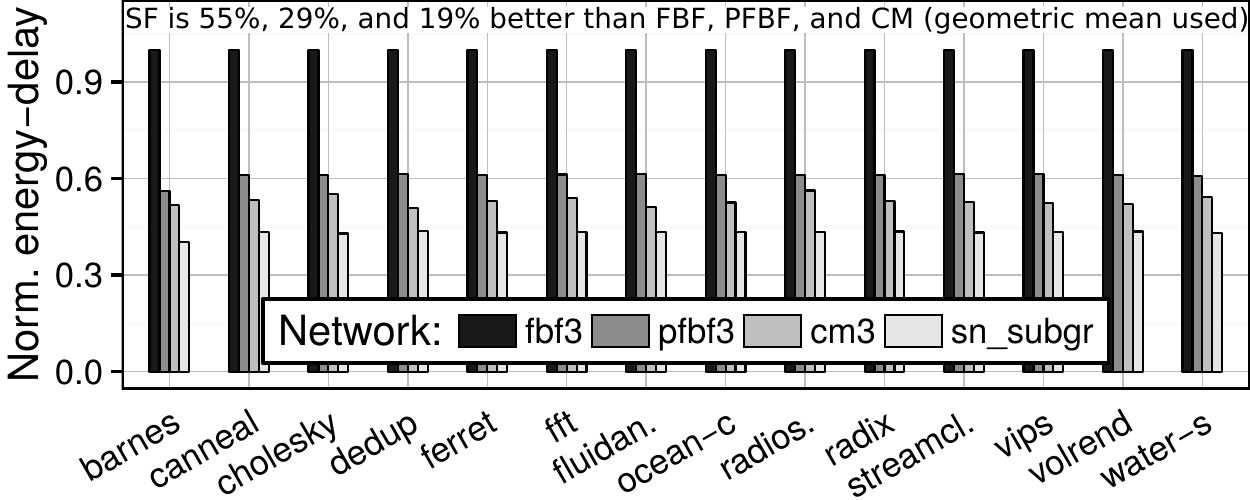}
 \captionof{figure}{(\cref{sec:real_traces_eval}) Energy-Delay Product analysis (with SMART).}
 \label{fig:traces-power-perf}
\end{figure}

\macb{Energy-Delay}
%
%
\hl{
Figure~\ref{fig:traces-power-perf} shows the normalized energy-delay product (EDP) results (for PARSEC/SPLASH traces)
with respect to \texttt{FBF}.
\texttt{SN} reduces EDP by $\approx$55\% on average (geometric mean)
compared to \texttt{FBF} as it 
consumes less static and dynamic power.
\texttt{SN}'s EDP is also $\approx$29\% smaller than
that of \texttt{PFBF} due to the latter's higher latencies and higher power
consumption.
Similarly, \texttt{SN} reduces EDP by $\approx$19\% compared to \texttt{CM}.
%
%
}

%
%



\subsection{Further Analysis: A Summary}
\label{sec:area_power_others}

\change{(§7.6) As in other parts of §7, we mainly: (1) improve the writing
style, (2) make the text less dense, (3) make the discussion of the results
more precise (avoiding vague and non-numerical adjectives).}

\change{(§7.6) We refer readers to our technical report for additional
results.}


We summarize our analysis of the influence of other parameters. 
%


\macb{Hierarchical NoCs}
\hlgreen{ Although we focus on direct symmetric topologies, we also compare
\texttt{SN} to a folded Clos~\cite{Scott:2006:BHC:1135775.1136488} that
represents hierarchical indirect networks such as fat trees or
Kilo-core~\cite{abeyratne2013scaling}.
\texttt{SN} retains its lower area benefits. For example, its area
is $\approx$24\% and $\approx$26\% smaller for $N=200$ and $N=1296$, respectively.  }

\macb{Other Network Sizes}
\hlgreen{
In addition to $N \in \{200, 192, 1296\}$, we analyzed
other systems where $N \in \{588, 686, 1024\}$.
\texttt{SN}'s advantages are consistent.
}

\macb{Global vs. Intermediate Wires}
\hlgreen{
Both types of wires result in the same advantages of \texttt{SN} over other networks.
}

\macb{Injection Rate}
\hlgreen{
Consumed dynamic power is proportional to injection rates; \texttt{SN}
retains its advantages for low and high rates.
}

\macb{45nm vs. 22nm}
%
Both technologies entail similar trends;
the only difference is that wires use relatively more area and power in 22nm than in
45nm (see Figures~\ref{fig:area_power_analysis_small}--\ref{fig:area_power_analysis_large}).


\macb{Concentration}
\hlgreen{
\texttt{SN} outperforms other designs for various $p$ 
($p \in \{3,4\}$ for $N \in \{192,200\}$ and $p \in \{8,9\}$ for $N \in \{1024,1296\}$).
}





\subsection{Analysis of Today's Small-Scale Designs}
\label{sec:tiny_analysis}



\begin{figure}[h!]
\centering
 \subfloat[Latency]{
  \includegraphics[width=0.14\textwidth]{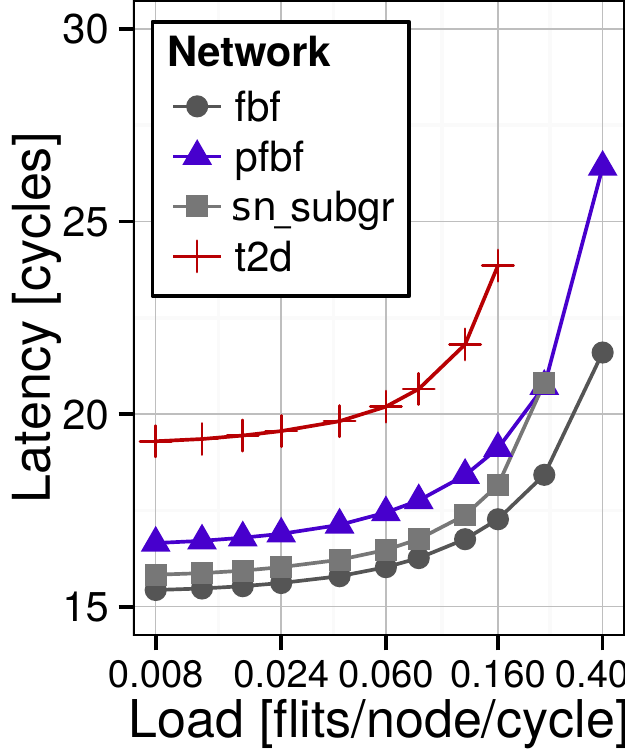}
  \label{fig:tiny_lat}
 }
 \subfloat[Area]{
  \includegraphics[width=0.14\textwidth]{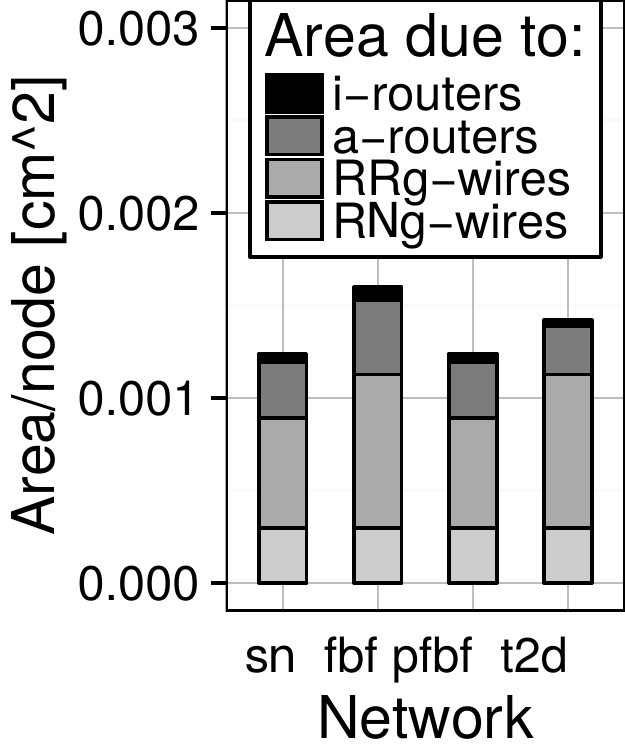}
  \label{fig:tiny_area}
  }
 \subfloat[Dynamic power]{
  \includegraphics[width=0.14\textwidth]{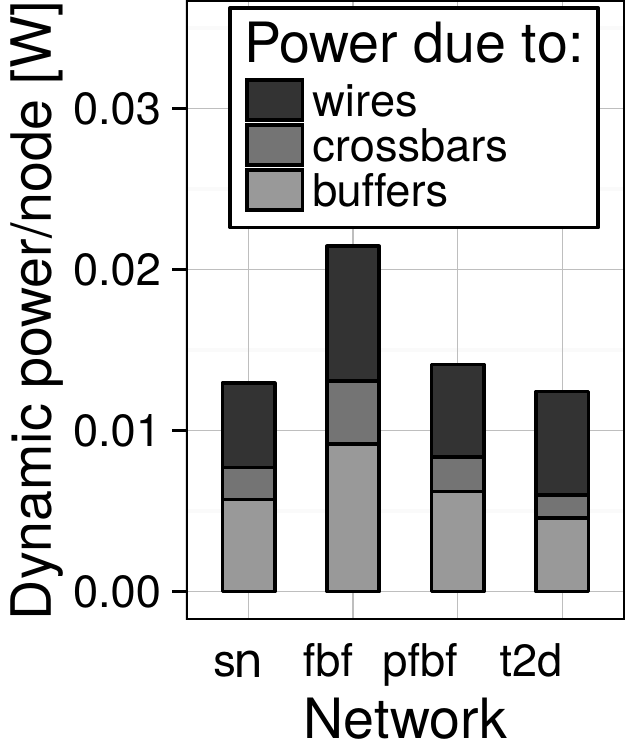}
  \label{fig:tiny_dyn}
 }
\caption{(\cref{sec:tiny_analysis}) Analysis of small-scale designs (for $N=54$).}
\label{fig:tiny_analysis}
\end{figure}

\texttt{SN} specifically targets massively parallel chips. Yet, we
also briefly discuss its advantages in today's small-scale designs
($N=54$), used in, e.g., Intel's Knights Landing
(KNL)~\cite{sodani2015knights}. See Figure~\ref{fig:tiny_analysis} for
representative results (45nm, SMART).
The power/performance tradeoff is similar to that of higher $N$.
\texttt{SN} has lower latency than \texttt{T2D} (by $\approx$15\%) and
\texttt{PFBF} (by $\approx$5\%). It uses less power (by $\approx$40\%) and
area (by $\approx$22\%) than
\texttt{FBF} and has advantages over
\texttt{PFBF}/\texttt{T2D} by $\approx$1-5\% in both metrics.

\macb{Conclusion}
Slim NoC retains advantages in small-scale systems where $N$ is small.  Slim
NoC's advantages become larger as $N$ grows (cf.~\mbox{\cref{sec:evaluation}}).
Thus, Slim NoC is likely to become an even more competitive NoC in the
foreseeable future.

\section{SUMMARY OF RESULTS \& DISCUSSION}
\label{sec:DISNUS}


\commt{(§8) In this section, we clarify the type of synthetic traffic (random uniform
traffic) that is used to obtain the data for our discussion in Section 8. We
also provide a brief conclusion of our findings at the end of the section.}

We now summarize \texttt{SN}'s advantages for $N \in
\{192,200\}$ at 45nm. We use the uniform random traffic for illustrating
latency and throughput, and PARSEC/SPLASH benchmarks for analyzing EDP.

%



%

\macb{SN vs.~Low-Radix Networks}
\hlgreen{
\texttt{SN} uses more area ($>$27\%) and static/dynamic power
($>$40/60\%) than \texttt{T2D} and \texttt{CM}, but significantly
lowers latency ($>$30\%) and increases throughput (3x). Thus,
\texttt{SN} improves throughput/power ratio and ED product of 
selected low-radix designs by $>$95\% and $\approx$19\%.
}

\macb{SN vs.~High-Radix Networks}
\hlgreen{
Compared to \texttt{FBF}, \texttt{SN} has lower bisection bandwidth ($\approx$60\%) and
has similar latency, but it significantly reduces area ($>$36\%), static power
($>$49\%), and dynamic power ($>$39\%).  Thus, it improves the
throughput/power ratio ($\approx$5\%) and especially EDP
($\approx$55\%).
}

\macb{SN vs.~Same-Radix Networks}
\hl{
\texttt{SN} delivers a better power/performance tradeoff than
\texttt{PFBF} that has comparable bisection bandwidth and radix. \texttt{SN} reduces latency
($\approx$13\%), area
($>$9\%), static power ($>$25\%), and dynamic power ($>$9\%), thereby improving 
throughput/power ($\approx$15\%) and EDP ($\approx$15\%).
}

\macb{Impact of SMART, CBR}
\hl{
Relative differences between \texttt{SN} and other networks are \emph{not}
vastly affected by SMART/CBR. For example, without SMART, \texttt{SN}
uses $\approx$42\% less static power than \texttt{FBF}
(see Figure~\ref{fig:nsap3}), similarly to the difference in static power with 
SMART (see Figure~\ref{fig:ap2}).
We observe (in Table~\ref{tab:sm-vs-nosm}) that the average
(geometric mean) gain from SMART in the average packet latency of each topology is $\approx$7.6\%
(\texttt{FBF}), $\approx$0\% (\texttt{CM}), $\approx$8\% (\texttt{PFBF}), and $\approx$11.3\%
(\texttt{SN}).
We conclude that 
\texttt{SN} is very synergistic with SMART and CBRs.
}

\begin{table}[h!]
\centering
\footnotesize
\sf
 \setlength{\tabcolsep}{1pt} 
     \renewcommand{\arraystretch}{0.9} 
\begin{tabular}{l||llllllllllllll}
\toprule
    & bar. & can. & cho. & de. & fer. & fft & fl. & oc. & radio. & radi. & str. & vip. & vol. & wat. \\ 
\midrule
fbf3 & 7.7 & 8.1 & 6.6 & 7.3 & 7.3 & 8.5 & 7.3 & 8.5 & 9.1 & 8.2 & 7.2 & 8 & 7.4 & 6.9 \\ 
pfbf3 & 9.2 & 8.7 & 6.8 & 7.5 & 7.6 & 9.2 & 7.3 & 7.8 & 9.8 & 8.6 & 7.4 & 8.3 & 7.6 & 7.3 \\ 
cm3 & 0 & 0 & 0 & 0 & 0 & 0 & 0 & 0 & 0 & 0 & 0 & 0 & 0 & 0 \\ 
sn & 13.2 & 11.7 & 9.8 & 10.5 & 10.6 & 12.6 & 10.4 & 10.9 & 13.6 & 11.8 & 10.6 & 11.6 & 10.7 & 10.3 \\ 
%
%
\bottomrule
\end{tabular}
\caption{(\cref{sec:DISNUS}) Percentage decrease in the average packet latency due to SMART, $N = 192$, PARSEC/SPLASH, \texttt{SN} uses \textsf{sn\_subgr}.}
\label{tab:sm-vs-nosm}
\end{table}




\macb{High-Level Key Observations}
\hl{
Slim NoC achieves a sweetspot in the combined power/performance metrics.
It is comparable to or differs negligibly from each compared topology
in some metrics (area and power consumption for low-radix and performance (average packet latency, throughput) for
high-radix comparison points). It outperforms every other topology 
in other metrics (performance for low-radix and area as well as
power consumption for high-radix topologies). \emph{\texttt{SN}
outperforms the other networks in combined power/performance 
metrics, i.e., throughput/power and EDP.}
%
}


\hlgreen{
The reasons for \texttt{SN}'s advantages are as stated in~\mbox{\cref{sec:intro}}. First, it
minimizes $k'$ (thus reducing buffer space) for fixed $D=2$ (ensuring low latency) and $N$ (enabling high scalability).
Next, it uses \emph{non-prime finite fields} (enabling more configurations).
Third, it offers optimized layouts (reducing wire lengths and buffer areas).
Finally, it incorporates
mechanisms such as SMART or CBR (further reducing buffer areas).
We conclude that a combination of all these benefits, 
enabled by Slim NoC, leads to a highly-efficient and scalable 
substrate as our evaluations demonstrate.
}

\begin{figure}[h!]
\centering
 \subfloat[Uniform Random]{
  \includegraphics[width=0.23\textwidth]{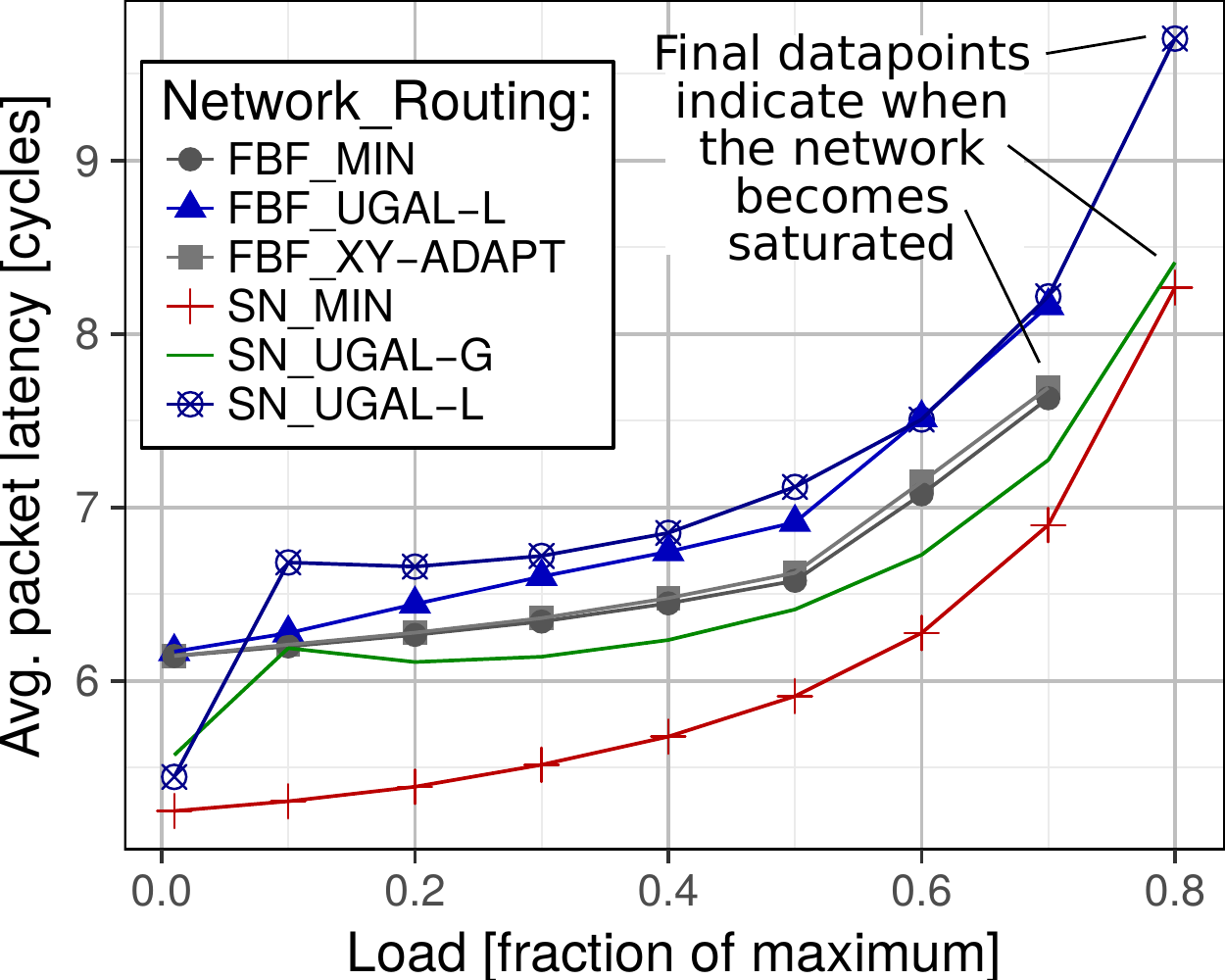}
  \label{fig:adapt_uniform}
 }
 \subfloat[Asymmetric]{
  \includegraphics[width=0.23\textwidth]{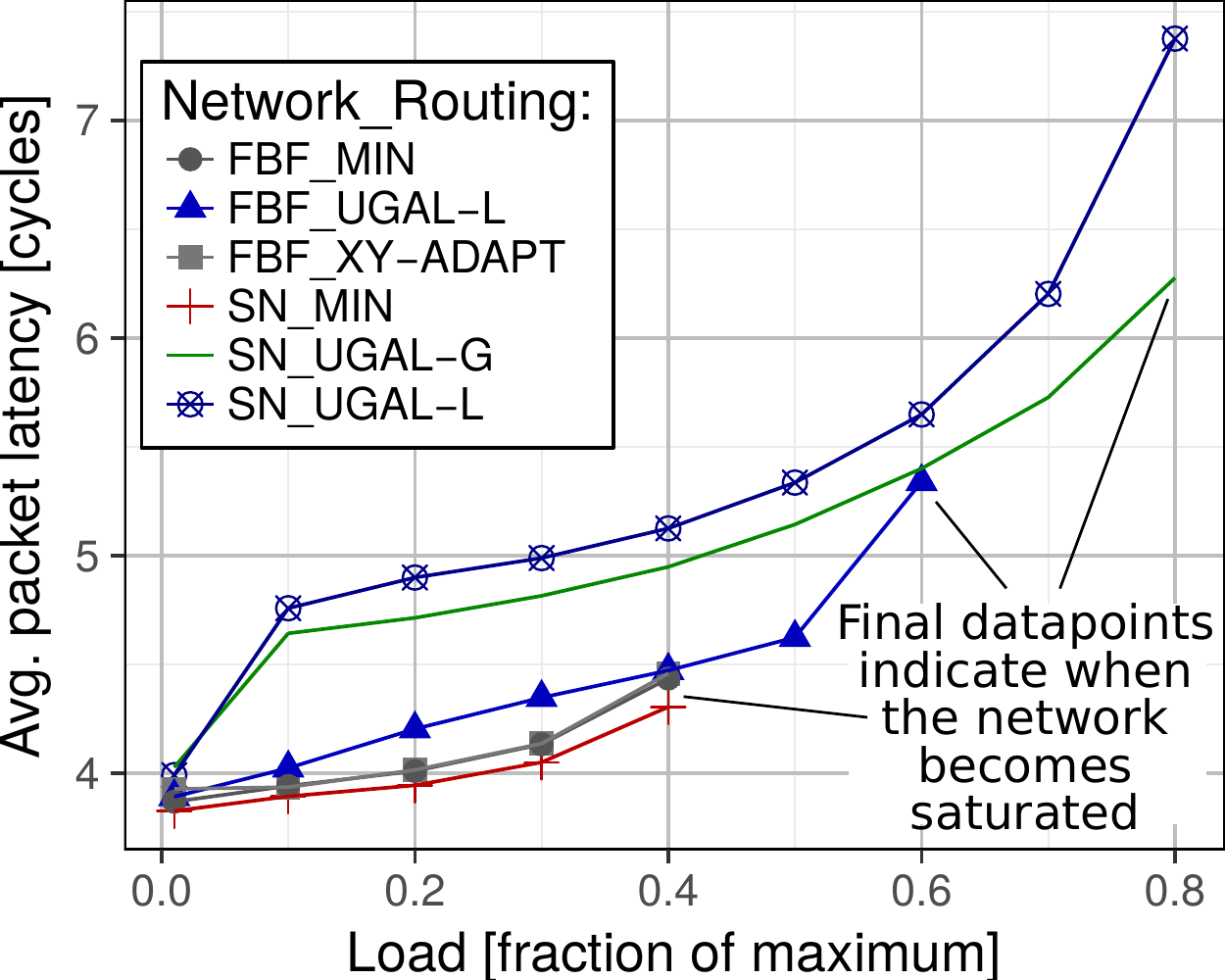}
  \label{fig:adapt_asymmetric}
  }
\caption{\hlcyan{Preliminary analysis of adaptive routing performance (for $N=200$). Each considered
position in a legend has the form ``A\_B'', where ``A'' is the network acronym while ``B'' is 
the routing scheme acronym.
The final data points in each considered combination of network and routing scheme indicate the last
evaluated scenario before network
saturation.
}}
\label{fig:adaptive_analysis}
\end{figure}

\macb{Adaptive Routing}
\hlcyan{
%
%
We conduct a preliminary analysis of adaptive routing.
%
%
%
For
this analysis, we use the Booksim simulator~\cite{jiang2010booksim}
that provides full support for adaptive routing.
Both \texttt{SN} and \texttt{FBF} use simple input-queued
routers and do not use any additional mechanisms such as Central Buffers,
SMART, or Elastic Links. The simulations use 200 nodes.
We analyze \texttt{SN}'s performance with the UGAL protocol~\cite{ugal2005scheme}. We consider
two UGAL variants, local (UGAL-L) and global (UGAL-G). In the former, routers can only access the sizes
of their local queues. In the latter, routers have access to the sizes of all the queues in the network.
We compare \texttt{SN} to \texttt{FBF} that uses two different adaptive schemes~\cite{dally07}: UGAL (a global variant)
and an XY adaptive protocol (denoted as XY-ADAPT) that adaptively selects one of available shortest paths~\cite{dally07}.
%
%
For an additional
comparison, we also plot the minimum static routing latency (MIN). Two traffic
patterns are used: \emph{uniform random} and \emph{asymmetric}, where, for source endpoint
$s$, destination $d$ is (with identical probabilities of $\frac{1}{2}$) equal
to either $d = (s \mod \frac{N}{2}) + \frac{N}{2}$ or $d = (s \mod
\frac{N}{2})$. The results are shown in Figure~\ref{fig:adaptive_analysis}.
For the uniform random traffic, \texttt{SN}'s UGAL-G and MIN outperform their
corresponding schemes in \texttt{FBF} for each injection rate. UGAL-L in
\texttt{SN} provides lower (by $\approx$12\%) latency for the injection rate of
1\%. It is slightly outperformed by \texttt{FBF}'s adaptive schemes for higher
loads (by $\approx$1-2\%). When the load is very high ($\approx$60\%), the
protocols in both topologies become comparable. \texttt{SN} offers negligibly higher
throughput.
For the asymmetric traffic, the performance trends are similar, with the
difference that \texttt{SN}'s UGAL schemes have comparable or higher (by
$\approx$10\%) latency than those of \texttt{FBF} but they provide higher (by
$>$100\%) throughput.

We conclude that, with the UGAL adaptive routing, \texttt{SN} trades latency
for higher throughput over \texttt{FBF} with the asymmetric traffic. Under the random traffic, its latency is better 
than that of \texttt{FBF} under very low ($\approx$1\%) injection rates and becomes higher
for higher injection rates. 
%
%
Finally, in our evaluation, \texttt{SN} uses a general unoptimized UGAL scheme while
\texttt{FBF} incorporates a tuned XY-adaptive scheme.
This suggests that developing optimized adaptive routing protocols
for \texttt{SN} is a productive area of future research.
%
%
}

\section{RELATED WORK}

\commt{(§9) We provide additional citations to relevant works.}

To our knowledge, this is the first work to design a highly scalable
and energy-efficient on-chip network topology 
by solving the problems of adapting state-of-the-art off-chip topologies
to the on-chip context, using key notions from graph theory and number theory.
We discuss how Slim NoC (\texttt{SN}) differs from major related works.

\macb{SN vs Slim Fly}
\hlgreen{
\texttt{SN} is inspired by the rack-level Slim Fly~\cite{slim-fly} (\texttt{SF}) in that it
approaches the optimal tradeoff between radix, network size, and diameter by
incorporating the underlying MMS graphs~\cite{mckay98, slim-fly}. In contrast
to \texttt{SF}, \texttt{SN}: (1) 
uses \emph{non-prime
finite fields} for more viable NoC configurations, (2) provides cost models and
layouts suitable for NoC settings, (3) takes advantage of various modern
architectural optimizations such as Central Buffers~\cite{6558397}, and (4)
resolves deadlock-freedom in NoC settings.  Consequently, \texttt{SN} exploits 
\texttt{SF}'s topological advantages and enables its adaptation to the on-chip constraints for an effective
on-chip network design.
}

\sloppypar{
\macb{SN vs Other NoCs}
\hlgreen{
Other topologies that reduce area and power consumption or maximize performance
were proposed, both low-radix (rings~\cite{hird-journal, hird},
tori~\cite{Alverson:2010:GSI:1901617.1902283},
meshes~\cite{balfour2006design}) and high-radix (Flattened
Butterflies~\cite{dally07}, fully-connected crossbars). Yet, the former have high 
latency while the latter are power-hungry. The rack-level
HyperX~\cite{Ahn:2009:HTR:1654059.1654101} network extends hypercubes and
Flattened Butterfly; it minimizes cost for fixed bisection bandwidth and radix while
\texttt{SN} fixes the diameter to two, lowering latency. Indirect networks
(Kilo-core~\cite{abeyratne2013scaling} with swizzle
switches~\cite{sewell2012swizzle}, CNoC~\cite{kao2011cnoc}, and hierarchical
NoCs~\cite{hierarchical,hird,hird-journal}) differ from \texttt{SN}, which can be manufactured
easier as a direct and symmetric network with identical routers. Various
fundamentally low-radix designs (EVCs~\cite{express-vc}, MECS~\cite{4798251},
Kilo-NoC~\cite{kilo-noc}, Dodec~\cite{yang2014dodec}, schemes for 3D networks~\cite{xu2009low}, and others~\cite{jain2014high}) limit
throughput at high injection rates; \texttt{SN} ensures a close-to-optimal
radix and diameter tradeoff, ensuring low cost and high performance for both
low and high loads. Finally, an on-chip Dragonfly topology was only considered in the
nanophotonics~\cite{pan2009firefly} and in-memory processing~\cite{ahn2016scalable} contexts.}
}

%

\macb{Optimized NoC Buffers}
Buffer space can be reduced in many ways (sharing among VCs~\cite{damq, damq2}
or ports~\cite{dist-shared-buf, RoShaq, sp2}, reducing VC
count~\cite{bubble,bcs,crit-bubble}, using bubble flow
control~\cite{bubble-sharing,worm-bubble}, using scalable networks within
switches~\cite{Ahn:2008:SHR:2509420.2512433}, or removing buffers
altogether~\cite{craik2011investigating, bless-rtr, chipper, fallin2014bufferless, nychis2010next, nychis2012chip, minbd, xiang2017carpool, xiang2016model, cai2015comparative}). These schemes
are largely orthogonal to \texttt{SN} but they may decrease performance at high
loads~\cite{chipper,bless-rtr,hat-sbac-pad12, nychis2010next, nychis2012chip, xiang2017carpool}. 
%
Our
Elastic Buffer-based Central Buffer routers 
eliminate the non-determinism and extra link traversals due to deflection-based bufferless routing.
%


\macb{Single-Cycle Wires}
\hlgreen{
ViChaR~\cite{nicopoulos2006vichar}, iDEAL~\cite{kodi2008ideal}, or other
schemes for long single-cycle
wires~\cite{manevich2014designing,Chen:2013:SSR:2485288.2485371} or
deadlock-free multi-VC elastic links~\cite{4798250,
Seitanidis:2014:EEB:2616606.2616900} can also enhance \texttt{SN}.
%
%
}

\section{CONCLUSION}


\commt{(§10) We significantly enhance the writing style and improve clarity.}

We introduce Slim NoC (SN), a new family of low-diameter on-chip networks (NoCs) that minimize area and
power consumption while providing high performance at both low and high loads.
Slim NoC extends the state-of-the-art rack-level Slim Fly topology to the on-chip setting. We
identify and preserve Slim Fly's attractive properties and develop mechanisms to overcome its
significant overheads in the NoC setting. In particular,
we introduce mathematically rigorous router placement schemes,
use non-prime finite fields to generate underlying graphs, thereby producing feasible on-chip layouts, and shift the
optimization goal to minimizing radix for a fixed core count.
Finally, we augment \texttt{SN} with state-of-the-art mechanisms such as Central
Buffer routers, ElastiStore, and SMART links.

We show that Slim NoC can be an effective and feasible on-chip network design for both small-scale and large-scale 
future chips with tens, hundreds, and thousands of cores.
Our evaluations show 
that Slim NoC significantly improves both performance and energy efficiency 
for regular and irregular
workloads~\cite{besta2018log, besta2017slimsell, tate2014programming, schweizer2015evaluating} over cutting-edge network topologies.
We believe and hope that our approach based on combining mathematical optimization with
state-of-the-art engineering will result in other highly-scalable
and energy-efficient on-chip network designs.

\section*{Acknowledgments}

We thank our shepherd, Abhishek Bhattacharjee, 
for his valuable comments.
We acknowledge insightful feedback from all the reviewers.
We thank Hussein Harake, Colin McMurtrie, and the whole CSCS team granting
access to the Greina, Piz Dora, and Daint machines, and for their excellent
technical support.
We acknowledge extensive support about the MIT-DSENT tool from Chen Sun.
%

{
\bibliographystyle{abbrv}
\bibliography{references_inprocs_to_articles}

\begin{thebibliography}{10}

\bibitem{abeyratne2013scaling}
N.~Abeyratne, R.~Das, Q.~Li, K.~Sewell, B.~Giridhar, R.~G. Dreslinski,
  D.~Blaauw, and T.~Mudge.
\newblock {Scaling Towards Kilo-Core Processors with Asymmetric High-Radix
  Topologies}.
\newblock {\em HPCA}, 2013.

\bibitem{sp2}
T.~Agerwala, J.~Martin, J.~Mirza, D.~Sadler, D.~Dias, and M.~Snir.
\newblock {SP2 System Architecture}.
\newblock {\em IBM Systems Journal}, 1995.

\bibitem{ahn2016scalable}
J.~Ahn, S.~Hong, S.~Yoo, O.~Mutlu, and K.~Choi.
\newblock {A Scalable Processing-in-Memory Accelerator for Parallel Graph
  Processing}.
\newblock {\em ISCA}, 2015.

\bibitem{Ahn:2009:HTR:1654059.1654101}
J.~H. Ahn, N.~Binkert, A.~Davis, M.~McLaren, and R.~S. Schreiber.
\newblock {HyperX: Topology, Routing, and Packaging of Efficient Large-Scale
  Networks}.
\newblock {\em SC}, 2009.

\bibitem{Ahn:2008:SHR:2509420.2512433}
J.~H. Ahn, Y.~H. Son, and J.~Kim.
\newblock {Scalable High-Radix Router Microarchitecture Using a Network Switch
  Organization}.
\newblock {\em ACM TACO}, 2008.

\bibitem{Alverson:2010:GSI:1901617.1902283}
R.~Alverson, D.~Roweth, and L.~Kaplan.
\newblock {The Gemini System Interconnect}.
\newblock {\em HOTI}, 2010.

\bibitem{hird}
R.~Ausavarungnirun, C.~Fallin, X.~Yu, K.~Chang, G.~Nazario, R.~Das, G.~H. Loh,
  and O.~Mutlu.
\newblock {Design and Evaluation of Hierarchical Rings with Deflection
  Routing}.
\newblock {\em SBAC-PAD}, 2014.

\bibitem{hird-journal}
R.~Ausavarungnirun, C.~Fallin, X.~Yu, K.~Chang, G.~Nazario, R.~Das, G.~H. Loh,
  and O.~Mutlu.
\newblock {A Case for Hierarchical Rings with Deflection Routing}.
\newblock {\em PARCO}, 2016.

\bibitem{balfour2006design}
J.~Balfour and W.~J. Dally.
\newblock {Design Tradeoffs for Tiled CMP On-Chip Networks}.
\newblock {\em ICS}, 2006.

\bibitem{besta2014fault}
M.~Besta and T.~Hoefler.
\newblock Fault tolerance for remote memory access programming models.
\newblock In {\em Proceedings of the 23rd international symposium on
  High-performance parallel and distributed computing}, pages 37--48, 2014.

\bibitem{slim-fly}
M.~Besta and T.~Hoefler.
\newblock {Slim Fly: A Cost Effective Low-Diameter Network Topology}.
\newblock {\em SC}, 2014.

\bibitem{besta2015accelerating}
M.~Besta and T.~Hoefler.
\newblock Accelerating irregular computations with hardware transactional
  memory and active messages.
\newblock In {\em Proceedings of the 24th International Symposium on
  High-Performance Parallel and Distributed Computing}, pages 161--172, 2015.

\bibitem{besta2015active}
M.~Besta and T.~Hoefler.
\newblock Active access: A mechanism for high-performance distributed
  data-centric computations.
\newblock In {\em Proceedings of the 29th ACM on International Conference on
  Supercomputing}, pages 155--164, 2015.

\bibitem{besta2017slimsell}
M.~Besta, F.~Marending, E.~Solomonik, and T.~Hoefler.
\newblock Slimsell: A vectorizable graph representation for breadth-first
  search.
\newblock In {\em 2017 IEEE International Parallel and Distributed Processing
  Symposium (IPDPS)}, pages 32--41. IEEE, 2017.

\bibitem{besta2018log}
M.~Besta, D.~Stanojevic, T.~Zivic, J.~Singh, M.~Hoerold, and T.~Hoefler.
\newblock Log (graph) a near-optimal high-performance graph representation.
\newblock In {\em Proceedings of the 27th International Conference on Parallel
  Architectures and Compilation Techniques}, pages 1--13, 2018.

\bibitem{cai2015comparative}
Y.~Cai, K.~Mai, and O.~Mutlu.
\newblock {Comparative Evaluation of FPGA and ASIC Implementations of
  Bufferless and Buffered Routing Algorithms for On-Chip Networks}.
\newblock {\em ISQED}, 2015.

\bibitem{6831831}
A.~Ceyhan, M.~Jung, S.~Panth, S.~K. Lim, and A.~Naeemi.
\newblock {Impact of Size Effects in Local Interconnects for Future Technology
  Nodes: A Study Based on Full-Chip Layouts}.
\newblock {\em IITC/AMC}, 2014.

\bibitem{hat-sbac-pad12}
K.~K.-W. Chang, R.~Ausavarungnirun, C.~Fallin, and O.~Mutlu.
\newblock {HAT: Heterogeneous Adaptive Throttling for On-Chip Networks}.
\newblock {\em SBAC-PAD}, 2012.

\bibitem{Chen:2013:SSR:2485288.2485371}
C.-H.~O. Chen, S.~Park, T.~Krishna, S.~Subramanian, A.~P. Chandrakasan, and
  L.-S. Peh.
\newblock {SMART: A Single-Cycle Reconfigurable NoC for SoC Applications}.
\newblock {\em DATE}, 2013.

\bibitem{worm-bubble}
L.~Chen and T.~M. Pinkston.
\newblock Worm-bubble flow control.
\newblock {\em HPCA}, 2013.

\bibitem{crit-bubble}
L.~Chen, R.~Wang, and T.~Pinkston.
\newblock {Critical Bubble Scheme: An Efficient Implementation of Globally
  Aware Network Flow Control}.
\newblock {\em IPDPS}, 2011.

\bibitem{craik2011investigating}
C.~Craik and O.~Mutlu.
\newblock {Investigating the Viability of Bufferless NoCs in Modern Chip
  Multi-Processor Systems}.
\newblock {\em CMU Safari Technical Report}, 2011.

\bibitem{Dally:2003:PPI:995703}
W.~Dally and B.~Towles.
\newblock {\em {Principles and Practices of Interconnection Networks}}.
\newblock Morgan Kaufmann Publishers Inc., 2003.

\bibitem{hierarchical}
R.~Das, S.~Eachempati, A.~Mishra, V.~Narayanan, and C.~Das.
\newblock {Design and Evaluation of a Hierarchical On-Chip Interconnect for
  Next-Generation CMPs}.
\newblock {\em HPCA}, 2009.

\bibitem{das_micro09}
R.~Das, O.~Mutlu, T.~Moscibroda, and C.~Das.
\newblock {Application-Aware Prioritization Mechanisms for On-Chip Networks}.
\newblock {\em MICRO}, 2009.

\bibitem{aergia}
R.~Das, O.~Mutlu, T.~Moscibroda, and C.~R. Das.
\newblock {A\'{e}rgia: Exploiting Packet Latency Slack in On-Chip Networks}.
\newblock In {\em ISCA}, 2010.

\bibitem{dongarra1979linpack}
J.~J. Dongarra, C.~B. Moler, J.~R. Bunch, and G.~W. Stewart.
\newblock {\em {LINPACK Users' Guide}}.
\newblock SIAM, 1979.

\bibitem{tile-mx100}
{EZchip Semiconductor Ltd.}
\newblock {EZchip Introduces TILE-Mx100 World’s Highest Core-Count ARM
  Processor Optimized for High-Performance Networking Applications}.
\newblock {\em http://www.tilera.com/News/PressRelease/?ezchip=97}, 2015.

\bibitem{chipper}
C.~Fallin, C.~Craik, and O.~Mutlu.
\newblock {CHIPPER: A Low-Complexity Bufferless Deflection Router}.
\newblock {\em HPCA}, 2011.

\bibitem{minbd}
C.~Fallin, G.~Nazario, X.~Yu, K.~Chang, R.~Ausavarungnirun, and O.~Mutlu.
\newblock {MinBD: Minimally-Buffered Deflection Routing for Energy-Efficient
  Interconnect}.
\newblock {\em NOCS}, 2012.

\bibitem{fallin2014bufferless}
C.~Fallin, G.~Nazario, X.~Yu, K.~Chang, R.~Ausavarungnirun, and O.~Mutlu.
\newblock {Bufferless and Minimally-Buffered Deflection Routing}.
\newblock {\em Routing Algorithms in Networks-on-Chip}, 2014.

\bibitem{fu2016sunway}
H.~Fu, J.~Liao, J.~Yang, L.~Wang, Z.~Song, X.~Huang, C.~Yang, W.~Xue, F.~Liu,
  F.~Qiao, et~al.
\newblock {The Sunway TaihuLight Supercomputer: System and Applications}.
\newblock {\em Science China Information Sciences}, 2016.

\bibitem{gerstenberger2013enabling}
R.~Gerstenberger, M.~Besta, and T.~Hoefler.
\newblock Enabling highly-scalable remote memory access programming with mpi-3
  one sided.
\newblock In {\em Proceedings of the International Conference on High
  Performance Computing, Networking, Storage and Analysis}, pages 1--12, 2013.

\bibitem{4798251}
B.~Grot, J.~Hestness, S.~Keckler, and O.~Mutlu.
\newblock {Express Cube Topologies for On-Chip Interconnects}.
\newblock {\em HPCA}, 2009.

\bibitem{kilo-noc}
B.~Grot, J.~Hestness, S.~Keckler, and O.~Mutlu.
\newblock {Kilo-NoC: A Heterogeneous Network-on-Chip Architecture for
  Scalability and Service Guarantees}.
\newblock {\em ISCA}, 2011.

\bibitem{6558397}
S.~Hassan and S.~Yalamanchili.
\newblock {Centralized Buffer Router: A Low Latency, Low Power Router for High
  Radix NoCs}.
\newblock {\em NOCS}, 2013.

\bibitem{bubble-sharing}
S.~Hassan and S.~Yalamanchili.
\newblock {Bubble Sharing: Area and Energy Efficient Adaptive Routers using
  Centralized Buffers}.
\newblock {\em NOCS}, 2014.

\bibitem{jain2014high}
A.~Jain, R.~Parikh, and V.~Bertacco.
\newblock {High-Radix On-Chip Networks with Low-Radix Routers}.
\newblock {\em ICCAD}, 2014.

\bibitem{jiang2010booksim}
N.~Jiang, G.~Michelogiannakis, D.~Becker, B.~Towles, and W.~J. Dally.
\newblock {Booksim 2.0 User’s Guide}.
\newblock {\em Standford University}, 2010.

\bibitem{kao2011cnoc}
Y.-H. Kao, M.~Yang, N.~S. Artan, and H.~J. Chao.
\newblock {CNoC: High-Radix Clos Network-on-Chip}.
\newblock {\em TCAD}, 2011.

\bibitem{kim2009low}
J.~Kim.
\newblock {Low-Cost Router Microarchitecture for On-Chip Networks}.
\newblock {\em MICRO}, 2009.

\bibitem{dally07}
J.~Kim, W.~J. Dally, and D.~Abts.
\newblock {Flattened Butterfly: A Cost-Efficient Topology for High-Radix
  Networks}.
\newblock {\em ISCA}, 2007.

\bibitem{dally08}
J.~Kim, W.~J. Dally, S.~Scott, and D.~Abts.
\newblock {Technology-Driven, Highly-Scalable Dragonfly Topology}.
\newblock {\em ISCA}, 2008.

\bibitem{kodi2008ideal}
A.~K. Kodi, A.~Sarathy, and A.~Louri.
\newblock {iDEAL: Inter-Router Dual-Function Energy and Area-Efficient Links
  for Network-on-Chip (NoC) Architectures}.
\newblock {\em ISCA}, 2008.

\bibitem{express-vc}
A.~Kumar, L.-S. Peh, P.~Kundu, and N.~Jha.
\newblock {Toward Ideal On-Chip Communication Using Express Virtual Channels}.
\newblock {\em IEEE Micro}, 2008.

\bibitem{leiserson1985fat}
C.~E. Leiserson.
\newblock {Fat-Trees: Universal Networks for Hardware-Efficient
  Supercomputing}.
\newblock {\em IEEE TC}, 1985.

\bibitem{lidl1997finite}
R.~Lidl and H.~Niederreiter.
\newblock {Finite Fields: Encyclopedia of Mathematics and Its Applications.}
\newblock {\em Comp. \& Math. with Applications}, 33(7):136--136, 1997.

\bibitem{damq2}
J.~Liu and J.~G. Delgado-Frias.
\newblock {A DAMQ Shared Buffer Scheme for Network-on-Chip}.
\newblock {\em CSS}, 2007.

\bibitem{manevich2014designing}
R.~Manevich, L.~Polishuk, I.~Cidon, and A.~Kolodny.
\newblock {Designing Single-Cycle Long Links in Hierarchical NoCs}.
\newblock {\em Microprocessors and Microsystems}, 2014.

\bibitem{mckay98}
B.~D. McKay, M.~Miller, and J.~\v{S}ir\'{a}n.
\newblock {A Note on Large Graphs of Diameter Two and Given Maximum Degree}.
\newblock {\em Journal of Combinatorial Theory, Series B}, 1998.

\bibitem{4798250}
G.~Michelogiannakis, J.~Balfour, and W.~Dally.
\newblock {Elastic-Buffer Flow Control for On-Chip Networks}.
\newblock {\em HPCA}, 2009.

\bibitem{bless-rtr}
T.~Moscibroda and O.~Mutlu.
\newblock {A Case for Bufferless Routing in On-Chip Networks}.
\newblock {\em ISCA}, 2009.

\bibitem{nicopoulos2006vichar}
C.~Nicopoulos, D.~Park, J.~Kim, N.~Vijaykrishnan, M.~S. Yousif, and C.~R. Das.
\newblock {ViChaR: A Dynamic Virtual Channel Regulator for Network-on-Chip
  Routers}.
\newblock {\em MICRO}, 2006.

\bibitem{nychis2010next}
G.~Nychis, C.~Fallin, T.~Moscibroda, and O.~Mutlu.
\newblock {Next Generation On-Chip Networks: What Kind of Congestion Control Do
  We Need?}
\newblock In {\em HotNets}, 2010.

\bibitem{nychis2012chip}
G.~P. Nychis, C.~Fallin, T.~Moscibroda, O.~Mutlu, and S.~Seshan.
\newblock {On-Chip Networks from a Networking Perspective: Congestion and
  Scalability in Many-Core Interconnects}.
\newblock {\em SIGCOMM}, 2012.

\bibitem{olofsson2016epiphany}
A.~Olofsson.
\newblock {Epiphany-V: A 1024 Processor 64-bit RISC System-on-Chip}.
\newblock {\em arXiv preprint arXiv:1610.01832}, 2016.

\bibitem{pan2009firefly}
Y.~Pan, P.~Kumar, J.~Kim, G.~Memik, Y.~Zhang, and A.~Choudhary.
\newblock {Firefly: Illuminating Future Network-on-Chip with Nanophotonics}.
\newblock {\em ISCA}, 2009.

\bibitem{Peh:2001:DMS:580550.876446}
L.-S. Peh and W.~J. Dally.
\newblock {A Delay Model and Speculative Architecture for Pipelined Routers}.
\newblock {\em HPCA}, 2001.

\bibitem{pezy}
{Pezy Computing}.
\newblock {PEZY-SC2}.
\newblock {\em http://pezy.jp}.

\bibitem{Pippenger:1992:FCN:140901.141867}
N.~Pippenger and G.~Lin.
\newblock {Fault-Tolerant Circuit-Switching Networks}.
\newblock {\em SPAA}, 1992.

\bibitem{bubble}
V.~Puente, R.~Beivide, J.~Gregorio, J.~Prellezo, J.~Duato, and C.~Izu.
\newblock {Adaptive Bubble Router: A Design to Improve Performance in Torus
  Networks}.
\newblock {\em ICPP}, 1999.

\bibitem{dist-shared-buf}
R.~Ramanujam, V.~Soteriou, B.~Lin, and L.-S. Peh.
\newblock {Design of a High-Throughput Distributed Shared-Buffer NoC Router}.
\newblock {\em NOCS}, 2010.

\bibitem{Rosenfeld:2011:DCA:1999163.1999216}
P.~Rosenfeld, E.~Cooper-Balis, and B.~Jacob.
\newblock {DRAMSim2: A Cycle Accurate Memory System Simulator}.
\newblock {\em IEEE CAL}, 2011.

\bibitem{schmid2016high}
P.~Schmid, M.~Besta, and T.~Hoefler.
\newblock High-performance distributed rma locks.
\newblock In {\em Proceedings of the 25th ACM International Symposium on
  High-Performance Parallel and Distributed Computing}, pages 19--30, 2016.

\bibitem{schweizer2015evaluating}
H.~Schweizer, M.~Besta, and T.~Hoefler.
\newblock Evaluating the cost of atomic operations on modern architectures.
\newblock In {\em 2015 International Conference on Parallel Architecture and
  Compilation (PACT)}, pages 445--456. IEEE, 2015.

\bibitem{Scott:2006:BHC:1135775.1136488}
S.~Scott, D.~Abts, J.~Kim, and W.~J. Dally.
\newblock {The BlackWidow High-Radix Clos Network}.
\newblock {\em {ISCA}}, 2006.

\bibitem{Seitanidis:2014:EEB:2616606.2616900}
I.~Seitanidis, A.~Psarras, G.~Dimitrakopoulos, and C.~Nicopoulos.
\newblock {ElastiStore: An Elastic Buffer Architecture for Network-on-Chip
  Routers}.
\newblock {\em DATE}, 2014.

\bibitem{sewell2012swizzle}
K.~Sewell, R.~G. Dreslinski, T.~Manville, S.~Satpathy, N.~Pinckney, G.~Blake,
  M.~Cieslak, R.~Das, T.~F. Wenisch, D.~Sylvester, D.~Blaauw, and T.~Mudge.
\newblock {Swizzle-Switch Networks for Many-Core Systems}.
\newblock {\em Emerging and Selected Topics in Circuits and Systems}, 2012.

\bibitem{ugal2005scheme}
A.~Singh.
\newblock {\em {Load-Balanced Routing in Interconnection Networks}}.
\newblock PhD thesis, Stanford University, 2005.

\bibitem{skiena1990dijkstra}
S.~Skiena.
\newblock Dijkstra’s algorithm.
\newblock {\em Implementing Discrete Mathematics: Combinatorics and Graph
  Theory with Mathematica. Addison-Wesley}, 1990.

\bibitem{sodani2015knights}
A.~Sodani.
\newblock {Knights Landing (KNL): 2nd Generation Intel{\textregistered} Xeon
  Phi Processor}.
\newblock {\em HCS}, 2015.

\bibitem{solomonik2017scaling}
E.~Solomonik, M.~Besta, F.~Vella, and T.~Hoefler.
\newblock Scaling betweenness centrality using communication-efficient sparse
  matrix multiplication.
\newblock In {\em Proceedings of the International Conference for High
  Performance Computing, Networking, Storage and Analysis}, pages 1--14, 2017.

\bibitem{strang1993introduction}
G.~Strang.
\newblock {\em Introduction to Linear Algebra}.
\newblock Wellesley-Cambridge Press Wellesley, MA, 1993.

\bibitem{dsent}
C.~Sun, C.~O. Chen, G.~Kurian, L.~Wei, J.~E. Miller, A.~Agarwal, L.~Peh, and
  V.~Stojanovic.
\newblock {DSENT - A Tool Connecting Emerging Photonics with Electronics for
  Opto-Electronic Networks-on-Chip Modeling}.
\newblock {\em NOCS}, 2012.

\bibitem{damq}
Y.~Tamir and G.~Frazier.
\newblock {Dynamically-Allocated Multi-Queue Buffers for VLSI Communication
  Switches}.
\newblock {\em IEEE TC}, 1992.

\bibitem{tate2014programming}
A.~Tate, A.~Kamil, A.~Dubey, A.~Gr{\"o}{\ss}linger, B.~Chamberlain, B.~Goglin,
  C.~Edwards, C.~J. Newburn, D.~Padua, D.~Unat, et~al.
\newblock Programming abstractions for data locality.
\newblock 2014.

\bibitem{RoShaq}
A.~T. Tran and B.~M. Baas.
\newblock {RoShaQ: High-Performance On-Chip Router with Shared Queues}.
\newblock {\em ICCD}, 2011.

\bibitem{udipi2010towards}
A.~N. Udipi, N.~Muralimanohar, and R.~Balasubramonian.
\newblock {Towards Scalable, Energy-Efficient, Bus-Based On-Chip Networks}.
\newblock {\em HPCA}, 2010.

\bibitem{wang2014manifold}
J.~Wang, J.~Beu, R.~Bheda, T.~Conte, Z.~Dong, C.~Kersey, M.~Rasquinha,
  G.~Riley, W.~Song, H.~Xiao, P.~Xu, and S.~Yalamanchili.
\newblock {Manifold: A Parallel Simulation Framework for Multicore Systems}.
\newblock {\em ISPASS}, 2014.

\bibitem{bcs}
R.~Wang, L.~Chen, and T.~M. Pinkston.
\newblock {Bubble Coloring: Avoiding Routing- and Protocol-Induced Deadlocks
  with Minimal Virtual Channel Requirement}.
\newblock {\em ICS}, 2013.

\bibitem{xiang2016model}
X.~Xiang, S.~Ghose, O.~Mutlu, and N.-F. Tzeng.
\newblock {A Model for Application Slowdown Estimation in On-Chip Networks and
  Its Use for Improving System Fairness and Performance}.
\newblock {\em ICCD}, 2016.

\bibitem{xiang2017carpool}
X.~Xiang, W.~Shi, S.~Ghose, L.~Peng, O.~Mutlu, and N.-F. Tzeng.
\newblock {Carpool: A Bufferless On-Chip Network Supporting Adaptive Multicast
  and Hotspot Alleviation}.
\newblock {\em ICS}, 2017.

\bibitem{xu2009low}
Y.~Xu, Y.~Du, B.~Zhao, X.~Zhou, Y.~Zhang, and J.~Yang.
\newblock {A Low-Radix and Low-Diameter 3D Interconnection Network Design}.
\newblock {\em HPCA}, 2009.

\bibitem{yang2014dodec}
H.~Yang, J.~Tripathi, N.~E. Jerger, and D.~Gibson.
\newblock {Dodec: Random-Link, Low-Radix On-Chip Networks}.
\newblock {\em MICRO}, 2014.

\bibitem{yuan2011nonblocking}
X.~Yuan.
\newblock {On Nonblocking Folded-Clos Networks in Computer Communication
  Environments}.
\newblock {\em IPDPS}, 2011.

\end{thebibliography}
}

\end{document}